\documentclass[twocolumn]{aastex63}

\submitjournal{ApJ}

\shortauthors{Vokrouhlick{\' y} et~al.}
\graphicspath{{./}{figures/}}
\usepackage{tikz}
\usetikzlibrary{arrows,shapes,positioning,shadows,trees}

\tikzset{
  basic/.style  = {draw, text width=4cm, drop shadow, font=\sffamily, rectangle},
  root/.style   = {basic, rounded corners=2pt, thin, align=center, fill=blue!10},
  level 2/.style = {basic, rounded corners=6pt, thin,align=center, fill=pink!40, text width=11em},
  level 3/.style = {basic, thin, align=center, fill=white!60, text width=8em}
}
\usepackage[utf8]{inputenc}
\usepackage{dirtytalk}
\usepackage{natbib}
\bibpunct{(}{)}{;}{a}{}{,} % to follow the A&A style

\begin{document}

%\title{Testing MOND at the outskirts of the solar system}
\title{Testing MOND on small bodies in the remote solar system}

\correspondingauthor{David Vokrouhlick\'y}
\email{vokrouhl@cesnet.cz}

\author[0000-0002-6034-5452]{David Vokrouhlick\'y}
\affiliation{Astronomical Institute, Charles University, V Hole\v{s}ovi\v{c}k\'ach 2,
             CZ 18000, Prague 8, Czech Republic}
\author[0000-0002-4547-4301]{David Nesvorn{\' y}}
\affiliation{Department of Space Studies, Southwest Research Institute, 1050 Walnut St., Suite 300,
             Boulder, CO 80302, USA}
\author[0000-0002-0278-7180]{Scott Tremaine}
\affiliation{Canadian Institute for Theoretical Astrophysics, University of Toronto, 60 St. George Street, Toronto, ON M5S 3H8, Canada}
\affiliation{School of Natural Sciences, Institute for Advanced Study, Princeton, NJ 08540, USA}               

\begin{abstract}
Modified Newtonian dynamics (MOND), which postulates a breakdown of Newton’s laws of gravity/dynamics below some critical acceleration threshold, can explain many otherwise puzzling observational phenomena on galactic scales. MOND competes with the hypothesis of dark matter, which successfully explains the cosmic microwave background and large-scale structure. Here we provide the first solar-system test of MOND that probes the sub-critical acceleration regime. Using the Bekenstein–-Milgrom AQUAL formulation, we simulate the evolution of myriads of test particles (planetesimals or comets) born in the trans-Neptunian region and scattered by the giant planets over the lifetime of the Sun to heliocentric distances of $10^2$--$10^5$~au. We include the effects of the Galactic tidal field and passing stars. While Newtonian simulations reproduce the distribution of binding energies of long-period and Oort-cloud comets detectable from Earth, MOND-based simulations do not. This conclusion is robust to plausible changes in the migration history of the planets, the migration history of the Sun, the MOND transition function, effects of the Sun’s birth cluster, and the fading properties of long-period comets. For the most popular version of AQUAL, characterized by a gradual transition between the Newtonian and MOND regimes, our MOND-based simulations also fail to reproduce the orbital distribution of trans-Neptunian objects in the detached disk (perihelion $q>38$~au). Our results do not rule out some MOND theories more elaborate than AQUAL, in which non-Newtonian effects are screened on small spatial scales, at small masses, or in external gravitational fields comparable in strength to the critical acceleration. 

\end{abstract}


% ............................................................................
\section{Introduction} \label{intro}
It has been forty years since the modified Newtonian dynamics (MOND) paradigm was introduced by \citet{m1983a}\footnote{For historical and philosophical perspectives see \citet{s2015} and \citet{mil2020}.}. MOND was primarily developed as an alternative to the dark-matter hypothesis in galactic and
extragalactic astronomy \citep[see, e.g., a thorough review by][]{fm2012}.  At the same time, it is generally accepted in physics that the validity of any theory stems in part from its universality. For this reason astronomers have sought to test MOND on smaller scales, and in particular in the solar system, where dark matter is absent, observations can be done with much higher accuracy, and the circumstances are often more controlled than in galaxies. However, this task is not straightforward since the modifications of Newtonian dynamics \citep[with appropriate relativistic corrections, e.g.,][]{will2014} originating from MOND arise in the regime of very small accelerations ($\leq10^{-10}$ m~s$^{-2}$). These are much smaller than the accelerations in most parts of the solar system that are directly accessible to observations (for comparison, the heliocentric acceleration of the Earth is $6\times 10^{-3}$ m~s$^{-2}$).

A promising option for testing MOND in the solar system is to look at the remote trans-Neptunian region, beyond a heliocentric distance of several thousand astronomical units (au), where MOND predicts an order-unity breakdown of the Newtonian description.%
\footnote{For the sake of completeness, we mention the work of \citet{bma2006} who discuss the emergence of MOND effects in spatially limited regions of the inner solar system near saddle points of the gravitational potential (see also \citealt{m2012,p2020b}).}
Unfortunately, there are no solar-system objects that can currently be detected at these distances, and there is little hope of improvement even from the next-generation sky surveys scheduled for the next decade. Fortunately, there is also the possibility of testing MOND in the solar system through observations of long-period comets (LPCs) and distant trans-Neptunian objects (TNOs). For sake of simplicity, we follow common practice and define LPCs as all comets with heliocentric orbital periods longer than $200$~yr, though we shall be primarily interested in the sub-class with much longer periods (Secs.~\ref{data} and \ref{res2}). For the distant TNOs, objects orbiting beyond Neptune, we specify our adopted criteria in Secs.~\ref{data} and \ref{res1}.

LPCs are believed to be planetesimals that were formed mainly in a massive disk just exterior to the planet Neptune at the birth of the solar system 4.5~Gyr ago (a smaller fraction of them may also originate from the zone between the orbits of the giant planets). Following Neptune's outward radial migration early in the history of the solar system \citep[e.g.,][]{nes2018}, most of the primordial trans-Neptunian disk
was ejected from the solar system. However,
a small fraction was gradually transported to orbits at very large heliocentric distances
by the combined effects of gravitational scattering by the outer planets, the smooth Galactic tidal field, and direct gravitational tugs from stars passing through the solar-system neighborhood \citep[see, e.g.,][]{vnd2019}. This reservoir of planetesimals is stable but leaky on Gyr timescales, and the leak produces the population of LPCs observed today, which set off on their deadly journey to visit the inner parts of
the solar system a few Myr ago. The LPCs that visit the inner solar system directly from this reservoir retain evidence of their original heliocentric distance in their orbits. In particular, the distribution of their binding energies, inversely proportional to the orbital semimajor axis before they enter the region of the giant planets, shows a striking peak in the range $0$ to $\simeq 0.75\times 10^{-4}$ au$^{-1}$ \citep[e.g.,][and Fig.\  \ref{fig_lpc_oort_newt}]{warsaw2020}. This is the so-called Oort peak, whose discovery and interpretation by \citet{o1950} was one of the landmarks in cometary science in the past century. LPCs may return to the inner parts of the solar system a few more times before they are ejected from the solar system by planetary encounters or fade to invisibility, producing a distinct tail in their energy distribution at semimajor axes smaller than those in the Oort peak. 

Seventy years ago only a handful of LPCs had accurate orbits, just enough to allow Oort to formulate his hypothesis, but the situation has changed dramatically over the intervening decades. Hundreds of LPCs with well-determined orbits are known today. Matching the distribution of LPC orbits is therefore a challenging task for any theory that aims at a description of their dynamical evolution. Because LPCs spend most of their lifetime roaming in the weak-field regime at the outskirts of the solar system, they are promising candidates to look for MOND\-ian effects. While this insight is not new, it has not been properly explored. This is because it requires a significant effort to accurately model MOND\-ian dynamics within a properly formulated simulation of the orbital evolution of LPCs on Gyr timescales. The primary goal of this paper is to carry out this modeling. Our methodology enables any combination of MOND\-ian parameters to be tested through confrontation with the LPC orbital data. It can also be used to test any variant of MOND, although we restrict our attention in this paper to the AQUAL formulation of the theory \citep{bm1984}. 

A second method for testing MOND is based on the distribution of orbits of TNOs. This is because some planetesimals that were born in the trans-Neptunian disk could have been transferred by the non-spherical components of the MOND potential to stable orbits with semimajor axes of a few hundred au and perihelion distances $> 38$ au (well outside Neptune's orbit at 30 au, where the orbits are unaffected by gravitational interactions with  Neptune and the other planets). A significant number of objects have been found on such orbits \citep[e.g.,][]{b2022}, with many more promised by upcoming survey programs; this population is known as the detached disk 
\citep[e.g.,][]{glad2008,pk2016,p2017}. While there are
other possibilities for populating the detached disk \citep[e.g.,][]{bra2006,gladman2006,kq2008,ketal2011,silsbee2018,p92019}, our simulations can be used to test whether MOND already {\it over}-populates the detached disk compared to the currently available data (Sec.~\ref{res1}).

Before we set out on these tasks, we briefly overview previous studies that had a similar goal in mind.
\smallskip

\noindent{\it Attempts to test the weak acceleration domain of MOND in the solar system.-- }The pioneering work of \citet{m1983a}, while primarily oriented toward galactic applications of the MOND hypothesis, mentions LPC dynamics as a test arena for MOND. The author suggests that comets that appear to come from the Oort cloud (semimajor axis $\sim 30,000$ au) actually come from the MOND characteristic distance $r_{\rm M}\simeq 7,000$~au (Eq.\ \ref{eq:rmdef}). This brief discussion seeded the idea
of LPCs as an important population for testing MOND.

In MOND\-ian dynamics, in contrast to Newtonian dynamics, the solar system's acceleration in the gravitational field of the Milky Way affects the gravitational potential (i.e., the strong equivalence principle is violated). This is the so-called external field effect (EFE), briefly mentioned already in \citet{m1983a}, rigorously described by \citet{bm1984}, and applied to the inner solar system by \citet{m2009}. The latter work mostly considers  the imprint of the EFE on the motion of the planets, but the author returns to his call for
testing MOND with comet dynamics in the last paragraph of Sec.\ 7.

\citet{maquet2015} used a quadrupole approximation for the gravitational potential due to MOND (see Eq.\ \ref{quadr} and Fig.\ \ref{fig_dg_local}), which is valid for sufficiently small heliocentric distances,  and computed its effects on the orbits of three comets with semimajor axes $\lesssim 50$ au. They
concluded that the effects of MOND could rival those of non-gravitational effects due to outgassing. However, no significant test of the validity of MOND could be made with this analysis, mostly because the semimajor axes of the comets considered are much smaller than the MOND characteristic radius $r_\mathrm{M}$.

\begin{figure*}[t!]
% \plottwo{.eps}{.eps}
 \begin{center}
 \begin{tabular}{cc}
  \includegraphics[width=0.47\textwidth]{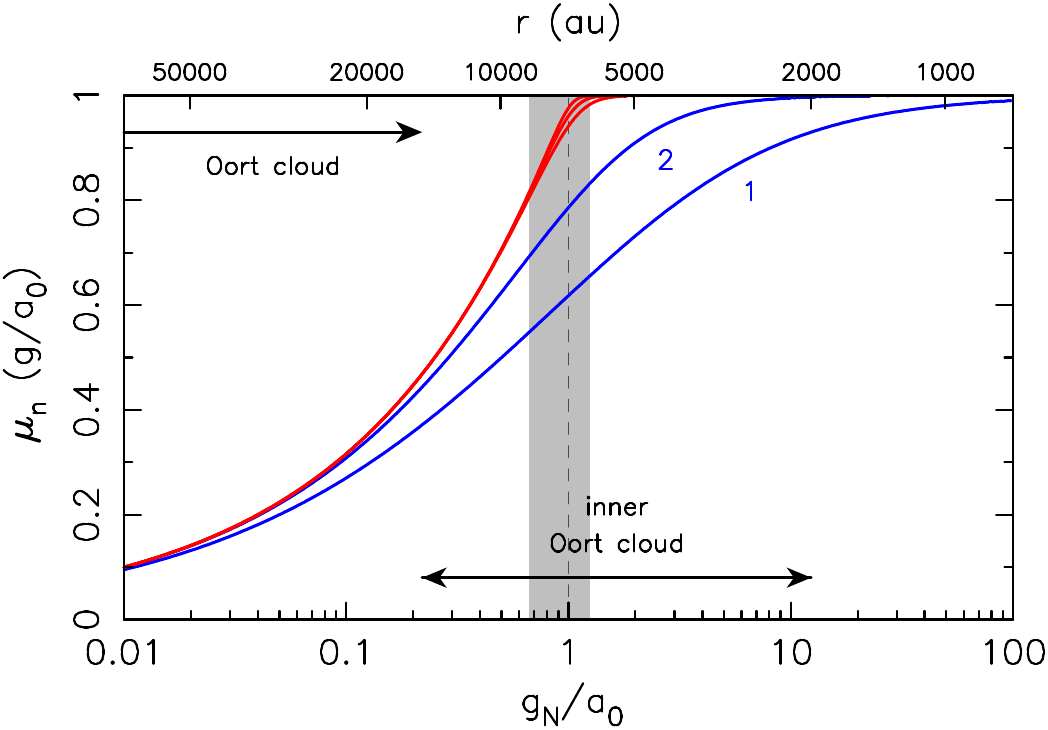} &
  \includegraphics[width=0.47\textwidth]{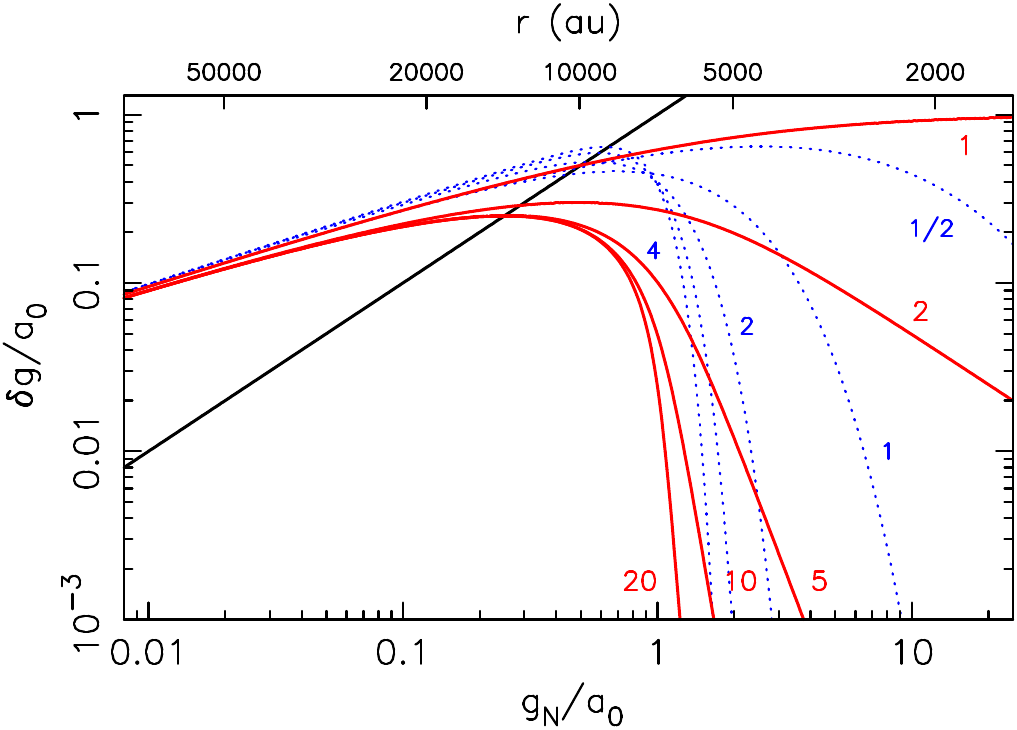} \\ 
 \end{tabular}
 \end{center}  
 \caption{Left panel: The family of MOND transition functions $\mu_n(x)$ (Eq.\ \ref{e2}) with $n=1$,
  $n=2$ (blue curves), and $n=8$, $n=12$ and $n=20$ (red curves); the abscissa is the Newtonian
 acceleration $g_{\rm N}$ in units of $a_0$, the ordinate is $\mu_n(g/a_0)=1/\nu_n(g_{\rm N}/a_0)$
 (see Eq.~\ref{eq:invert}). The upper abscissa translates the Newtonian acceleration into heliocentric
 distance $r$ in astronomical units: the vertical dashed line marks the MOND scale where $g_{\rm N}=a_0$ for the nominal value of $a_0=1.2\times 10^{-10}$ m~s$^{-2}$, the gray rectangle shows the uncertainty of $a_0$ based on values obtained in \citet{heetal2016}.
The arrows indicate the location of the outer Oort cloud ($15,000$--$100,000$~au) and the inner Oort cloud
  ($2,000$--$15,000$~au). Right panel: Difference between the MOND and Newtonian accelerations $\delta g=g-g_{\rm N}$ scaled by $a_0$, as a function of $g_{\rm N}/a_0$. The red curves are for five examples of the $\mu_n(x)$ family of MOND transition functions: $n=1$, $n=2$, $n=5$, $n=10$ and $n=20$ (see the labels). The blue dotted lines are for an alternative class of transition functions
  $\nu_\alpha(y)=[1-\exp(-y^\alpha)]^{-1/2\alpha}+(1-1/2\alpha)\exp(-y^\alpha)$ with $\alpha=1/2$ and $\alpha=1,\ldots,4$ (blue labels) -- the relation between $\nu(y)$ and $\mu(x)$ is given by Eq.~(\ref{eq:invert}). The cases $\alpha\geq 2$ were found by \citet{heetal2016} to satisfy observational tests from galactic rotation curves and planetary ephemerides. The black line is $g_{\rm N}/a_0$ itself for reference. The upper abscissa translates $g_{\rm N}/a_0$ into heliocentric distance $r$ in astronomical units. Beyond $\simeq 20,000$~au, the non-Newtonian acceleration $\delta g$ exceeds $g_{\rm N}$ by at least a factor of two for all transition functions shown.}
 \label{fig1}
\end{figure*}

\citet{pk2016} discussed the structure of the Oort cloud in MOND\-ian dynamics \citep[see also a similar work by][]{p2020a}. They considered a group of observed LPCs and integrated them numerically backward in time for couple of heliocentric revolutions (tens of Myr at maximum) to infer the parameters of the MOND\-ian Oort cloud from which they came. We have several comments. (i) This is a much more limited investigation than the one in the present paper, in which we follow comets that are initially formed in a heliocentric disk, scattered by giant planets into the forming Oort cloud, and subsequently returned back to the inner parts of the solar system. (ii) \citet{pk2016} mostly used a MOND transition function (Eq.\ \ref{e1}) that is in conflict with planetary ephemerides (Eq.\ \ref{eq:trans2} with $\alpha=\frac{1}{2}$ and Fig.~\ref{fig1}; see \citealt{heetal2016}, who argue that $\alpha\ge 2$ is required).  Upon choosing a transition function that meets the planetary constraints, they concluded that the structure of the Oort cloud would be similar to that produced by the Newtonian model (our more detailed analysis, however, allows us to identify significant differences; see Sec.\ \ref{res2}). (iii) However, \citet{pk2016} pointed out an interesting and novel aspect of MOND, namely the possibility that the non-spherical component of the MOND\-ian potential arising from the EFE could decouple objects from the Neptunian scattered disk onto detached orbits, that is, orbits with perihelia $q> 38$~au that are large enough to be immune to planetary perturbations. The ability of MOND to store objects in this way was further described in \citet{p2017}. This effect is the basis for one of our tests of MOND, and several results drawn from these papers agree with ours (see Sec.\ \ref{res1}).

We also note a recent paper by \citet{bm2023}, who describe in more detail how MOND can populate the detached disk from the scattered disk. In particular, they promote the idea that MOND not only injects planetesimals to the detached population, but at the same time makes many of these orbits apsidally aligned with the direction to the Galactic center. Previously this non-axisymmetry of the orbital architecture of distant trans-Neptunian objects (TNOs) has been taken as a sign of an unknown distant planet \citep[see the review by][]{p92019}. We believe that this interesting and novel evidence of MOND is flawed, and that the conclusions of \citet{bm2023} are incorrect, for several reasons. (i) The transition function these authors use is too slow to be consistent with constraints from planetary ephemerides, falling between the $\mu_1$ and $\mu_2$ cases in Eq.\ (\ref{e2}). While admitting this fact, Brown and Mathur argue that their general conclusions will still hold for sharper transition functions that satisfy solar-system constraints. However, they do not support this claim by any numerical experiments, relying instead on an incomplete analysis of the secular quadrupole model. The secular timescales become very long for sharp transition functions, such that the comparably slow motion of the solar system about the Galactic center would erase the desired orbital confinement. (ii) The conclusions of \citet{bm2023} are based on the existence of fixed points in the secular Hamiltonian. With the lack of obvious dissipation mechanism, a fixed point (even if surrounded by a macroscopic stable region) would not attract the majority of orbits. (iii) Apsidal alignment was not present in our numerical simulations of TNOs (Sec.\ \ref{res1}).

\cite{miga2023} considers a similar problem to \cite{bm2023}, but calculates the MOND\-ian potential with more care. Instead of using the quadrupole potential, he finds the MOND\-ian
potential (using the QUMOND formulation; see Sec.\ \ref{theory1}) on a large heliocentric grid. The external Galactic field is modeled simply as that of a point mass at the Galactic center. He demonstrates by backward integrations that the EFE can populate the detached disk. However, his conclusions are compromised somewhat by his choice of a gradual transition function ($n=2$ in Eq.~\ref{e2}), which is inconsistent with planetary ephemerides (see Sec.\ \ref{efe1}). Also, his backward integration of the orbits of a few distant TNOs does not illuminate the fundamental result we obtain from our forward modeling
of a large statistical sample of orbits (Sec.~\ref{res1}): if this variant of MOND were correct, current sky surveys would have detected too many TNOs with the wrong distribution of orbital elements. 
\smallskip

\noindent{\it The strong acceleration domain tests of MOND in the solar system.-- } The properties of MOND -- or more precisely the nature of the transition function between the strong- and weak-field regimes in MOND -- can be quantitatively constrained using planetary ephemerides, even though all of the planets have accelerations at least $5\times 10^4$ times larger than the MOND acceleration scale $a_0$ (Eq.\ \ref{a0def}). This is because the EFE can penetrate deep into the strong-field regime. This possibility was first pointed out by \citet{m2009}, who 
realized that the EFE can induce a significant  contribution to the precession rate of the longitude of perihelion of the planets. Thanks to radar ranging and a
multitude of spacecraft missions, planetary ephemerides have become extremely accurate over the past several decades. Thus, the analyses of \citet{bn2011}, \citet{hetal2014}, and \cite{f2018}, whose results will be reviewed in Sec.~\ref{internal}, have strongly constrained the behavior of the transition function in the strong-field regime. 
\smallskip

The paper is organized as follows. In Sec.~\ref{theory}, we summarize our description of
MOND and the EFE, paying particular attention to its dependence on the choice of the transition function
and the magnitude of the external acceleration. Our main focus is on a thorough description of
the gravitational field experienced by LPCs, but we also develop an approximate model for the tidal field of the Galaxy as seen in the solar system. In Sec.~\ref{met}, we describe our 
numerical model for the orbital evolution of planetesimals that are born in the trans-Neptunian region and scattered by the giant planets onto extreme heliocentric orbits, some of which  return to the inner solar system as LPCs. We also describe the observational data on LPCs that will be used to test the model. In Sec.~\ref{res} we report the results of our simulations and compare them with observations of both TNOs and LPCs. Conclusions from our work are presented in Sec.~\ref{concl}.

% ----------------------------------------------------------------------------
% SEC ???
\section{Theory}\label{theory}

We review the basic concepts of MOND in Sec.~\ref{internal}. In this paper, following most of its up-to-date models applied to astronomical systems, we adopt the modified gravity approach to the MOND \citep[see, e.g.,][and Sec.~\ref{concl} for further discussion]{bm1984,fm2012}. This leaves aside the modified inertia variant of MOND.
In Sec.~\ref{theory1} we use the Bekenstein--Milgrom AQUAL approach to model the EFE on the dynamics of planetesimals scattered by the giant planets to semimajor axes far beyond Neptune. 

% FIG 1 %%%%%%%%%%%%%%%%%%%%%%%%%%%%%%%%%%%%%%%%%%%%%%%%%%%%%%%%%%%%%%%%%%%%%%%%%%%%%%%%%%%%%%%%%%%%%%%
%%%%%%%%%%%%%%%%%%%%%%%%%%%%%%%%%%%%%%%%%%%%%%%%%%%%%%%%%%%%%%%%%%%%%%%%%%%%%%%%%%%%%%%%%%%%%%%%%%%%%%%

\subsection{Phenomenological origins of MOND}\label{internal}

In a non-relativistic MOND scheme, the radial heliocentric acceleration  $g_{\rm N}=GM/r^2$ of an isolated, spherically symmetric source like the Sun is replaced with its effective or actual value $g$ determined by the relation \citep[e.g.,][]{m1983a,bm1984}
\begin{equation}
 \mu\left(g/a_0\right)\,g = g_{\rm N}\; . \label{e1}
\end{equation}
The theory does not provide a hint of what should be (i) the fundamental acceleration scale $a_0$, and (ii) the shape of the transition function $\mu(x)$. One only knows that in the strong-field (large acceleration) limit $x\gg 1$, $\mu(x) \rightarrow 1$ is required to restore the well-tested Newtonian description. If in the weak-field (small acceleration) limit $x\ll 1$, we have  $\mu(x) \rightarrow x + O(x)$ (as is usually assumed), one opens the interesting door to explain the flat rotation curves of galaxies without any need for dark matter. This is because for $\mu(x)=x$ we have $g=\sqrt{g_{\rm N} a_0}$, and thus the Keplerian circular velocity $v$ about a center with Gaussian constant $GM$ is $v=(GM a_0)^{1/4}$, independent of distance $r$, thereby explaining the asymptotically flat rotation curves of disk galaxies. At the same time, the correlation $v^4\propto M$ between the asymptotic constant rotation velocity $v$ in galaxies and their baryonic mass content $M$ explains the empirically established Tully--Fisher law for disk galaxies. The observed normalization of the Tully--Fisher law dictates the acceleration scale $a_0$, resulting in 
\begin{equation}
   a_0\simeq 1.2\times 10^{-10}\mbox{~m s}^{-2}
   \label{a0def}
\end{equation}
\citep[e.g.,][]{m1983b,m1983c,fm2012}, the value we shall use throughout this paper. There is a possible connection to
cosmology since this value of $a_0$ of the same order of magnitude as $cH=6.80\times 10^{-10}\;\mbox{m s}^{-2}\,
(H/70\,{\rm km~s}^{-1}\,\mbox{Mpc}^{-1})$, where $c$ is the speed of light and $H$ the Hubble constant. This 
relation can be further developed in a fully relativistic approach to MOND that allows the construction of 
cosmological models and has the non-relativistic MOND theory (Sec.~\ref{theory1}) as a limiting case
 \ \citep[see, e.g.,][]{bek2004,sko2009,fm2012,m2015}. Updated fits to astronomical observations, such as galactic
rotation curves, have led to $a_0$ values within the range $0.7\times 10^{-10}$ m~s$^{-2}$ to $1.6\times 10^{-10}$ m~s$^{-2}$, depending on the choice of transition function  
\citep[e.g.,][]{heetal2016}.

In Eq.\ (\ref{e1}) the relation between the Newtonian acceleration $g_{\rm N}$ and the actual acceleration $g$ is algebraic, so it is relatively simple to solve for $g$. This approach is over-simplified and can only be applied to systems with a high degree of symmetry (e.g., spherical or slab). Nevertheless, Eq.~(\ref{e1}) introduces the useful concept of the MOND scale $r_{\rm M}$ in the context of solar-system studies. This is where $g_{\rm N}= a_0$, or
\begin{equation}
 r_{\rm M}=\sqrt{\frac{GM_\odot}{a_0}}=7,030\mbox{\,au}\left(\frac{1.2\times 10^{-10}
  \mbox{\,m s}^{-2}}{a_0}\right)^{1/2}\; .
  \label{eq:rmdef}
\end{equation}
Thus MOND begins to influence the dynamics strongly at a few thousand au from the Sun. This is the solar-system backyard
 that only LPCs visit. Since $r_{\rm M}\propto a_0^{-1/2}$, the effects of MOND on the outer solar system do not depend strongly on the choice of $a_0$.
 For instance, changing $a_0$ within the above-mentioned range would make $r_{\rm M}$ span values from $6,290$~au to $8,610$~au, and this has no effect on our conclusions (Fig.~\ref{fig1}).

Once the acceleration scale $a_0$ has been set, the MOND practitioner still needs to specify the transition or interpolating function
$\mu(x)$ in Eq.\ (\ref{e1}). The literature offers many suggestions that satisfy the above-mentioned asymptotic behavior of $\mu(x)$ at $x\ll 1$ and $x\gg 1$ \citep[see, e.g.,][]{fm2012}.
Here we consider a family of transition functions 
\begin{equation}
 \mu_n\left(x\right)= \frac{x}{(1+x^n)^{1/n}}  \; , \label{e2}
\end{equation}
parametrized by a positive integer $n$.  One often needs to express $g$ as a function of $g_{\rm N}$
from Eq.~(\ref{e1}), or in other words, to invert
\begin{equation}
 x\,\mu(x)=y \Leftrightarrow x=y\,\nu(y)\; . \label{eq:invert}
\end{equation}
The $\nu_n(y)$ functions that correspond to $\mu_n(x)$ in Eq.~(\ref{e2}) read
\begin{equation}
 \nu_n\left(y\right)= \left[\frac{1+\left(1+4y^{-n}\right)^{1/2}}{2}\right]^{1/n}
  \; . \label{e3}
\end{equation}
In the weak-field limit ($y\ll 1$) one has $\nu_n(y)\rightarrow y^{-1/2}$, such that the gravitational acceleration $g\rightarrow
\sqrt{g_{\rm N} a_0}$, as required to have asymptotically constant rotation velocities in galaxies. In the strong-field limit ($y\gg 1$), the non-Newtonian acceleration $\delta g = g-g_{\rm N}= g_{\rm N}\left[\nu_n\left(y\right)-1\right]$
behaves as
\begin{equation}
 \delta g\rightarrow \frac{a_0}{n}\left(\frac{a_0}{g_{\rm N}}\right)^{n-1}
  \; . \label{e4}
\end{equation}
For the most gradual transition function, $n=1$, we would have $\delta g\rightarrow a_0$ (Fig.~\ref{fig1}). A constant acceleration of this magnitude conflicts with observations of the orbits of the inner planets \citep[e.g.,][]{bn2011,f2018,fienga2024}. For $n\geq 2$ the non-Newtonian acceleration $\delta g$ in the inner solar system is much smaller but can still be significant. Analyses of the precession of planetary perihelia require that $n\geq 6-8$ \citep[e.g.,][]{bn2011,hetal2014,heetal2016}. The most careful comparison to planetary ephemerides yields an even stronger constraint, $n\ge 20$ \citep{f2018}.

Figure~\ref{fig1} shows the behavior of the transition functions in the $\mu_n(x)$ family for various values of $n$ (left panel), and also displays the non-Newtonian acceleration $\delta g$ scaled by $a_0$ (right panel). The right panel also shows $\delta g/a_0$ for an alternative class of transition functions, 
\begin{equation}
 \nu_\alpha(y)= [1-\exp(-y^\alpha)]^{-1/2\alpha}+(1-1/2\alpha)\exp(-y^\alpha), \label{eq:trans2}
\end{equation}
that were found to be consistent with constraints from galactic dynamics and planetary ephemerides by \citet{heetal2016} only for $\alpha\geq 2$. They are consistent with planetary ephemerides because they decay suitably fast in the strong-field limit. However, in the Oort cloud zone, they produce a non-Newtonian acceleration $\delta g$ similar or larger than our tested class of transition functions (\ref{e2}). For this reason we do not consider the $\nu_\alpha(y)$ class of transition functions in our work.

The simple modification of Poisson's equation described by Eq.\ (\ref{e1}) can lead to physical inconsistencies. The
potential $u(r)$ associated with $g$, such that $g=du/dr$, may be obtained with a simple
quadrature
\begin{equation}
 u\left(r\right)=u_0+\int_{r_0}^r dr' \Phi\left(r'\right)\; , 
 \label{eq:isoa}
\end{equation}
where
\begin{equation}
 \Phi\left(r\right)=g_{\rm N}\left(r\right)\,\nu_n\left[g_{\rm N}\left(r
  \right)/a_0\right]\; .
  \label{eq:isob}
\end{equation}
At small radii, where $\nu_n\simeq 1$, this implies the usual Newtonian behavior  $u(r)\propto 1/r$. At large radii, where $\nu_n(y)\propto y^{-1/2}$, we have 
\begin{equation}
 u\left(r\right)\simeq \sqrt{GM a_0}\, \ln r + \mbox{const}\; . \label{emondinf}
\end{equation}
Therefore, unlike in the Newtonian case, the potential does not approach a constant at large distances, but has a logarithmic divergence. If taken literally, such a behavior would bind all test bodies to the system. In practice, this unphysical behavior is remedied by noting that no system is isolated to infinity. This leads us to a more realistic prescription for MOND\-ian effects, in which an arbitrary, but non-zero, external field exists (Sec.~\ref{efe}).

\subsection{Non-relativistic field theory resulting in MOND}\label{theory1}
\citet{bm1984} anchored MOND in the standard concepts of theoretical physics by developing a non-relativistic Lagrangian approach to gravity from which MOND emerges (this is known as the AQUAL theory of gravity, short for ``aquadratic Lagrangian''). The fundamental field equation for the gravitational potential $U$, yielding the acceleration ${\bf g}=\nabla U$ of a test particle, reads
\begin{equation}
 \nabla\cdot\left(\mu\,{\bf g}\right) = - 4\pi G \rho \;  \label{e5}
\end{equation} 
in the Bekenstein--Milgrom theory. Here $\rho$ is the density of matter, $g=|{\bf g}|$, and $\mu=\mu(g/a_0)$ is the MOND transition function. When $g\gg a_0$, $\mu(g/a_0)\rightarrow 1$ and Eq.\ (\ref{e5}) becomes the familiar Poisson equation of Newtonian gravity. \citet[][Appendix A]{bm1984} show that Eq.~(\ref{e1}) is recovered from the more general (\ref{e5}) at large distances from a bound system of total mass $M$, since
\begin{equation}
 \mu\left(g/a_0\right)\,g = g_{\rm N} + O\left(1/r^3\right) \label{e1bis}
\end{equation}
follows as a first integral. For isolated systems of high symmetry, such as the spherical case, the second term on the right-hand side of (\ref{e1bis}) vanishes. 

In general, though, the presence of $g=|{\bf g}|$ in the argument of $\mu$ makes the partial differential equation (\ref{e5}) non-linear. Therefore its solutions are difficult to find, even when some degree of symmetry is present (such as the axisymmetry that applies in the case examined here; see Sec.\ \ref{efe}), and numerical methods are needed. Reorganizing terms in (\ref{e5}), one can also write
\begin{equation}
 \nabla\cdot{\bf g} = - 4\pi G \left(\rho + \rho_{\rm pm}\right)
  \; , \label{e6}
\end{equation}  
introducing the density of phantom matter by
\begin{equation}
 \rho_{\rm pm} = \frac{1}{4\pi G}\nabla\cdot\left(\chi\, {\bf g}\right)
  \; , \label{e7}
\end{equation}  
with $\chi(g/a_0)=\mu(g/a_0) - 1$. The MOND\-ian effects are now expressed as a matter density. Obviously, this reshuffling of terms does not make the non-linear problem easier, but it provides a new way to look at the effects of MOND, and it may offer a starting point for some approximations. For instance, if the MOND\-ian effects are small, one may plug the Newtonian acceleration ${\bf g}_{\rm N}$ due to the mass distribution $\rho$ (given by the solution of the linear problem $\nabla\cdot {\bf g}_{\rm N} = -4\pi G \rho$) into the arguments of $\rho_{\rm pm}$ \citep[e.g.,][]{m2009}. The simple Poisson equation for the potential $U$ is thus restored, and the whole theory is easier to apply.%

This approximation, proposed in \citet{m2009}, can be generalized to a rigorous non-relativistic theory of gravity known as QUMOND \citep[see][]{m2010}. The analog of Eq.\ (\ref{e5}) for QUMOND is
\begin{equation}
    \nabla\cdot{\bf g}=\nabla\cdot(\nu\,{\bf g}_{\rm N}), \quad \nabla\cdot{\bf g}_{\rm N}=-4\pi G\rho,
    \label{eq:qqq}
\end{equation}
where ${\bf g}_{\rm N}$ is the Newtonian acceleration, $g_{\rm N}=|{\bf g}_{\rm N}|$, and $\nu=\nu(g_{\rm N}/a_0)$ is related to the transition function $\mu(g/a_0)$ by Eq.\ (\ref{eq:invert}). QUMOND requires the solution of two linear differential equations, whereas AQUAL requires the solution of one non-linear differential equation. In this paper we restrict ourselves to the AQUAL formulation of MOND. We believe that our conclusions would be very similar in QUMOND (compare for example the distribution of non-Newtonian acceleration in AQUAL, from Fig.\ \ref{fig_dg_global1}, with the distribution in QUMOND, from Fig.\ 1 of \citealt{miga2023}).

This brings us to the task of implementing a numerical solution for Eq.\ (\ref{e5}), in which we follow the methods of \citet{bn2011}. A choice of boundary conditions on the potential $U$ is also required \citep[see discussion in][for context]{m1986}.
The boundary conditions play a double role, namely to make the setup physically realistic (dropping the assumption of the system being isolated), while at the same time removing the unphysical logarithmic divergence of the gravitational potential at infinity (Eq.\ \ref{emondinf}).

\subsubsection{The external field effect (EFE) in MOND}\label{efe}

The solar system is embedded in a larger structure, namely the Galaxy. The Galactic mass distribution generates a gravitational field which, for a solar-system observer, has two components in the Galactic reference frame (which we assume to be inertial, i.e., not accelerated): (i) a uniform acceleration ${\bf g}_{\rm e}$, and (ii) a tidal field ${\bf g}_{\rm tide}$. In Newtonian dynamics, the first component has no effect on the internal dynamics of the solar system because it is freely falling in the Galactic field (the strong equivalence principle), but the second affects the motion of bodies (especially those at large heliocentric distances such as distant TNOs or LPCs) bound to the solar system. The effects of both (i) and (ii) are modified by MOND. In our initial discussion in this Section we ignore the tidal field (ii); this is actually the level at which the EFE has been described in previous literature. However, the tidal field (ii), together with its MOND\-ian correction, is needed for an accurate description of the orbits of objects in the inner and outer Oort clouds ($\sim 2,000$ to $100,000$ au). This description is developed in  Sec.~\ref{efe3}.

\begin{figure*}[t!]
% \plottwo{.eps}{.eps}
 \begin{center}
 \begin{tabular}{cc}
  \includegraphics[width=0.47\textwidth]{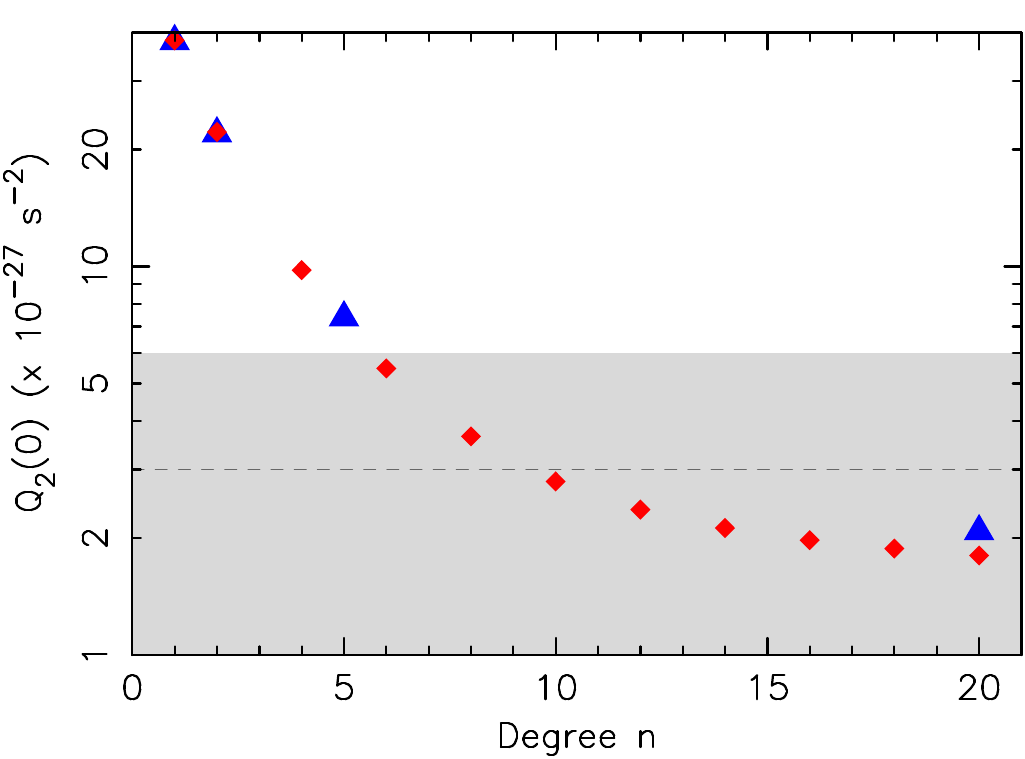} &
  \includegraphics[width=0.47\textwidth]{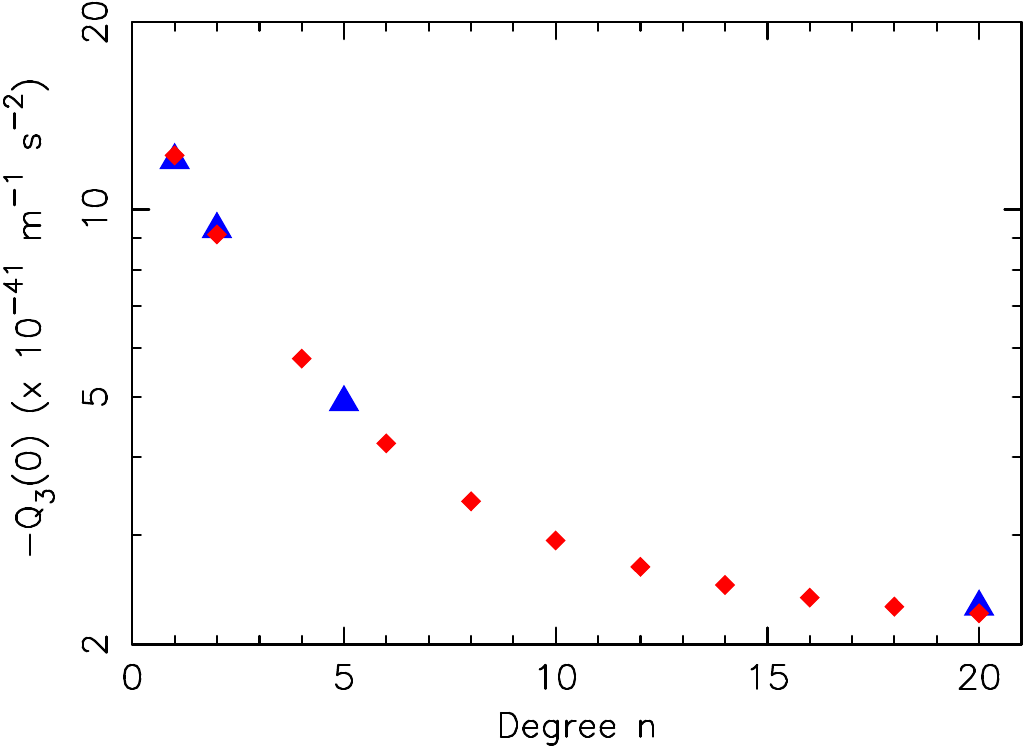} \\ 
 \end{tabular}
 \end{center}  
 \caption{The quadrupole and octupole moment values $Q_2(0)$ (left) and $-Q_3(0)$ (right) for different degrees $n$ (abscissa) of the family of transition functions $\{\mu_n(x)\}$ \citep[because this figure provides an opportunity to compare our results with those of][we computed these parameters for their assumed value of the external acceleration,
 $g_{\rm e}=1.9\times 10^{-10}$~m s$^{-2}$]{bn2011}. Red diamonds show our values, blue triangles are from Table~1 in
 \citet{bn2011}. Radar ranging to the Cassini spacecraft constrains $Q_2(0)= (3\pm 3)\times 10^{-27}$~s$^{-2}$ 
 \citep{hetal2014} as indicated by the gray area (with the mean value shown by the horizontal dashed line). Analyses
 of planetary motion \citep{bn2011,f2018,fienga2024} require $n\geq 6$, or even $n\geq 20$ \citep{f2018}.}
 \label{fig_q230}
\end{figure*}

The following analysis of Eq.\ (\ref{e5}) is different from but equivalent to that of \citet{bn2011}. We assume an exterior field ${\bf g}_{\rm e}$ that is independent of position, which is going to be the asymptotic value of the MOND acceleration ${\bf g}=\nabla U$ at large distances $r\rightarrow \infty$ from the solar system. The acceleration vector ${\bf g}_{\rm e}$ points toward the center of the Galaxy and is fixed in magnitude, i.e., we assume that the Sun is on a circular orbit around the Galactic center and ${\bf g}_{\rm e}$ is the centripetal acceleration. \citet{bn2011} consider $g_{\rm e}=1.9\times 10^{-10}$~m s$^{-2}$. In this paper, however, we adopt a somewhat larger value, $g_{\rm e} = 2.32\times 10^{-10}$ m~s$^{-2}$, based on a recent analysis of Gaia observations \citep{klio2021}; the only exception is the results shown in Fig.~\ref{fig_q230}, in which we use the Blanchet--Novak value for a direct
comparison with their work. We recall that the acceleration scale $a_0$ is a free parameter in MOND, while $g_{\rm e}$ is an observable feature of the astronomical system that we are examining. We briefly explore the dependence of our results on the value of $g_{\rm e}$ in Sec.~\ref{efe2}.

For simplicity, we neglect the planetary masses and thus assume a spherical model for the mass distribution in the solar system about the origin representing the Sun (\citealt{m2012} discusses the effects of the planetary contribution to the mass distribution, but we neglect them in this work as they are small at the heliocentric distance we are interested in). In principle, the density $\rho(r)$ inside the solar radius $R$ should be determined from a numerical solar model, but \cite{bn2011} showed that the EFE does not depend significantly on the solar internal structure so it is sufficient to assume a homogeneous model in which the density is constant.%

Following \citet{bn2011}, in the region inside the Sun ($r\leq R$, where for simplicity we set $R=0.01$~au), we impose $\mu(x)=1$; this results in a traditional Newtonian potential $u_{\rm N}(r)$ that is characterized by the boundary condition $\partial u_{\rm N}/\partial r=0$ at $r=0$. This approximation is well justified since the accelerations near and in the Sun are much larger than $a_0$, except in a very small zone near the center \citep[the corresponding small error has been estimated by][]{bn2011}. Thus we can write
 \begin{eqnarray}
  u_{\rm N} & = & \frac{GM}{2R}\left(3-\frac{r^2}{R^2}\right)\; , \qquad {\rm for}\;
   r\leq R\; , \label{e8a} \\
  u_{\rm N} & = & \frac{GM}{r}\; , \quad\qquad\qquad\qquad {\rm for}\; r\geq R\; , \label{e8b}
 \end{eqnarray}
where $M$ is the solar mass. Obviously these values are defined up to an arbitrary constant. 

We next solve for the MOND\-ian gravitational field outside the Sun, which is connected to the interior solution (\ref{e8a}) via the appropriate boundary conditions at the solar surface (potential and its radial derivative continuous at $r=R$). The gravitational potential defining the heliocentric motion -- as opposed to the motion in the Galactic or inertial frame -- of solar-system bodies (planets, asteroids, comets, etc.) is defined by $u({\bf r})=U({\bf r})-{\bf g}_{\rm e}\cdot {\bf r}$, which now has ${\rm lim}_{r\rightarrow \infty} u = 0$. We may also define the non-Newtonian potential $\delta u=u-u_{\rm N}$, where $u_{\rm N}$ is the conventional Newtonian potential generated by the Sun, Eq.\ (\ref{e8b}). 

Following \citet{bn2011}, we now rewrite the MOND field equation (\ref{e5}) in the form (recall ${\bf g}=\nabla u + {\bf g}_{\rm e}$)
\begin{equation}
 \nabla^2 u = -\frac{1}{\mu}\left(4\pi G\rho +\frac{\mu^\prime}{a_0}
  \,g^i \partial_i g\right) \equiv\sigma\left(u,\rho\right)\; , \label{e9}
\end{equation}
%\begin{equation}
% \Delta u = -\frac{1}{\mu(g/a_0)}\left[4\pi G\rho + \frac{\mu^\prime}{a_0 g}
%   g^i g^j \partial_i h_j\right] \; , \label{e9}
%\end{equation}
where $\mu(x)$ is the chosen (arbitrary) MOND transition function and $\mu^\prime(x) = d\mu / dx$ ($x=g/a_0$). This way (\ref{e9}) resembles a traditional Poisson equation, with source term on the right-hand side depending non-linearly on the potential $u$ itself. Outside the Sun, $r>R$, the density $\rho=0$. 

Our goal is to solve for the potential $u$ at an arbitrary position ${\bf r}$ between zero at origin and infinity. Equation~(\ref{e9}) is solved numerically using successive iterations $\{u_n\}$, such  that the new value $u_{n+1}$ is computed by the linear Poisson equation
\begin{equation}
 \nabla^2{\tilde u}_{n+1} = \sigma\left(u_n,\rho\right) \; , \label{e10}
\end{equation}
with relaxation $u_{n+1}=\lambda {\tilde u}_{n+1} + (1-\lambda) u_n$ ($\lambda \in (0,1]$, equal to $0.5$ in our runs); $\sigma\left(u_n,\rho\right)$ is the right-hand side of (\ref{e9}), with $u_n$ defining the inertial acceleration ${\bf g}=\nabla u_n + {\bf g}_{\rm e}$ at the $n^{\rm th}$ iterative step. The initial guess for the solution is simply the Newtonian solution, $u_0=u_{\rm N}$. 

In order to simplify the functional form of the solution, we assume that the coordinate system has the $z$-axis parallel to the external field ${\bf g}_{\rm e}$. Using spherical coordinates, all potentials are then axisymmetric (independent of $\phi$) and thus depend only on the radial distance $r$ from the origin and the colatitude $\theta$. We use $\tau=\cos\theta$
instead of $\theta$ itself to remove the coordinate singularity at the poles. Obviously, the source term $\sigma$ is also axisymmetric, so it can be decomposed in Legendre polynomials,
\begin{equation}
 r^2\,\sigma\left(u_n,\rho\right) = \sum_{\ell=0}^\infty S^{(n)}_\ell(r)\,P_\ell(\tau)
  \; . \label{e11}
\end{equation}
In practice, we restrict ourselves to a finite maximum degree $L$ in the representation (\ref{e11}) (we tested $L$ between $15$ and $30$, using $L=27$ in the final simulations; see also Sec.~\ref{efe1}). The coefficients 
$S^{(n)}_\ell(r)$ are easily computed by Gaussian quadrature once the source $r^2\,\sigma\left(u_n,\rho\right)$ is known on a predefined coordinate grid. Then, if  ${\tilde u}_{n+1}
= \sum_{\ell=0}^L \upsilon^{(n)}_\ell(r)\,P_\ell(\tau)$, the radial functions $\upsilon^{(n)}_\ell(r)$ satisfy
\begin{equation}
 r^2 \frac{d^2 \upsilon^{(n)}_\ell}{dr^2}+ 2r\, \frac{d\upsilon^{(n)}_\ell}{dr} -
  \ell(\ell+1) \upsilon^{(n)}_\ell = S^{(n)}_\ell(r) \; . \label{e12}
\end{equation}
To solve this linear inhomogeneous ordinary differential equation, we implement spectral decomposition of $\upsilon^{(n)}_\ell(r)$ and $S^{(n)}_\ell(r)$ in Chebyshev polynomials \citep[see, e.g.,][and Appendix \ref{details}]{gn2009}. 
In order to keep the maximum degree of the Chebyshev polynomials reasonably small (say between $50$ and $100$), we must split the radial domain from $R=0.01$~au to $R'=500,000$~au into multiple zones, each having inner boundary $r_{\rm min}^i$ and outer boundary
$r_{\rm max}^i$ ($i=2,\ldots, I$, saving index $i=1$ for the innermost radial  zone inside the Sun), such that $r_{\rm min}^2=R$ and $r_{\rm max}^I=R'$. We use zones with a constant stretching factor $f=r_{\rm max}^i/r_{\rm min}^i\simeq 3$--$5$, and employ $I\simeq 12$--$17$ radial zones. In each of these zones we map the radial coordinate $r$ onto an interval $(-1,1)$ on which the Chebyshev polynomials are defined, using a linear transformation: $r=\xi\, r^i_- + r^i_+$ and $\xi=(r-r^i_+)/r^i_-$, with $r^i_\pm=\frac{1}{2} (r_{\rm max}^i \pm r^i_{\rm min})$ ($\xi$ is the needed argument of the Chebyshev polynomials). At the boundaries of the adjacent radial domains, we require that the potential $u$ and its radial derivative $\partial u/\partial r$ are continuous.

The general solution of Eqs.~(\ref{e12}) is composed of a particular solution of the inhomogeneous problem, plus an arbitrary combination of the two fundamental solutions of the homogeneous equation. The contributions of these superposed terms are determined by the boundary conditions at the edges $R$ and $R'$ of the radial coordinate domain. At the inner boundary, $r=R$, we simply require that the potential and its derivative are continuous with the spherically symmetric potential of the Sun (Eq.\ \ref{e8a}). The outer boundary condition is that the potential $u$ must vanish as $r\rightarrow \infty$. For that reason we must add a final radial zone extending from $R'$ to infinity, in which the variable $r$ is replaced by $s=1/r$; then, as for the other zones, $s$ is mapped by a linear transformation onto the interval $(-1,1)$ and the functions $\upsilon_\ell^{(n)}(s)$ and $S_\ell^n(s)$ are expanded in Chebyshev polynomials.  Care must be paid to (i) ensuring the continuity and differentiability of the potential at the boundary of this last radial zone, and (ii) the form of Eq.~(\ref{e12}) that emerges when using $1/r$ instead of $r$ as the independent variable.

At convergence, which is attained when the fractional change between the values $u_{n+1}$ and $u_{n}$ is smaller than a specified limit at all grid points (we use a threshold of $10^{-8}$), we thus have
\begin{equation}
 u(r,\tau) = {u}_{n+1} = \sum_{\ell=0}^L \upsilon_\ell(r)\,P_\ell(\tau)\; . \label{e14}
\end{equation}
The typical number of iterations in our runs was between 30 and 200, depending on the degree $n$ of the transition function (\ref{e2}).
% FIG 1 %%%%%%%%%%%%%%%%%%%%%%%%%%%%%%%%%%%%%%%%%%%%%%%%%%%%%%%%%%%%%%%%%%%%%%%%%%%%%%%%%%%%%%%%%%%%%%%
%%%%%%%%%%%%%%%%%%%%%%%%%%%%%%%%%%%%%%%%%%%%%%%%%%%%%%%%%%%%%%%%%%%%%%%%%%%%%%%%%%%%%%%%%%%%%%%%%%%%%%%

In order to compare our results with \citet{bn2011}, we also define multipole-moment radial functions $Q_\ell(r)$ using ($\ell\geq 1$)
\begin{equation}
 Q_\ell(r) = (-1)^\ell \left(2\ell-1\right)!!\,\frac{\upsilon_\ell(r)}{r^\ell}\;,
  \label{e15}
\end{equation}
of which the first functions are
\begin{eqnarray}
 Q_1(r) & = & -\frac{\upsilon_1(r)}{r}\; , \label{e15a} \\
 Q_2(r) & = & 3\,\frac{\upsilon_2(r)}{r^2}\; , \label{e15b} \\
 Q_3(r) & = & -15\,\frac{\upsilon_3(r)}{r^3}\; , \label{e15c} 
\end{eqnarray}
\citep[see Eq.~29 in][]{bn2011}. The functions $Q_\ell(r)$ are bounded as $r\to 0$. In the case of sharp transition functions (e.g., $\mu_n(x)$ with $n\gtrsim 10$), they are roughly constant as long as $r\lesssim r_{\rm e}=r_{\rm M}/\sqrt{g_{\rm e}/a_0}=(GM/g_{\rm e})^{1/2}$, but for the most gradual transition functions (e.g., $\mu_1(x)$ or $\mu_2(x)$) they change even on a scale as small as a few hundred au (see Fig.~\ref{fig_dg_local}). The behavior of $\upsilon_\ell(r)$ as $r\rightarrow\infty$ is discussed in Appendix~\ref{check}.

As a technical detail, we found it convenient to work with non-dimensional, scaled variables when solving Eq.~(\ref{e9}). The potential  $u$ is normalized using $u_{\rm norm}=GM/{\rm au}$, where $M$ is the solar mass and au is the value of the astronomical unit, and we denote ${\bar u}=u/u_{\rm norm}$. Similarly, the radial functions $\upsilon_\ell(r)$  have their scaled counterparts ${\bar \upsilon}_\ell(r)=\upsilon_\ell(r)/u_{\rm norm}$, and the acceleration $g$ is normalized with $g_{\rm norm}=GM
/{\rm au}^2$.

We denote the direction of ${\bf g}_e$ by ${\bf e}={\bf g}_e/|{\bf g}_e|$ and the direction
of ${\bf r}$ by ${\bf n} = {\bf r}/r$. Then  $\tau={\bf e}\cdot {\bf n}$. Given the potential representation (\ref{e14}), the MOND heliocentric acceleration of a body orbiting the Sun is
\begin{eqnarray}
 \nabla u &=& \frac{1}{r} \sum_{\ell=0}^L \Bigl\{\upsilon^\prime_\ell(r)\,P_\ell(\tau)
   \,{\bf r} +  \nonumber \\
  & & \qquad\quad \upsilon_\ell(r)\,P^\prime_\ell(\tau)\left[{\bf e}-\left({\bf e}\cdot{\bf n}
  \right){\bf n}\right]\Bigr\}  \; , \label{hdef}
\end{eqnarray}
where $\upsilon^\prime_\ell(r)$ and $P^\prime_\ell(\tau)$ denote derivatives with respect
to the corresponding argument. Note that the monopole term, $\ell=0$, also contains the
Newtonian acceleration due to the Sun. Therefore, in order to represent the
non-Newtonian acceleration $\delta{\bf g}$ due to MOND, ${\bf g}_{\rm N}$ must be subtracted from (\ref{hdef}).

\subsubsection{Nominal parameter set}\label{efe1}

The nominal value of the external acceleration of the solar system in the Galaxy adopted in this paper is $g_{\rm e}=2.32\times 10^{-10}$~m s$^{-2}$ \citep{klio2021}. This value is determined with a formal uncertainty of $0.16\times 10^{-10}$~m s$^{-2}$ and independent estimates are consistent with this value within about $\pm 0.2\times 10^{-10}$~m s$^{-2}$ \citep[e.g.,][]{mcmil2017}. However, larger variations of $g_{\rm e}$ may have occurred in the past due to radial migration of the Sun in the Galaxy. We explore the dependence of our results on this parameter in the next subsection.
% FIG 1 %%%%%%%%%%%%%%%%%%%%%%%%%%%%%%%%%%%%%%%%%%%%%%%%%%%%%%%%%%%%%%%%%%%%%%%%%%%%%%%%%%%%%%%%%%%%%%%
\begin{figure*}[t!]
% \plottwo{.eps}{.eps}
 \begin{center}
 \begin{tabular}{c}
  \includegraphics[width=0.9\textwidth]{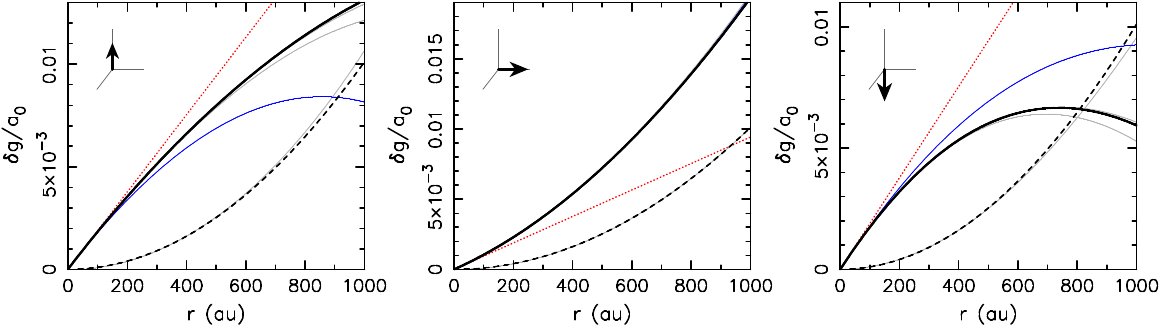} \\
  \includegraphics[width=0.9\textwidth]{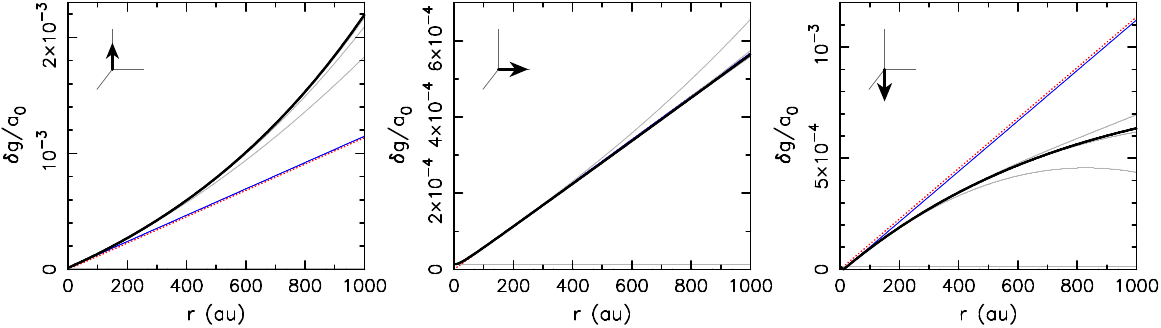} \\ 
 \end{tabular}
 \end{center}  
 \caption{The magnitude of the non-Newtonian acceleration $\delta g$ (normalized by $a_0$) along three spatial directions: (i) parallel to the direction of the exterior acceleration ${\bf e}={\bf g}_{\rm e}/|{\bf g}_{\rm e}|$ (left column), (ii) perpendicular to the exterior acceleration (middle column), and (iii) antiparallel to the direction of the exterior acceleration (right column); the little embedded diagrams show the sampled direction assuming the exterior acceleration is along the positive $z$-axis. The top row of panels is computed for the gradual transition function $\mu_2(x)$, the bottom row of panels for the sharp transition function $\mu_{10}(x)$. The curves show different values of the maximum order $L$ used in the multipole expansion (\ref{hdef}): (i) the monopole term $L=0$ is given by the black dashed line, (ii) the quadrupole representation with $L=2$ is shown in blue; (iii) the representation used in our models, $L=27$, is shown as a black solid line; (iv) some intermediate values of $L$ are shown in gray (the blue lines in the middle panels are overlaid by the black lines). The red-dotted straight line is the prediction from the formula $\delta {\bf g}=\nabla U_2({\bf r})$ (Eq.\ \ref{quadr}) often used to represent the effects of MOND in the inner solar system. For the transition function $\mu_2(x)$, when $r\simeq 200$~au the octupole and even $\ell=4$ multipole should be taken into account, as well as the variations in the multipole components $Q_\ell(r)$ with radius.}
 \label{fig_dg_local}
\end{figure*}
%%%%%%%%%%%%%%%%%%%%%%%%%%%%%%%%%%%%%%%%%%%%%%%%%%%%%%%%%%%%%%%%%%%%%%%%%%%%%%%%%%%%%%%%%%%%%%%%%%%%%%%
% FIG 1 %%%%%%%%%%%%%%%%%%%%%%%%%%%%%%%%%%%%%%%%%%%%%%%%%%%%%%%%%%%%%%%%%%%%%%%%%%%%%%%%%%%%%%%%%%%%%%%
\begin{figure*}[t!]
% \plottwo{.eps}{.eps}
 \begin{center}
 \begin{tabular}{cc}
  \includegraphics[width=0.47\textwidth]{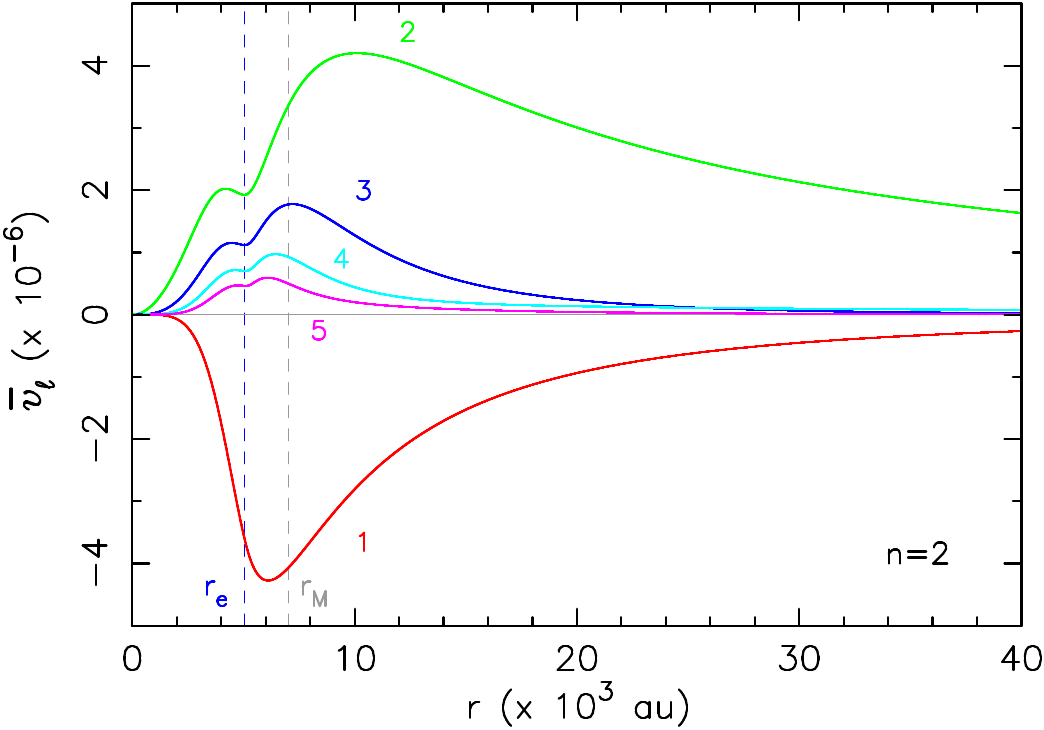} &
  \includegraphics[width=0.47\textwidth]{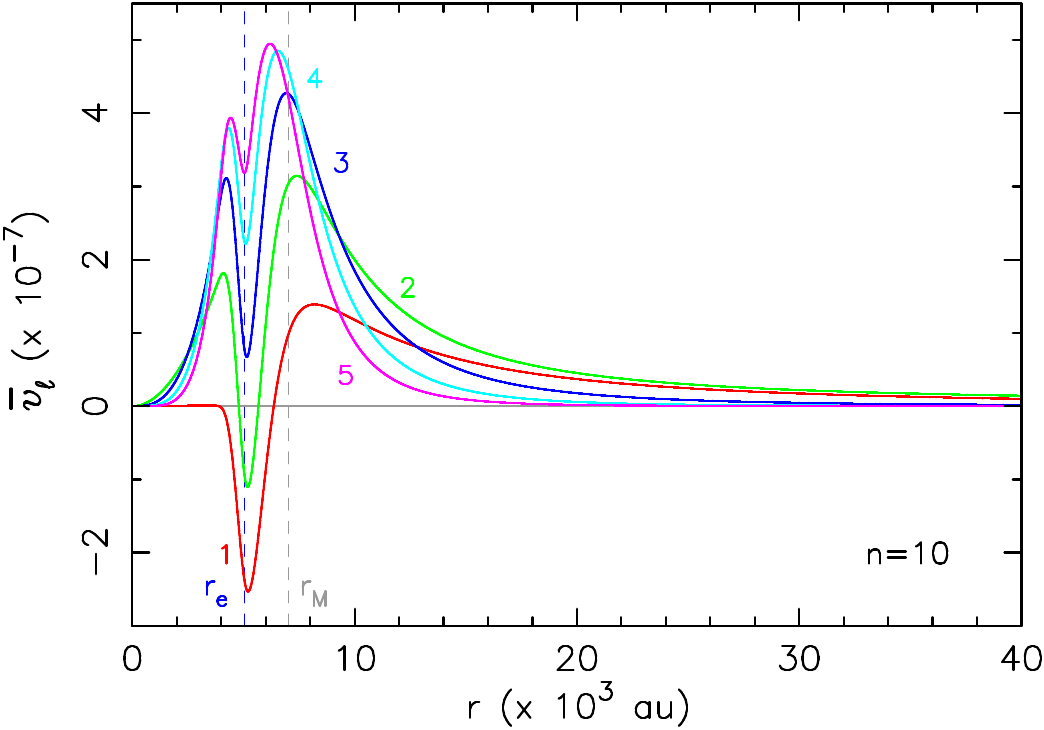} \\ 
 \end{tabular}
 \end{center}  
 \caption{Radial potential functions ${\bar \upsilon}_\ell(r)= \upsilon_\ell(r)/u_{\rm norm}$
  for the multipoles $\ell=1,\ldots,5$ (labeled) of the potential $u(r,\theta)$ from Eq.\ (\ref{e14}). The normalization
 $u_{\rm norm}=GM_\odot/{\rm au}$ makes the ordinate dimensionless. The vertical gray dashed line shows the MOND
 scale $r_{\rm M}=\sqrt{GM_\odot/a_0}$. The vertical blue dashed line at a heliocentric distance $r_{\rm e} =
 r_{\rm M}/\sqrt{\eta}$ ($\eta=g_{\rm e}/a_0$) refers to the approximate location of the singularity in the phantom mass density
 (see the discussion in the main text, as well as Fig.~\ref{fig_dg_global} and Appendix~\ref{details}). Left panel: The gradual transition function $\mu_2(x)$
 is used. The dipole and quadrupole potentials dominate in this case. Right panel: The sharp transition function
 $\mu_{10}(x)$ is used. In this case higher multipoles contribute substantially to the potential, and their
 contributions are more tightly concentrated around the region between $r_{\rm e}$ and $r_{\rm M}$.}
 \label{fig_pot}
\end{figure*}
%%%%%%%%%%%%%%%%%%%%%%%%%%%%%%%%%%%%%%%%%%%%%%%%%%%%%%%%%%%%%%%%%%%%%%%%%%%%%%%%%%%%%%%%%%%%%%%%%%%%%%%

A lot of attention has been focused on testing MOND in the inner solar system (i.e., the strong acceleration regime). Therefore we can check some of our numerical results by comparing them to previously published solutions in this regime. Of particular interest are the values of the lowest-degree multipole moments $Q_\ell(r)$ at the spatial origin $r=0$. First, we verified that in all our solutions the dipole vanishes, namely $Q_1(0)=0$ to
numerical accuracy\footnote{In quantitative terms, our solutions obey $|Q_1(0)|\leq 10^{-15}$ m~s$^{-2}$
 \citep[see also Table~1 and the discussion in Sec.~3.3 of][]{bn2011}. We note that $Q_\ell$ is expected to be $\sim a_0/r_{\rm M}^{\ell-1}$,
 which for $\ell=1$ implies $\sim a_0$. The numerical value of  $|Q_1(0)|$ is therefore
 satisfactorily smaller. \label{foot:q10}} (see \citealt{bn2011} for a proof of this property). Many of the previous studies considered only the quadrupole term evaluated at $r=0$ \citep[e.g.,][]{hetal2014,maquet2015,f2018}, and at best included the octupole term \citep[e.g.,][]{bn2011}. At the quadrupole level the MOND\-ian perturbation is simply expressed as $\delta {\bf g}=\nabla U_2({\bf r})$, with
\begin{equation}
 U_2\left({\bf r}\right)=\frac{1}{6} Q_2\left(0\right)\left[3\left({\bf r}\cdot{\bf e}\right)^2-
  r^2\right]\; , \label{quadr}
\end{equation}
and similarly for the octupole part. Figure~\ref{fig_q230} shows the values of the quadrupole and octupole moments $Q_2(0)$ and $Q_3(0)$ (Eqs.~\ref{e15b} and \ref{e15c}) for various transition functions $\mu_n(x)$. In order to profit from the opportunity to compare our results to those in \citet{bn2011} (their Table~1 and Fig.~7), we use their original $g_{\rm e}=1.9\times 10^{-10}$~m s$^{-2}$ value here (but not elsewhere in the paper). We find good agreement, within $3$\% for $n\leq 6$ (even better for the $n=1$ and $n=2$ cases). The agreement is slightly worse, within $6$\%, for $8\leq n \leq 12$. Only at $n\simeq 20$ does the difference increase to about $9$\%. The source of this mismatch is not known to us, but note that we conduct a number of other internal tests of our numerical results below and in Appendix~\ref{check}. 
 
\citet{hetal2014} analyzed a decade of ranging data to the Cassini spacecraft at Saturn. By approximating the potential using Eq.\ (\ref{quadr}), they derived a limit 
\begin{equation}
    Q_2(0)= (3\pm 3)\times 10^{-27}\mbox{\, s}^{-2} \; . \label{eq:q20}
\end{equation}
This range is marked in gray in the left panel of Fig.\ \ref{fig_q230}. We conclude that the transition functions $\mu_n(x)$ must have $n\ge 6$ \citep[see also][]{bn2011}, i.e., the transition function must rise steeply at $x\simeq 1$ and stay near unity for $x>1$. In other words, the interior parts of the solar system must be strongly shielded from the EFE. \citet{f2018} analyzed a global fit to a large dataset of planetary observations and provide an even stronger constraint, $n\geq 20$.

Restricting $u(r,\tau)$ from Eq.~(\ref{e14}), or acceleration $\nabla u(r,\tau)$ from Eq.~(\ref{hdef}), to the quadrupole and octupole terms with their multipole moments evaluated
at the origin is adequate at the distances of the planets, but cannot be used to describe dynamics in the distant trans-Neptunian zone -- especially for gradual transition functions. To illustrate this, we analyze the behavior of $\delta {\bf g} = \nabla u(r,\tau) - {\bf g}_{\rm N}$ at heliocentric distances $r\leq 1,000$~au, including the radial dependence of the multipole moments $Q_\ell(r)$. In Fig.~\ref{fig_dg_local} we show the results for the transition functions $\mu_2(x)$ (top)
and $\mu_{10}(x)$ (bottom) for position vectors ${\bf r}$ along three directions relative to the external acceleration ${\bf e}={\bf g}_{\rm e}/|{\bf g}_{\rm e}|$: (i) ${\bf r}\parallel {\bf e}$ (left), (ii) ${\bf r}\perp {\bf e}$ (middle), and (iii) $-{\bf r}\parallel {\bf e}$ (right). The little embedded diagrams show
the sampled direction assuming the exterior acceleration is along the positive z-axis.
These results confirm that for transition functions like $\mu_2(x)$, beyond $\simeq 200$~au the evaluation of the MOND\-ian potential must include a large number of multiples and account properly for the variation of the multipole moments with radius.
% FIG 1 %%%%%%%%%%%%%%%%%%%%%%%%%%%%%%%%%%%%%%%%%%%%%%%%%%%%%%%%%%%%
%%%%%%%%%%%%%%%%%%%%%%%%%%%%%%%%%%%
\begin{figure*}[t!]
% \plottwo{.eps}{.eps}
 \begin{center}
 \begin{tabular}{cc}
  \includegraphics[width=0.45\textwidth]{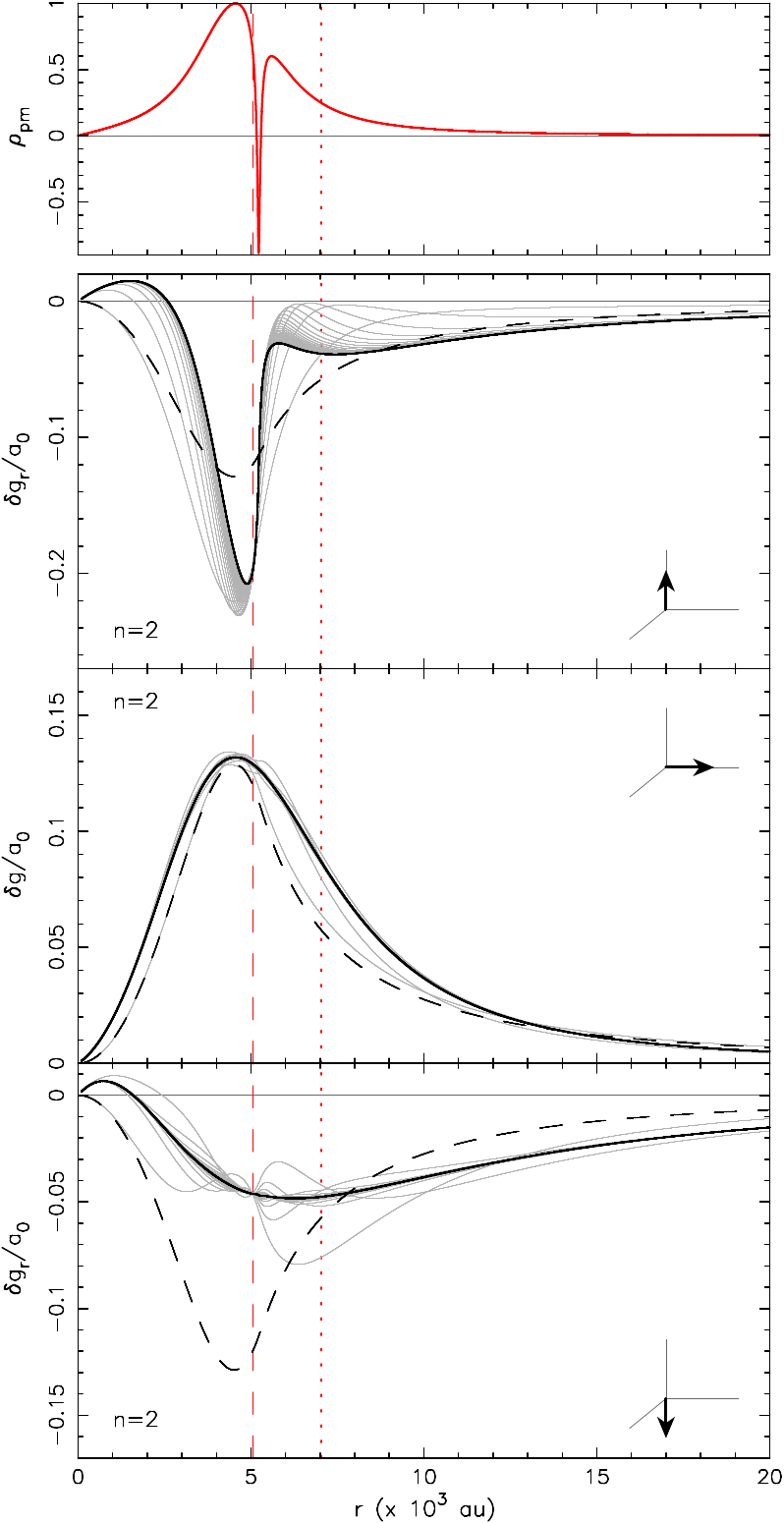} &
  \includegraphics[width=0.45\textwidth]{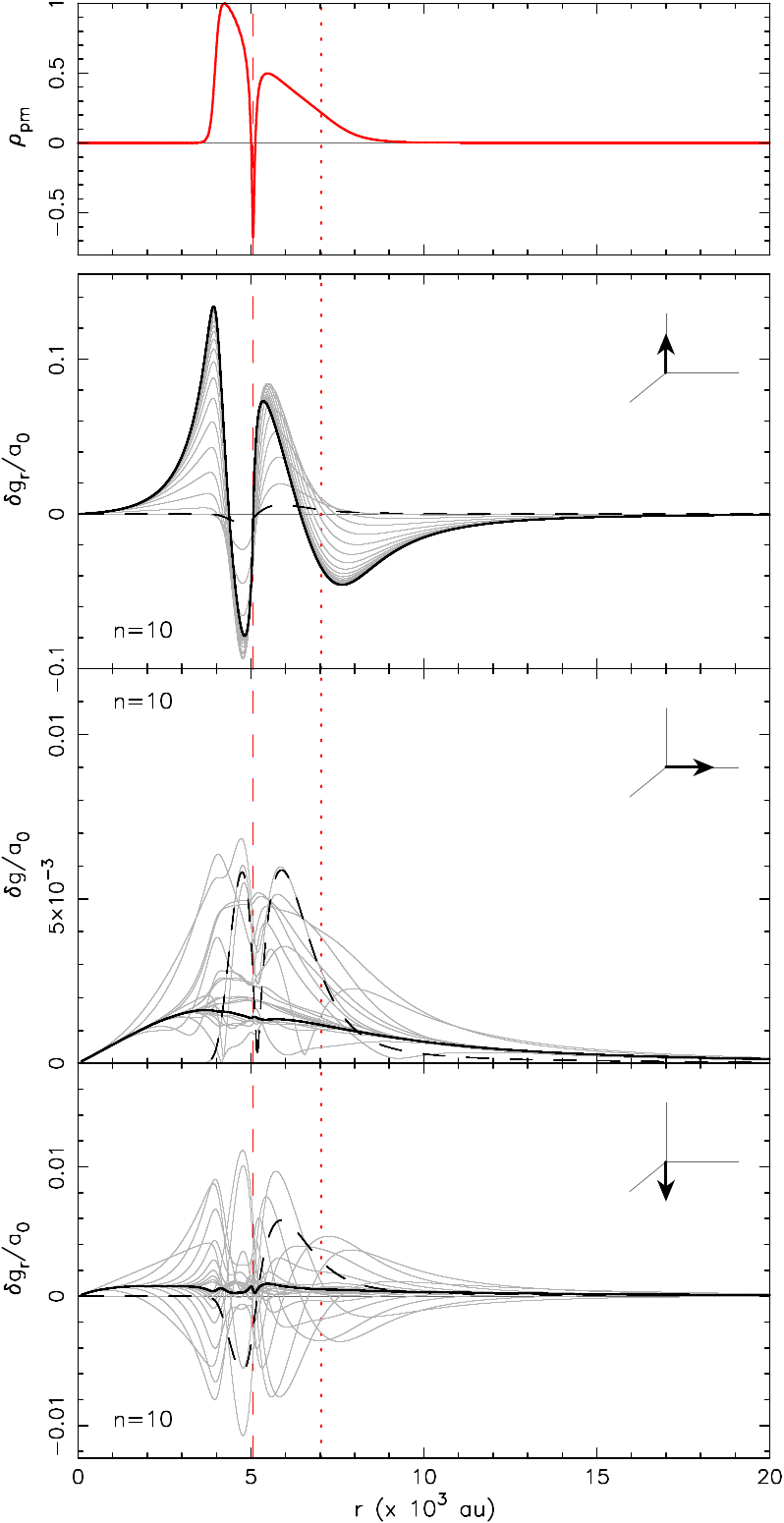} \\ 
 \end{tabular}
 \end{center}  
 \caption{Dependence of the non-Newtonian acceleration, $\delta g/a_0$, on the maximum  multipole $L$ in the expansion (\ref{hdef}). The left panels are for the gradual transition function $\mu_2(x)$, the right panels for the sharp transition function $\mu_{10}(x)$. The three rows show the perturbation to the acceleration (normalized by $a_0$) along three spatial directions: (i) parallel to the direction of the exterior acceleration ${\bf g}_{\rm e}$ (in this case the radial acceleration component $\delta g_r$ is shown), (ii) perpendicular to the exterior acceleration (in this case the magnitude $\delta g
  =|\delta {\bf g}|$ is shown), and (iii) antiparallel to the exterior acceleration (in this case the radial acceleration component $\delta g_r$ is shown); the little embedded diagrams show the sampled direction assuming the exterior acceleration is along the positive $z$-axis. Curves show different cutoff values $L$ of the multipole representation (\ref{hdef}): (i) $L=0$ (black dashed line), (ii) maximum $L=27$ (black solid line), and (iii) intermediate choices in gray. The radial distribution of the phantom matter density $\rho_{\rm pm}$ (Eq.\ \ref{e7}) on the positive $z$-axis, normalized to a maximum absolute value of unity, is shown by the red line in the top row of panels; note that $\rho_{\rm pm}$ may be negative. The dotted red line is located at the MOND scale $r_{\rm M}$ (Eq.\ \ref{eq:rmdef}), and the dashed red line at $r_{\rm e} = r_{\rm M}/\sqrt{\eta}$ ($\eta=g_{\rm e}/a_0$) is the approximate location of the integrable singularity in the phantom matter distribution on the positive $z$-axis, as discussed in the text and in \cite{m1986a} and \cite{bm2023}. The behavior of the radial acceleration $\delta g_r$ broadly reflects the radial distribution of $\rho_{\rm pm}$. Fig.~\ref{fig_dg_local} is similar to this Figure, but zoomed in to the region $r<1,000$ au.}
 \label{fig_dg_global}
\end{figure*}
%%%%%%%%%%%%%%%%%%%%%%%%%%%%%%%%%%%%%%%%%%%%%%%%%%%%%%%%%%%%%%%%%%%%%%%%%%%%%%%%%%%%%%%%%%%%%%%%%%%%%%%
% FIG 1 %%%%%%%%%%%%%%%%%%%%%%%%%%%%%%%%%%%%%%%%%%%%%%%%%%%%%%%%%%%%%%%%%%%%%%%%%%%%%%%%%%%%%%%%%%%%%%%
\begin{figure*}[t!]
% \plottwo{.eps}{.eps}
% \begin{center}
% \begin{tabular}{c}
%  \includegraphics[width=0.85\textwidth]{f6a.eps} \\ [2pt]
%  \includegraphics[width=0.85\textwidth]{f6b.eps} \\ 
% \end{tabular}
% \end{center}  
 \begin{center}
 \begin{tabular}{c}
  \includegraphics[width=0.85\textwidth]{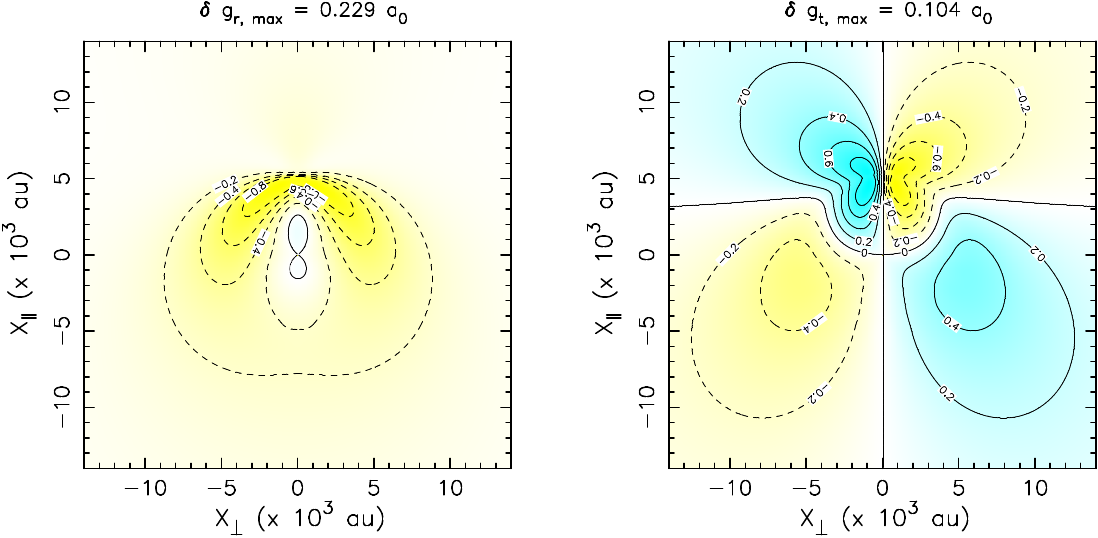} \\ [2pt]
  \includegraphics[width=0.85\textwidth]{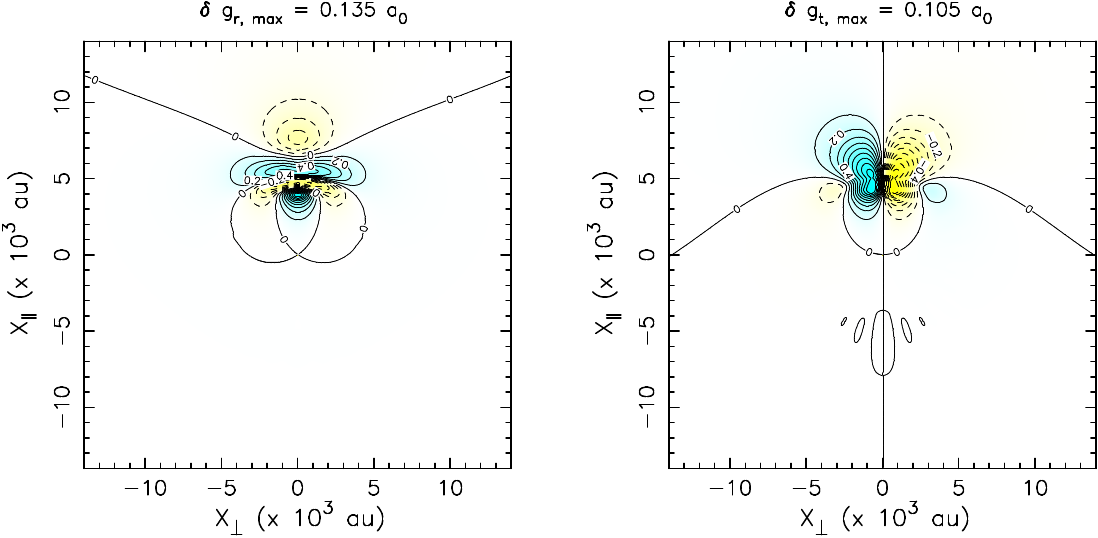} \\ 
 \end{tabular}
 \end{center}
 \caption{Global maps of the radial $\delta g_r = \delta {\bf g}\cdot {\bf n}$ (left panels)
  and transverse $\delta g_t = \delta {\bf g}\cdot [{\bf n}({\bf e}\cdot {\bf n})-{\bf e}]/
  \sqrt{1-({\bf e}\cdot {\bf n})^2}$ (right panels) components of the perturbation to the acceleration due to the EFE. The coordinate $X_\parallel$ is parallel to the direction ${\bf e}$ of
   he external acceleration field and $X_\perp$ is perpendicular to ${\bf e}$. Yellow sectors are for negative values, blue sectors for positive values of the displayed quantity. The color intensity is proportional to the corresponding value of $\delta g_r$ or $\delta g_t$. The maximum
  values, in units of $a_0$, are shown at the top of each panel, and selected contours referred to
  this maximum value are labeled. The top panels are for the gradual transition function $\mu_2(x)$, and the bottom panels for the sharp transition function $\mu_{10}(x)$. The origin of the coordinates is the Sun, and the characteristic radius $r_{\rm M}\simeq 7$ in thousands of au, the units of the axes (Eq.\ \ref{eq:rmdef}). See Fig.\ 1 of \citet{miga2023} for similar plots.}
 \label{fig_dg_global1}
\end{figure*}
%%%%%%%%%%%%%%%%%%%%%%%%%%%%%%%%%%%%%%%%%%%%%%%%%%%%%%%%%%%%%%%%%%%%%%%%%%%%%%%%%%%%%%%%%%%%%%%%%%%%%%%

At still greater distances, in particular beyond the MOND scale $r_{\rm M}$ of 
Eq.\ (\ref{eq:rmdef}), the solution becomes even more complicated, because (i) the boundary 
condition at spatial infinity strongly affects the behavior of the potential functions, and (ii) there is a point near the radius $r_{\rm e}=r_{\rm M}/\sqrt{g_{\rm e}/a_0}=(GM/g_{\rm e})^{1/2}$ at which the phantom mass density diverges, leading to non-analytic behavior of the acceleration field (see below for more detail). In order to see how the solutions 
look, we show in Fig.~\ref{fig_pot} the radial potential functions $\upsilon_\ell(r)$, $\ell=1,\ldots,5$, for two different MOND transition functions: (i) the gradually varying $\mu_2(x)$ (left panel), and (ii) the sharply varying $\mu_{10}(x)$ (right panel)%
\footnote{As we have argued, the $\mu_2(x)$ transition function is excluded by observations of planetary ephemerides, but we use it here to illustrate the behavior of the
 radial potential functions.}.
As a rule of thumb, $|\upsilon_\ell(r)|$ has a global maximum near the MOND scale $r_{\rm M}$, but the function often exhibits an additional extremum near $r_{\rm e} = r_{\rm M}/\sqrt{\eta}=\sqrt{GM/g_{\rm e}}$ ($\eta=g_{\rm e}/a_0$) as discussed below (and further in Appendix~\ref{details}). The limiting behavior at the origin, $v_\ell(r)\propto r^\ell$, corresponds to constant values of the multipole functions $Q_\ell(r)$ (see Eq.\ \ref{e15} and \citealt{bn2011}), while the asymptotic behavior at infinity is briefly discussed in Appendix~\ref{check}. The gradual transition characterized by $\mu_2(x)$ is reflected in the broad distribution of $\upsilon_\ell(r)$, with the lowest multipoles decaying only slowly at heliocentric distances up to many tens of thousands of au, within the region of the traditional Oort cloud. There is also a well-defined hierarchy in the magnitudes of the successive multipoles, with the lowest degree ($\ell=1$ and $\ell=2$) multipoles dominating over the higher degrees. The dipole $\ell=1$ perturbation is very small in the inner parts of the solar system (in fact $Q_1(0)=0$), but its contribution to the total perturbation is comparable to that of the quadrupole ($\ell=2$) near $r_{\rm M}$. The situation is quite different for the sharp transition function $\mu_{10}(x)$ shown in the right panel of Fig.~\ref{fig_pot}. The non-zero contributions of the multipoles are more confined to the region around the MOND scale $r_{\rm M}$. The low-degree multipoles are no longer dominant over the high-degree multipoles, so an accurate description of cometary dynamics requires us to include terms up to large $L$ in (\ref{hdef}). In our simulations below we take $L=27$ (see later in this subsection for justification of this choice).

The acceleration (\ref{hdef}) contains a contribution from the monopole $\ell=0$ term as well. This is simply a radial acceleration of magnitude $d\upsilon_0/dr$. Note that
$\upsilon_0(r)$ replaces the badly behaved $u(r)$ from Eq.\ (\ref{emondinf}); the latter's logarithmically divergent asymptotic behavior is now regularized, with $\upsilon_0(r)\propto 1/r$, although its magnitude is $GMw_0/r$ rather than the Newtonian value $GM/r$, with $w_0$ given by Eq.~(\ref{w0asym}). An additional difference from the Newtonian case is that the asymptotic potential is not spherically symmetric (see Eq.\ \ref{e13}). 

The non-Newtonian radial acceleration $\delta g/a_0$ is shown by the red curves in the top panels of Fig.~\ref{fig_ge_b}, and this can be compared to the same quantity described in Sec.~\ref{internal} for an isolated system (shown by the blue curves in  Fig.~\ref{fig_ge_b}). If we choose the gradual transition function $\mu_2(x)$ (left panels), $\delta g/a_0$ reaches a maximum at a heliocentric distance of about $5,000$~au and then declines, eventually following the usual $\delta g \propto 1/r^2$ asymptotic behavior (rather than the anomalous $\propto 1/r$ for isolated systems, seen in the blue curve). At maximum, $\delta g/a_0 \simeq 0.14$, namely $\delta g \simeq 1.68\times 10^{-11}$ m~s$^{-2}$. In the case of the sharp transition function $\mu_{10}(x)$ (right panels in Fig.~\ref{fig_ge_b}), the radial acceleration $\delta g$ is significantly smaller (the maximum $\delta g/a_0 \simeq 0.01$). Interestingly, it also flips from positive to negative  near $r_{\rm M}$, a reminder that the MOND\-ian phantom mass may be negative \citep[see also][]{m1986a}.
% FIG 1 %%%%%%%%%%%%%%%%%%%%%%%%%%%%%%%%%%%%%%%%%%%%%%%%%%%%%%%%%%%%%%%%%%%%%%%%%%%%%%%%%%%%%%%%%%%%%%%
\begin{figure}[t!]
% \plottwo{.eps}{.eps}
 \begin{center}
  \includegraphics[width=0.47\textwidth]{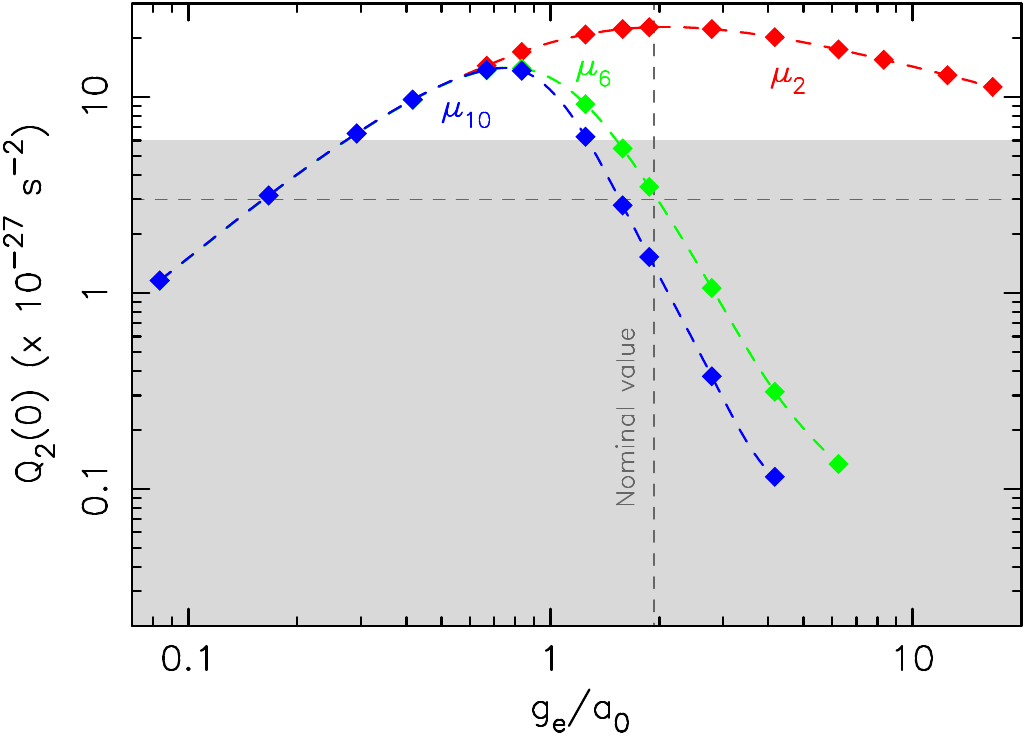} 
 \end{center}  
 \caption{The quadrupole moment $Q_2(0)$ at the origin $r=0$ for different values of the external acceleration $g_{\rm e}$;  $a_0=1.2\times 10^{-10}$ m~s$^{-2}$ is assumed. Solutions are shown for three choices of the MOND transition function (\ref{e2}): (i) $\mu_2(x)$ (red), $\mu_6(x)$ (green), and $\mu_{10}(x)$ (blue). Diamonds represent individual numerical models, and the dashed connecting lines are cubic spline interpolations. The vertical dashed line shows the nominal value $\eta_\star = 1.93$, corresponding to $g_{\rm e} = 2.32\times 10^{-10}$ m~s$^{-2}$. The constraint $Q_2(0)= (3\pm 3)\times 10^{-27}$~s$^{-2}$ from the Cassini spacecraft \citep{hetal2014} is indicated by the gray area (the mean value shown by the horizontal dashed line).}
 \label{fig_ge_a}
\end{figure}
%%%%%%%%%%%%%%%%%%%%%%%%%%%%%%%%%%%%%%%%%%%%%%%%%%%%%%%%%%%%%%%%%%%%%%%%%%%%%%%%%%%%%%%%%%%%%%%%%%%%%%%
% FIG 1 %%%%%%%%%%%%%%%%%%%%%%%%%%%%%%%%%%%%%%%%%%%%%%%%%%%%%%%%%%%%%%%%%%%%%%%%%%%%%%%%%%%%%%%%%%%%%%%
\begin{figure*}[t!]
% \plottwo{.eps}{.eps}
 \begin{center}
 \begin{tabular}{cc}
  \includegraphics[width=0.47\textwidth]{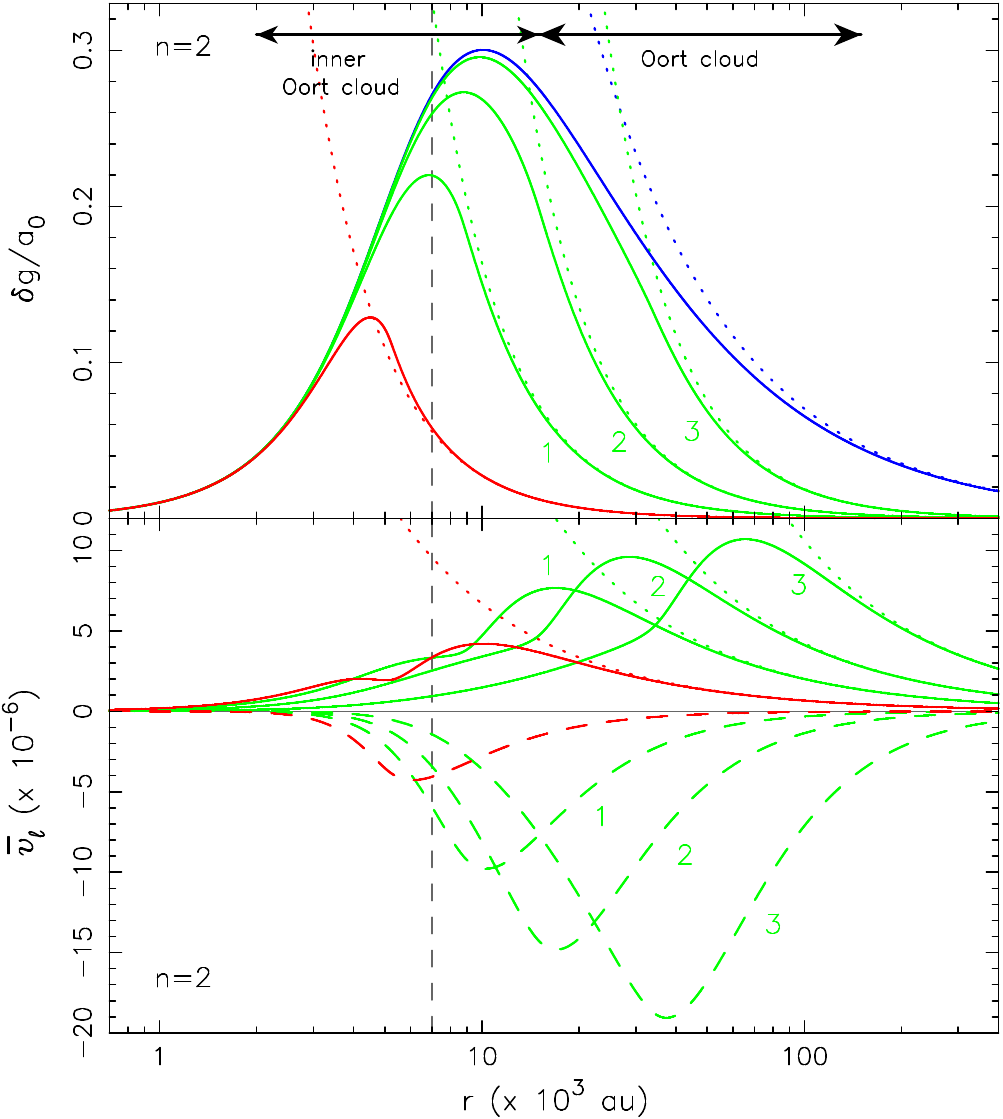} &
  \includegraphics[width=0.47\textwidth]{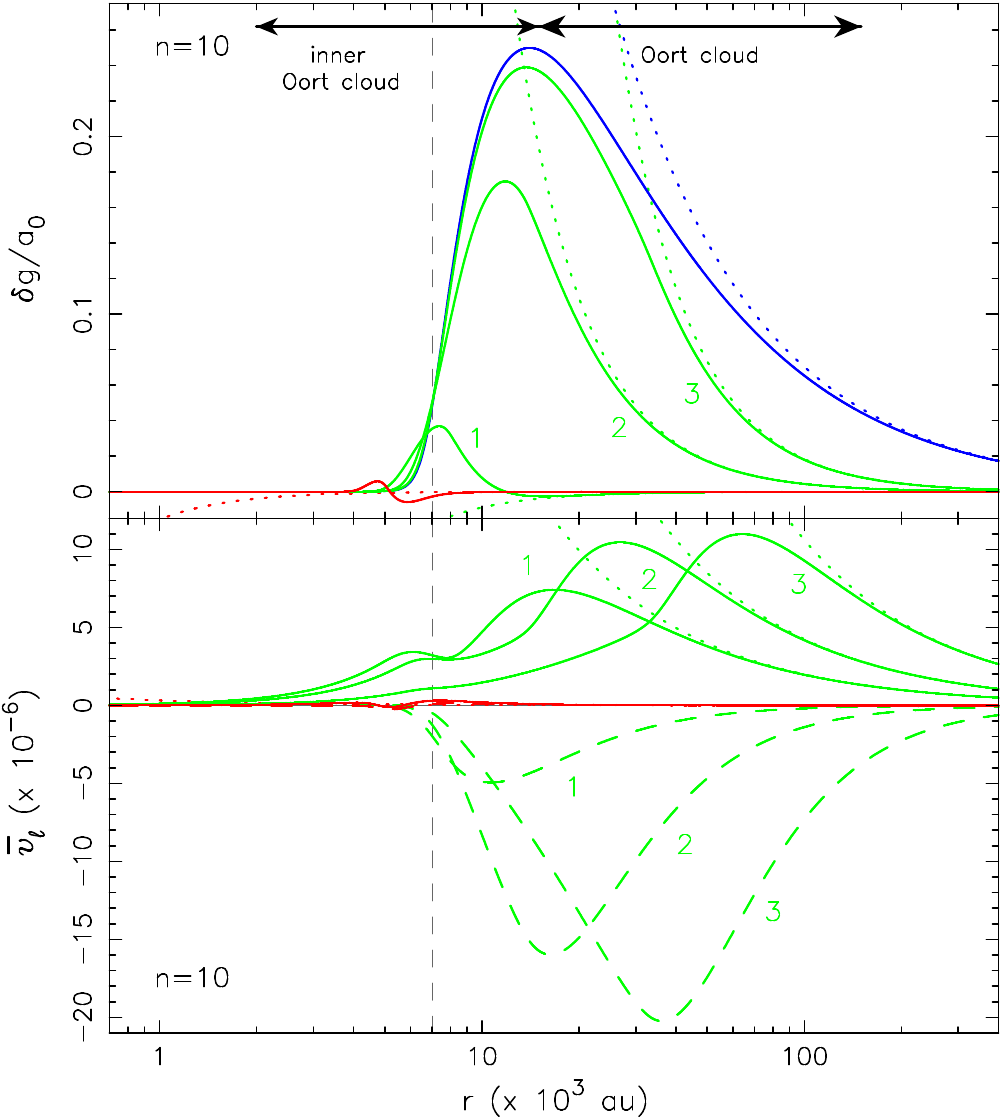} \\ 
 \end{tabular}
 \end{center}  
 \caption{Dependence of the perturbation to the acceleration on the value of $\eta=g_{\rm e}/a_0$ for two choices of the MOND\-ian transition function: (i) $\mu_2(x)$ (left panels), and (ii) $\mu_{10}(x)$ (right panels). The top panels show the perturbation to the radial acceleration from the monopole term (the $l=0$ term in Eq.~\ref{hdef}, with the Newtonian acceleration $g_{\rm N}$ subtracted). The blue line is the (unphysical) case of an isolated system ($\eta=0$), in which $\delta g_{\rm iso} = g_{\rm N}\left[\nu_n\left(y\right)-1\right]$ (Eqs.\ \ref{eq:isoa} and \ref{eq:isob}). The red line corresponds to the nominal choice of the external acceleration $\eta_\star=1.93$ from Sec.~\ref{efe1}. The three green lines are for smaller external accelerations: (i) $\eta=0.83$ (label 1), (ii) $\eta=0.42$ (label 2), and (iii) $\eta=0.17$ (label 3). The bottom panels show the normalized radial functions ${\bar \upsilon}_\ell(r)= \upsilon_\ell(r)/u_{\rm norm}$ for the multipoles $\ell=1$ (dashed lines) and $\ell=2$ (solid lines). The colors and labels are the same as in the top panels (red for $\eta_\star$, green for the three choices of $\eta<\eta_\star$). The dotted lines in all panels show the asymptotic behavior as $r\rightarrow \infty$, from Appendix~\ref{check}. The vertical dashed line shows the MOND scale $r_{\rm M}=\sqrt{GM_\odot/a_0}$, and the traditional locations of the inner and outer Oort cloud regions are indicated by the arrows in the top panels. Smaller external accelerations push the strongest MOND effects to larger heliocentric distances. As the external acceleration shrinks, the monopole radial acceleration $\delta g/a_0$ follows the isolated-system result more closely to larger $r$, but it always eventually approaches zero more rapidly than the isolated-system result, due to the regular boundary conditions at spatial infinity for solutions that include the EFE.}
 \label{fig_ge_b}
\end{figure*}
%%%%%%%%%%%%%%%%%%%%%%%%%%%%%%%%%%%%%%%%%%%%%%%%%%%%%%%%%%%%%%%%%%%%%%%%%%%%%%%%%%%%%%%%%%%%%%%%%%%%%%%

In order to explore the total value of the MOND\-ian contribution to the acceleration, and at the same time to test what maximum multipole $L$ is required in Eq.\ (\ref{hdef}), we conducted the following numerical experiment whose results are summarized in Fig.~\ref{fig_dg_global}. We used the gradual transition function $\mu_2(x)$ in the left panels, and the sharp function $\mu_{10}(x)$ in the right panels. Given the symmetry axis ${\bf e}$ of the external field (approximately the direction to the Galactic
center), we chose three representative radial directions along which the heliocentric position ${\bf r}$ satisfies: (i) ${\bf r}\parallel {\bf e}$ (upper panel), (ii) ${\bf r} \perp {\bf e}$ (middle panel), and (iii) $-{\bf r}\parallel {\bf e}$ (lower panel); the little diagrams on the panel illustrate the geometry of ${\bf r}$ assuming ${\bf e}$ is directed upward along the $z$-axis. In cases (i) and (iii), symmetry requires that the MOND contribution to the acceleration is along the radial (or $z$-) direction and we denote this by $\delta g_r$.  In case (ii) the MOND contribution has both radial and transverse components, so we simply plot the magnitude $|\delta g|$. 

When evaluating the MOND acceleration using
(\ref{hdef}), we sum the contribution of all values of $\ell$ between 0 and $L$.  Two extreme cases are highlighted in Fig.~\ref{fig_dg_global}, namely (i) $L=0$ (black dashed line; this is just the radial monopole acceleration), and (ii) $L=L_{\rm max}=27$ (black solid line). Intermediate values of $L$ are shown by gray lines. Let us start with the left panels corresponding to the gradual transition function $\mu_2(x)$. The convergence of $\delta {\bf g}(L)$ with increasing $L$ is quite regular, and we are satisfied to stop at $L=L_{\rm max}=27$. The situation is more complicated in the case of the sharp transition function $\mu_{10}(x)$ in the right panels of Fig.~\ref{fig_dg_global}. The fundamental reason is that the phantom mass density is much more localized, and thus higher multipoles $L$ are needed to resolve its effects. The convergence with $L$ is slow especially in the configurations depicted on the middle and bottom panels (${\bf r}\perp {\bf e}$ and $-{\bf r} \parallel {\bf e}$). There remains still some jitter at $L=L_{\rm max}$, but we believe that this residual variation is acceptable given that the computational costs grow rapidly with $L_{\rm max}$.

% ----
The rapid evolution of the radial component $\delta g_r$ of the acceleration along the positive $z$-axis is related to the presence of a critical point $P$ near heliocentric distance $r_{\rm e} = r_{\rm M}/\sqrt{\eta}=\sqrt{GM/g_{\rm e}}$ ($\eta=g_{\rm e}/a_0$). The radius $r_{\rm e}$ corresponds to a saddle point of the Newtonian potential $u_{\rm N}$ or zero of the corresponding acceleration $g_{\rm N}$; the zero occurs when the acceleration due to the Sun cancels the external acceleration ${\bf g}_{\rm e}$. A number of studies have discussed the role of $P$ in relation to the phantom mass density distribution (see \citealt{m1986a,bma2006,p2020a,bm2023} and Appendix~\ref{details}). If the critical point $P$ is at $r_\star\sim r_{\rm e}$, then near this point the phantom mass density varies as $\rho_{\rm pm}\propto |r-r_{\star}|^{-\alpha}$, with $\alpha<1$ \citep[][Sec.~III, argue that $\alpha=1/2$ generically]{bma2006}. Considering for the sake of simplicity only the radial acceleration component, one obtains $\delta g_r\propto \pm |r-r_{\star}|^{1-\alpha}$ locally (opposite signs for $r<r_\star$ and $r>r_\star$, respectively). This is a bow-shock-type transition, implying that near $P$ the acceleration $\delta g_r$ undergoes a rapid change. The top panels on Fig.~\ref{fig_dg_global} reflect this phenomenon.

% ----
Figure~\ref{fig_dg_global1} shows maps of the radial and transverse components of the non-Newtonian accelerations $\delta {\bf g}$ for the two example transition functions $\mu_2(x)$ (top) and
$\mu_{10}(x)$ (bottom). As expected, the perturbation in the latter case is much more localized and centered
at a heliocentric distance of about $r_{\rm M}$ in the direction ${\bf e}$ of the exterior acceleration.

There are two take-away messages from this analysis. First, the non-Newtonian acceleration field $\delta {\bf g}$ has a complex geometry at heliocentric distances of a few thousand au, where it influences the Oort cloud structure and the orbital history of LPCs. Unlike in the inner zone of the solar system, the MOND\-ian potential in this region  
cannot be approximated by a simple analytic formula (e.g., Eq.\ \ref{quadr}), and must be evaluated numerically. Second, the maximum values of $\delta g$ represent only a fraction of $a_0$, and remain smaller than $g_{\rm N}$ (in contrast with the MOND\-ian acceleration of an isolated, spherically symmetric system from Fig.~\ref{fig1}). This means that we can introduce MOND as a small perturbation in an $N$-body integrator designed for Keplerian orbits, as described in Sec.~\ref{inte}.

\subsubsection{Effects of changing the exterior acceleration}\label{efe2}
We briefly consider the sensitivity of the results to changes in the magnitude of the
external acceleration $g_{\rm e}$, keeping $a_0$ fixed. There are three reasons for this exploration: (i) our knowledge of this parameter is not exact; (ii) the external acceleration might have been different in the past, in particular because of migration of the Sun in the Galactic disk; (iii) it is instructive to see the role
of $g_{\rm e}$ in general (in particular the limits of small and large $g_{\rm e}$).

First, we determined the value of the quadrupole moment $Q_2(0)$ at the origin as a function of $g_{\rm e}$ (since this is well-constrained from planetary observations; see Eq.\ \ref{eq:q20}). The results are shown in Fig.~\ref{fig_ge_a}, where we allow $\eta=g_{\rm e}/a_0$ to span a range from $\simeq 0.1$ to $\simeq 20$, with the nominal value from Sec.~\ref{efe1} equal to $\eta_\star\simeq 1.93$. We explore three example transition functions $\mu_2(x)$, $\mu_6(x)$ and $\mu_{10}(x)$. At the nominal value $\eta_\star$ the quadrupole moments differ by up to an order of magnitude (as already shown in Fig.~\ref{fig_q230}) and only $\mu_6(x)$ and $\mu_{10}(x)$ satisfy the constraints from planetary ephemerides (shaded region). However, when $g_{\rm e}$
is less than $\sim 0.6$ times the nominal value, the results for the different transition functions converge. For $\eta\leq 0.3$, an order of magnitude smaller than $\eta_\star$, the planetary ephemerides are consistent with all three transition functions. This is because when $g_{\rm e}$ is small the external effects do not penetrate deep into the planetary zone -- they are confined to larger heliocentric distances. Interestingly, $Q_2(0)$ also shrinks for external accelerations $g_{\rm e}$ larger than the nominal one. This is because the MOND\-ian effects disappear when $g_{\rm e}\gg a_0$ \citep[e.g.][]{fm2012}, making the multipole moments approach zero. This trend is rapid for steep transition functions, and somewhat slower for gradual transition functions -- Fig.\ \ref{fig_ge_a} shows that $Q_2(0)$ is only a factor of two smaller for the $\mu_2(x)$ case when $\eta\simeq 10\,\eta_\star$, consistent with Fig.~6 in \citet{bn2011}, while it drops much more rapidly for the $\mu_6(x)$ and $\mu_{10}(x)$ cases.

 Figure~\ref{fig_ge_b} illustrates the contribution of the lowest multipoles in Eq.\ (\ref{hdef}) to the acceleration when $g_{\rm e}$ is smaller than the nominal value (when $g_{\rm e}$ becomes larger than the nominal value, the MOND\-ian effects rapidly fade away, especially for sharp transition functions). We compare results for the gradual transition function $\mu_2(x)$  (left panels) and the sharp transition function $\mu_{10}(x)$ (right panels). The top panels show the perturbation to the acceleration of the monopole term, namely a purely radial acceleration $d\upsilon_0/dr$. We find it useful to compare this to the corresponding radial acceleration $\delta g_{\rm iso} = g_{\rm N}\left[\nu_n\left(y\right)-1\right]$ (Sec.~\ref{internal}) of the isolated
MOND\-ian system, shown by the blue curve. The red curve shows $\delta g$, normalized by $a_0$, for the nominal external acceleration, $\eta_\star$. The green lines are for three cases where $\eta<\eta_\star$. In the case of the weakest external acceleration shown here ($\eta=0.17$), the green curve (label 3) closely follows the isolated-system perturbation up to the beginning of the Oort cloud at $\simeq 15,000$~au heliocentric distance. The boundary conditions at spatial infinity impose $\delta g\propto 1/r^2$ as $r\to\infty$ for any non-zero $g_{\rm e}$, which contrasts with $\delta g_{\rm iso}\propto 1/r$ for an isolated MOND system. This explains the divergence of the green and blue curves at the largest heliocentric distances. 
The value of $\delta g/a_0$ is negligible up to heliocentric
distances of a few thousand au in the case of the sharp transition function $\mu_{10}(x)$ (right panels).
This result holds even for small values of $\eta$, implying that the monopole MOND\-ian
perturbation is efficiently expelled from radii $r\lesssim r_{\rm M}$. This is not true for the gradual transition function $\mu_2(x)$ (left panels), for which even the monopole perturbation penetrates well below $r_{\rm M}$.

The bottom panels in Fig.~\ref{fig_ge_b}  show the dipole and quadrupole potential functions $\upsilon_1(r)$ and $\upsilon_2(r)$. The red lines are for $\eta_\star$, and the three green lines for $\eta<\eta_\star$ as in the top panels. A smaller value of the external acceleration $g_{\rm e}$ generally repels the MOND\-ian effects to larger heliocentric distances. A characteristic property of these dipole and quadrupole potential functions is that they peak in absolute value near the heliocentric distance $r_{\rm e}=r_{\rm M}/\sqrt{\eta}$ that characterizes the zero point of the Newtonian acceleration, rather than at the MOND scale $r_{\rm M}$ (at the same time, they extend to smaller radii than the monopole perturbation $\delta g/a_0$). The magnitude of the extrema in $\upsilon_1(r)$ and $\upsilon_2(r)$ grows as $\eta$ shrinks, suggesting that there is more phantom mass beyond $r_{\rm M}$ when $\eta$ is small (see Appendix~\ref{check}).

\subsection{Tidal field of the Galaxy}\label{efe3}
We turn now to the effects of the Galactic gravitational field on the motion of bodies bound to the solar system, starting with the Newtonian description. As discussed in Sec.~\ref{efe}, the Galactic acceleration on a test particle near the Sun can be divided into two terms, namely ${\bf g}_{\rm ex} = {\bf g}_{\rm e} + {\bf g}_{\rm tide}$. The first term ${\bf g}_{\rm e}$
is the uniform acceleration of the solar system; in the heliocentric Newtonian description the effects of this term vanish because the solar system is freely falling in the Galactic field, but in the MOND\-ian description this is the source of the EFE, which has profound effects on the motion of LPCs and other bodies at large heliocentric distances (Sec.~\ref{efe}). 

The gravitational tidal effects ${\bf g}_{\rm tide}$ arise from a quadrupole tidal potential.  We assume that the solar orbit around the Galaxy is circular and lies in the Galactic midplane. Then the tidal field is most conveniently described with respect to a heliocentric, slowly rotating orthonormal reference frame $({\bf e}_x,{\bf e}_y, {\bf e}_z)$, such that ${\bf e}_x$ points away from the Galaxy center (therefore in this system ${\bf g}_{\rm e}=-g_{\rm e}\,{\bf e}_x$), ${\bf e}_y$ points in the direction of the solar motion in the Galaxy, and ${\bf e}_z$ is normal to the Galactic midplane. The tidal acceleration ${\bf g}_{\rm tide}$ is then a linear function
of the coordinates in this system \citep[e.g.,][]{ht1986,bt2008,rickman2010}\footnote{Some studies simplify ${\bf g}_{\rm tide}$ by considering only the first term in the bracket multiplying the vertical tide component (i.e., along ${\bf e}_z$). This is because it numerically dominates other contributions by an order of magnitude.}
\begin{equation}
 {\bf g}_{\rm tide} = \Omega_0^2\left[\left(1-2\delta\right)x\,{\bf e}_x-y\,{\bf e}_y
          -2\left(2 {\cal Q}-\delta\right)z\,{\bf e}_z
          \right]\, . \label{tide}
\end{equation}
The independent coefficients $(\Omega_0,\delta,{\cal Q})$ are often represented using the Oort constants $A$ and $B$ and the mass density in the solar neighborhood $\rho_0$: $\delta = -(A+B)/(A-B)\simeq -0.09$, $\Omega_0 = A-B \simeq 2.78\times 10^{-8}$~yr$^{-1}$, and ${\cal Q}=\pi G\rho_0/\Omega_0^2\simeq 2.75$. These values assume $A=14.82$ km~s$^{-1}$~kpc$^{-1}$ and $B=-12.37$ km~s$^{-1}$~kpc$^{-1}$ \citep{fw1997}, and $\rho_0\simeq 0.15$~M$_\odot$ pc$^{-3}$. The 
$(\Omega_0,\delta,{\cal Q})$ parameters are uncertain, in part because they are based on the assumption that the Galaxy is in an axisymmetric steady state, which is only approximately true. For example the model by \cite{mcmil2017} has $A=14.2$ km~s$^{-1}$~kpc$^{-1}$, $B=-14.2$ km~s$^{-1}$~kpc$^{-1}$ and $\rho_0\simeq 0.11$~M$_\odot$ pc$^{-3}$. In our simulations (Sec.~\ref{res}) we adopted these latter values and we keep them constant, neglecting the effects of any possible radial migration of the solar system in the Galaxy on Gyr timescales \citep[e.g.,][]{ketal2011,detal2015}. To investigate the effects of migration, we performed several simulations with larger accelerations $g_{\rm e}$, corresponding to a past solar-system location that was closer to the Galactic center (Sec.~\ref{res}), but (somewhat inconsistently) we kept $(\Omega_0,\delta,{\cal Q})$ the same in order to isolate the effects of changes in the acceleration. 

It is relevant now to estimate the expected acceleration regime of ${\bf g}_{\rm ex}$ up
to the heliocentric distance of the outer Oort cloud.
The uniform part is $g_{\rm e}=2.32\times 10^{-10}$ m~s$^{-2}\sim a_0 =1.2\times 10^{-10}$ m~s$^{-2}$,
if the nominal values are assumed. The tidal part ${\bf g}_{\rm tide}$ is largely subcritical,
because $g_{\rm tide}\lesssim 4\Omega_0^2\,{\cal Q}\,\times 10^5\; {\rm au}\simeq 10^{-3}\,a_0$. Thus we may describe the effects of MOND on the tidal acceleration ${\bf g}_{\rm tide}$ using perturbation theory, which we now do. 
\smallskip

\noindent{\it MOND\-ian correction of the quadrupolar tide.-- }A full-fledged MOND\-ian description
of the motion of distant solar-system bodies would require the inclusion of ${\bf g}_{\rm tide}$ in the solution presented in Sec.~\ref{efe}.%
\footnote{An alternative approach would be to adopt the Galactocentric reference frame for the description of the MOND\-ian corrections of the Galactic global potential. Next, introduce the heliocentric reference frame, and the Sun as a point mass, and consider the appropriate transformation of the equations of motion.} There, however, we kept only the
uniform component ${\bf g}_{\rm e}$ of the total Galactic acceleration ${\bf g}_{\rm ex}$. This allowed us to take advantage of the axisymmetry of the problem in developing a self-consistent numerical solution of the non-linear equation (\ref{e5}). Including ${\bf g}_{\rm tide}$ in our treatment of the EFE would have significantly complicated the problem. We choose instead to treat the MOND\-ian part of ${\bf g}_{\rm tide}$ in an approximate way. 

We start by noting that ${\bf g}_{\rm tide}=\nabla u_{\rm tide}$, where $u_{\rm tide}$ is a diagonal quadratic form in the reference frame $({\bf e}_x,{\bf e}_y, {\bf e}_z)$ introduced above, which can be derived from (\ref{tide}). At the Newtonian level it satisfies the Poisson equation $\nabla^2 u_{\rm tide}=-4\pi G \rho_0$, where $\rho_0$ is the local density of Galactic material (other stars and interstellar gas and dust). In MOND, we need to replace $u_{\rm tide}$ according to the steps outlined in Sec.~\ref{theory1}. The total acceleration ${\bf g}=\nabla U$, generalizing ${\bf g}_{\rm ex}$, satisfies Eq.~(\ref{e6}) with two components on the right-hand side, the homogeneous density $\rho=\rho_0$, and the phantom matter $\rho_{\rm pm}$ given by Eq.~(\ref{e7}). As before, $U={\bf g}_{\rm e}\cdot{\bf r}+w$, with ${\bf g}_{\rm tide}=\nabla w$ being the relevant acceleration component that perturbs the motion of solar-system bodies. Thus we obtain 
\begin{equation}
  \nabla^2 w = -4\pi G (\rho_0 + \rho_{\rm pm}) ,\label{ee1}
\end{equation}
where $\rho_{\rm pm}$ depends on $w$. 
We solve this non-linear equation approximately by replacing $w$ with $u_{\rm tide}$ in the expression (\ref{e7}) for the phantom matter density, which then takes the explicit form
\begin{equation}
  \rho_{\rm pm} = \frac{1}{4\pi G}\,\nabla\cdot\left(\chi_{\rm ex}\,{\bf g}_{\rm ex}\right)\; ,\label{ee2}
\end{equation}
with $\chi_{\rm ex}=\mu_{\rm ex}-1$, $\mu_{\rm ex}=\mu(g_{\rm ex}/a_0)$. The acceleration ${\bf g}_{\rm ex}$ is now the Newtonian approximation of the external field. Simple algebra results in
\begin{equation}
 \frac{\rho_{\rm pm}}{\rho_0} = \frac{\lambda_{\rm ex}\mu_{\rm ex}}{4{\cal Q}}\,\frac{{\bf g}_{\rm ex}\cdot
   \mathbb{T}\cdot {\bf g}_{\rm ex}}{g_{\rm ex}^2}-\chi_{\rm ex}\; ,\label{ee3}
\end{equation}
where $\lambda_{\rm ex} = d\log\mu(x)/d\log x|_{x=g_{\rm ex}/a_0}$, and the  matrix $\mathbb{T}$ takes the diagonal form $\mathbb{T}={\rm diag}
[1-2\delta,-1,-2(2{\cal Q}-\delta)]$. To estimate the order of magnitude of the MOND\-ian correction, one may further replace ${\bf g}_{\rm ex}$ by ${\bf g}_{\rm e}$ in (\ref{ee3}). For transition functions of the family $\mu=\mu_n(x)$ (Eq.\ \ref{e2}), we have a particularly simple form
\begin{equation}
 \lambda_n(x) = \frac{1}{1+x^n}\; . \label{ee4}
\end{equation}
It is straightforward to check that the right-hand side of Eq.\ (\ref{ee3}) is always small within the Oort cloud's heliocentric distance from the Sun. In particular, it reaches $\simeq 0.15$ for the gradual transition function characterized by $n=2$, but is smaller than $10^{-3}$ when $n\geq 10$ \citep[compare with QUMOND-based estimates of $\rho_{\rm pm}/\rho_0$ in][which are similar]{pk2016}. As a result the phantom matter density is always small compared to the matter density $\rho_0$. This finding already hints that the MOND\-ian correction to the
Galactic tide is small, and justifies the approximate method we have used to evaluate it. 

Define the MOND\-ian correction to the tidal field as $\delta u_{\rm tide}=w-u_{\rm tide}$,
corresponding to an acceleration $\delta {\bf g}_{\rm tide}= \nabla w - {\bf g}_{\rm tide}$. Using (\ref{ee1}) we note that
$\delta {\bf g}_{\rm tide}$ obeys $\nabla\cdot\left(\delta {\bf g}_{\rm tide}+\chi\,{\bf g}_{\rm ex}
\right)=0$, with a general solution
\begin{equation}
 \delta {\bf g}_{\rm tide}= -\chi_{\rm ex}\,{\bf g}_{\rm ex} + \nabla \Phi + \nabla \times {\bf A}\; .
   \label{ee5}
\end{equation}
where $\Phi$ is an arbitrary harmonic function, a solution of the Laplace equation $\nabla^2\Phi=0$, and ${\bf A}$ an arbitrary
vector field. The weak equivalence principle, satisfied in MOND \citep[e.g.,][]{bm1984,bn2011}, requires that the acceleration is zero at the origin. This requirement is easily satisfied by considering the appropriate linear field in $\Phi={\bf F}\cdot{\bf r}$. Other components
in $\Phi$, and ${\bf A}$, still offer a large degree of freedom. For instance, the exact solution of the non-linear MOND\-ian
equation for $w$ may generate higher-order tidal multipoles, whose presence already at the Newtonian level is related to
non-uniformity of the density term on the right-hand side of (\ref{ee1}). Here, however, we restrict ourselves to the
simple assumption that $w\rightarrow u_{\rm tide}$ at spatial infinity. This corresponds to
\begin{equation}
 \delta {\bf g}_{\rm tide}= -\left(\chi_{\rm ex}-\chi_{\rm e}\right)\,
  {\bf g}_{\rm e}-\chi_{\rm ex}\,{\bf g}_{\rm tide} \; ,   \label{ee6}
\end{equation}
where $\chi_{\rm e}=\mu(g_{\rm e}/a_0)-1$. The second term can be directly interpreted as an amplification of the usual tidal acceleration ${\bf g}_{\rm tide}$. This effect is small, less than $10^{-3}\, g_{\rm tide}$ for admissible transition functions having $n\geq 10$. The first term is also quite small, because $\chi_{\rm ex}-\chi_{\rm e}$ remains $\leq 10^{-6}$. Thus the first-order calculation we have used to evaluate the MOND\-ian change in ${\bf g}_{\rm tide}$ represents a small correction and should be quite accurate.

\section{Methods}\label{met}
Before presenting our results, we briefly outline the overall architecture of the numerical simulations we carry out. For the most part, our approach is very close to that of \citet{vnd2019}, and we refer the reader to that paper for more detail.

\subsection{Integration method}\label{inte}

Orbit integrations were performed using {\tt swift\_rmvs4}, part of the {\tt swift}
$N$-body package.%
\footnote{\citet{swift1994} and \url{http://www.boulder.swri.edu/~hal/swift.html}.}
The core of {\tt swift} is the implementation of the symplectic integrator developed by \citet{wh1991}, which describes in properly chosen canonical variables motion close to  conics in planetary systems. Specifically, denoting the timestep by $h$, the second-order scheme implemented in {\tt swift} begins with execution of the effects of the (mutual) planetary perturbations acting on
momenta for a duration $h/2$, followed by the free motion on heliocentric conics for $h$, and finally completing the timestep by again performing the planetary perturbations acting on momenta for $h/2$. This procedure is applied to $N$ massive bodies (planets, the Sun, and passing stars in our method) and $M$ massless particles (planetesimals/comets). An added feature of {\tt swift}, described by \citet{swift1994}, is the careful treatment of close encounters between planets and particles. This is achieved by an appropriate transformation from a heliocentric to a planet-centric description for particles within the Hill sphere of the planet. The code also applies special treatment to particles in an intermediate zone between 1 and $3.5$ Hill radii from a planet. In this case, the propagation remains in the heliocentric description but the timestep is decreased, as described by \citet{swift1994}.
      
We have extended the original version of {\tt swift\_rmvs4} by including additional perturbing accelerations arising from the Galactic tide, passing stars and MOND. The added accelerations, like the accelerations due to planetary perturbations, are implemented during the half-steps when positions are fixed and momenta changed.

In some of our simulations we also added artificial accelerations that allowed us to mimic the planetesimal-driven migration of the giant planets during the first tens to hundreds of Myr after the formation of the solar system. This is an approximate method that replaces the full-scale simulations presented in \citet{nm2012}, in which the planetesimals feel the planetary perturbations and at the same time make the planets migrate. The scheme implemented here is the one developed by \cite{nv2016}, which closely reproduces
the simulations from \citet{nm2012} and leads to a final state that matches the
current architecture of the giant planets. Following the currently emerging picture of planet formation in the protoplanetary nebula, the giant planets were left in a resonant and compact configuration at the moment of gas dispersal. This moment sets the
initial time in our simulations. The fine details of giant-planet migration determine the architecture of the populations of small bodies (Kuiper belt objects) currently found in the immediate trans-Neptunian region \citep[e.g.,][]{nes2018}.
Fortunately the populations currently located farther out, such the detached disk, the scattered disk, and the Oort cloud, are less sensitive to the details of the migration history (see Appendix \ref{jobs2p}). We may thus simplify the situation and let Jupiter and Saturn reside on their current orbits during the whole duration of our simulations. Uranus and Neptune are started on nearly circular orbits at $17$~au and $22$~au respectively, in the plane defined by the total orbital momentum of Jupiter and Saturn (the terrestrial
planets are not modeled). The artificial accelerations acting on Uranus and Neptune make them approach their current orbits with semimajor axes $19$~au and $30$~au. During the initial phase, the evolution is characterized by an $e$-folding timescale of $10$~Myr. When Neptune reaches $27.5$~au its orbit is allowed to jump
forward by $0.4$~au \citep[see][]{n2015}, the sole trace of a giant-planet instability, otherwise a major event of the inner solar-system dynamics. At that moment, Neptune's eccentricity is assumed to be excited to $0.1$. After the instability, the $e$-folding pace of the orbital evolution of Uranus and Neptune is slowed to $30$~Myr. Figure~1 in \cite{netal2020} illustrates the planetary evolution in our simple model, and its ability to reproduce the more realistic case in which planetesimals drive the planetary migration. We complete this evolution in the first $100$~Myr of our simulation. For the remaining 4.4 Gyr of the simulation, the planets are fixed on their current orbits. For readers who are skeptical of this migration model, we contrast the results with those from a model with no planetary migration in Appendix \ref{jobs2p}. 

The simulation also contains a population of planetesimals, modeled by massless particles. When Neptune sets off on its outward migration, it enters a disk of planetesimals (described later in Sec.~\ref{ini}), scattering them both inward (towards the Sun) and outward. This process is most intense during the initial $\simeq 100$~Myr in which the outer planets are migrating. At the end of this period, the initial planetesimal disk is dispersed and a tiny fraction ($\leq 1\%$) is transformed into a structure called the scattered disk (SD) (there are other, even smaller, orbitally stable populations of planetesimals as well). The SD is a collection of orbits extending beyond the orbit of Neptune, semimajor axes $\geq 30$~au, but for the most part still gravitationally communicating with this planet, perihelia $\leq 38$~au. Orbits in the SD are unstable because of their repeated interactions with Neptune, but the characteristic timescale involved in their evolution is hundreds of Myr to Gyr. Exterior resonances with Neptune may temporarily lock the SD orbits into quasi-stable islands
\citep[e.g.,][]{netal2016,ks2016}, but more efficient processes cause most of them to evolve into two quasi-stable populations:
\begin{itemize}

\item in MOND, the EFE may apply a torque that moves particles from the SD to larger perihelion distances ($> 38$ au), at which the orbits are largely unaffected by Neptune and the other planets; this population is called the detached disk \citep[see already][]{pk2016,p2017}; and

\item Galactic tides and stellar encounters, perhaps assisted by the EFE, transfer SD orbits beyond $\simeq 2,000$~au heliocentric distance to the Oort cloud. 
\end{itemize}
Current sky surveys have discovered dozens of objects in the detached disk, enabling us to probe the efficiency of the dynamical processes -- including MOND -- 
that feed the region of phase space occupied by the detached disk. At the same time, both the detached disk and the Oort cloud feed the population of LPCs, so the distribution 
of orbital properties of the LPCs offers a second potential test of the influence of MOND.
The Galactic tides implemented in our simulations (Sec.~\ref{efe3}) are computed for the present Galactic environment of
the solar system. They were likely stronger in the past for two reasons: (i) solar-system migration from smaller Galactocentric
distance, and (ii) there is a brief period when the solar system still resided in its natal cluster of stars. In a
previous work \citep{clu2023}, we studied (ii) and found that during this period the detached disk may be also efficiently populated beyond semimajor axes of a few hundred au by fierce cluster tides and perturbations by close stellar passages \citep[in accord with a number of previous studies, e.g.,][]{ml2004,bra2006}.
An example from these simulations is shown in Fig.~\ref{fig_detach}. One could thus wonder, whether this process in the Newtonian model may not mimic MOND\-ian perturbations, which are expected to produce a similar result. However, this is not the case. There is a fundamental
difference between the detached disk populations at hundreds to thousands of au that are produced by Newtonian cluster tides and encounters, as compared to  MOND. In the former case, the detached disk is a fossil structure, in which the orbits do not evolve and do not contribute to the current LPC population.
In contrast, in MOND the orbits in the detached disk are continually  exchanging places with orbits in the LPC population (Fig.\ \ref{fig_detach1}).

The MOND\-ian dynamics implemented in our simulations was described at length in
Sec.~\ref{theory}. The effects of Galactic tides and stellar encounters are briefly recalled in Sec.~\ref{tidstars}.

\subsection{Planetesimal disk}\label{ini}

Various lines of evidence indicate that the primordial planetesimal disk was divided in two parts. The first, extending in our model from the initial orbit of Neptune at 22 au to about $30$~au, had an estimated mass of $\simeq 15\mbox{--}20\,$~M$_\oplus$. This  mass, comparable to that of Neptune, is required to make Neptune migrate  some $\simeq 6\mbox{--}8$~au across the whole extent of the disk (as demanded by the orbital architecture of the Plutino population, planetesimals captured in the exterior 3/2 mean-motion resonance with Pluto). Figure~14 in \citet{spc2017} provides a hint about the size distribution of planetesimals in this part of the disk. Beyond $30$~au heliocentric distance, the disk mass dropped considerably, to only a few times $10^{-3}$~M$_\oplus$, even though this component of the disk extended to nearly $50$~au.  The initial radial profile of the
planetesimal surface density outside Neptune is not well-known, and we adopt a plausible model denoted as the hybrid model in Fig.~2 of \citet{netal2020}: (i) a simple power law $\propto 1/r$ at radii  $<28$~au, (ii) followed by an exponential decay with an $e$-folding length of $\simeq2$~au beyond 28 au.

Each of our simulations included initially one million zero-mass disk particles on nearly
circular and coplanar orbits. Their semimajor axis distribution matched the assumed
surface density of the disk, and their eccentricities and inclinations followed a Rayleigh distribution with scale parameters of $0.05$ in eccentricity and $2^\circ$ in
inclination. All other initial orbital elements of the disk particles were uniform in the interval from $0^\circ$ to $360^\circ$. The large number of test particles in our simulations is necessary to achieve adequate statistics to model accurately the distribution of particles in the detached disk and the LPCs
that visit the inner solar system from the distant Oort cloud. In practical terms, we performed each of the simulations on 2,000 cores of the NASA Pleiades Supercomputer, with each core following the massive bodies and 500 massless disk particles.

\subsection{Galactic tide and stellar encounters}\label{tidstars}
Two dynamical effects shape the structure of the Oort cloud at large heliocentric distances: (i) the Galactic tidal field, and (ii) the gravitational perturbations from individual stellar encounters with the solar system \citep[see reviews by][]{rickman2010,detal2015}.
Both are due to the gravitational effects of masses outside the solar system, but it is convenient to split them into a collective, smooth, steady-state component (i) and individual transitory effects (ii) \citep{ht1986,sari2010}.

Our model for the Galactic tide, including the MOND\-ian correction, was developed in Sec.~\ref{efe3}. We implemented this perturbation, as described by Equations (\ref{tide}) and (\ref{ee6}), in {\tt swift}. As in \citet{spc2017} and \citet{vnd2019}, we model stellar encounters by introducing passing stars in {\tt swift} as additional massive bodies at their entry to a sphere of 1 pc heliocentric radius, and removing them from the simulation at their exit from this sphere.  We use the statistical method discussed in \citet{retal2008} to generate a sequence of stellar encounters through the history of the solar system, with the distribution of stellar masses, velocity dispersion and frequency of encounters taken from \citet{gsetal2001}. These data are taken from observations by the Hipparcos spacecraft. Data releases 2 and 3 from the Gaia spacecraft 
\citep[e.g.,][]{bj2018,bjetal2018,bj2022} give more accurate estimates of the properties of the population of stars in the solar neighborhood. However, given the approximations and uncertainties elsewhere in our analysis, the somewhat older Hipparcos-based data are satisfactory for our purposes. Statistically there are $\simeq 12$ stellar encounters per Myr within the chosen 1~pc heliocentric distance. Most encounters (more than $65$\%) are with dwarf stars of subsolar mass ($\simeq 0.25$~M$_\odot$)
passing with relatively high speeds ($30$ km~s$^{-1}$ or so). A typical encounter in our simulations lasts from a few tens of thousands to hundreds of thousands of years. The characteristic distance of closest approach between a planetesimal roaming in the Oort cloud and a star passing through the cloud is several tens of thousands of au.
As a result, the induced acceleration to the heliocentric planetesimal motion from a stellar encounter is typically small, except during rare close encounters \citep[e.g.,][]{rickman2010}.

In an ideal model, we would treat the gravitational interaction between the passing stars and planetesimals on heliocentric orbits using MOND\-ian dynamics rather than Newtonian dynamics. We justify the Newtonian approximation by recalling the very different role of the Sun and the passing stars in the orbital history of
planetesimals. Those of them that eventually become observed LPCs travel around the Sun for Gyr, on orbits that typically reach far beyond the heliocentric MOND scale $r_{\rm M}$. It is thus essential to model the statistical parameters of this population in the context of MOND. The effect of passing stars is to introduce a time-localized, random component in the heliocentric motion of the planetesimals which leads to a random walk of their orbital elements (causing, for instance, the distribution of orbital planes in the outer Oort cloud to be isotropic in space). Provided the diffusion coefficients of this process are roughly correct (and we argue that they are), it would basically not even matter what is the origin of the diffusion. 

\subsection{LPC modeling}\label{lpc}

We use the same approach as in \citet{vnd2019} to obtain the synthetic population of LPCs. We record all particle approaches to the inner solar system with perihelia $\leq 15$~au during the last $500$~Myr of our simulations. At the moment of perihelion passage we save the heliocentric state vectors of the particle and all the planets. This information is used in the post-processing phase, in which we perform a short numerical integration of each such recorded comet visit backward in time, until the comet reaches a heliocentric distance of $250$~au. We use Newtonian mechanics for these runs. At this heliocentric distance we compute the barycentric orbital elements of the particle. This procedure is necessary to characterize the LPC orbits properly, because the heliocentric orbital elements at pericenter passage are perturbed by the solar motion about the barycenter of the solar system and the gravitational potentials of the planets. The binding energy of LPCs with orbits in the Oort cloud is so small that these corrections are significant. In particular, many heliocentric LPC orbits appear to be hyperbolic at perihelion even though their barycentric orbits
are elliptical as they approach the planetary system, with the corresponding semimajor axis termed
the ``original'' semimajor axis $a_{\rm ori}$. While the perihelion distance and inclination are little changed, as the maximum solar distance from the barycenter is only $\simeq 0.01$~au, $a_{\rm ori}$ 
deviates significantly from the heliocentric $a$. The same correction is carried out for observed comets before their orbits are recorded in catalogs; see \citet{warsaw2020} for the analysis of nearly parabolic LPCs, and \citet{mw2008} for all comets.

After the orbital elements of the observable LPCs have been processed to obtain their original, barycentric values, we need to account for the cometary fading process. The simulated comets in our simulations are indestructible and are removed from the simulation only if they physically hit one of the massive bodies or are ejected from the solar system (which we operationally define as reaching a heliocentric distance of $500,000$~au). Real comets, however, have lives limited primarily by a still debated physical process (i.e., they either disintegrate, or cease to be active and observable by exhausting their volatile components, or disappear through some unknown process). The need to include comet fading in models of the LPC distribution was already recognized in the pioneering work of \citet{o1950}. The fading phenomenon remains poorly understood, as it is not exactly known to what heliocentric distances it operates \citep[see, e.g.,][for a surprising result arguing for a comet fading well beyond the orbit of Saturn]{kaib2022}, or what is the minimum necessary set of parameters to describe it \citep[see, e.g.,][who argues that fading may be primarily dependent on cometary
  size in contrast to the more traditional description by number of returns to perihelion in the inner solar system]{jewitt2022}. Fortunately, our conclusions do not depend strongly on these details. 

We adopt an extension of the popular model by \citet{whipple1962}, who argues that most LPCs fade within the first few
returns to the inner solar system with about a 15\% fraction surviving much longer \citep[e.g.,][]{weiss1980}. The
fading process thus depends only on the number of LPC returns to the inner solar system. Following \citet{wt1999} we define $\Phi_m$, the probability that a given comet survives 
fading for at least $m$ perihelion passages, and adopt the offset power-law
\begin{equation}
 \Phi_m = \left(\frac{m+c}{1+c}\right)^{-\kappa}\; , \label{fading}
\end{equation}
with $\kappa$ and $c$ positive constants (\citealt{wt1999} also review a number of other fading models). The parameters $(\kappa,c)$ are adjusted to obtain the best fit to the observed distribution of
the original semimajor axes of LPCs. We consider the following range of values: $\kappa\in (0,1)$, and $c\in [0,10]$.

In order to verify the robustness of our results to the choice of the fading function, we also explored a two-population model consisting of (i) a population of LPCs that disappear with  probability $\lambda$ at each perihelion passage, and (ii) a population that does not fade at all \citep[fractionally expressed with a parameter $f$; e.g.,][]{weis1979,weiss1980}. This is again a two-parameter fading family given by
\begin{equation}
 \Phi_m = \left(1-f\right)\left(1-\lambda\right)^{m-1}+f\; . \label{fading1}
\end{equation}
The parameters $(\lambda,f)$ are adjusted in the intervals $\lambda\in(0,1)$ and $f\in [0,1]$ to obtain the best match between our model and the data in the Marsden--Williams catalog of LPCs.

\subsection{Datasets: distant TNOs and LPCs}\label{data}
The results of our simulations are compared with two observational datasets: the orbital elements of the distant TNOs and the binding energies (or original semimajor axes) of LPCs\footnote{In Appendix \ref{aph} we also compare with the distribution of aphelion directions of LPCs.}. The first of these datasets is relatively small, as not many TNOs have been detected at very large heliocentric distances. However, many new detections are expected from forthcoming large surveys such as those of the Vera C.\ Rubin Observatory. At present the distribution of original semimajor axes of the LPCs provides a stronger test of MOND. 
\smallskip

\noindent{\it Distant TNOs.-- }To generate input for the Dark Energy Survey (DES) simulator (see below), all numerical integrations are continued, with MOND effects included, from $t=4.5$ Gyr to $t=4.51$ Gyr. The orbits of all bodies are recorded with a $10^5$ yr cadence. The main purpose of this integration is to increase 
the statistics such that the DES simulator generates a large enough number of detections to be compared with actual DES detections. We rotate the reference system such that Neptune's position on the sky approximately corresponds to its position during DES observations, as needed to correctly model azimuthal biases related to Neptune resonances.  The orbits of the same TNO obtained at different times are supplied to the simulator as if they were orbits of different TNOs.  

We collect all orbits in this simulation with semimajor 
axes $100<a<500$ au and perihelion distances $q>35$ au, nearly 800,000 orbits in total. 
%\st{maybe it's worth saying how many objects are in this sample}. 
This selection corresponds to a class of TNOs known as
detached disk objects \citep[][all orbital elements are referred to the barycenter of the solar system]{glad2008}. The detached TNOs are largely immune to perturbations from the planets. Their orbits are therefore difficult to populate from a planetesimal disk in standard models of the evolution of TNOs, which makes them a good diagnostic of MOND. We also ignore
orbits with $a<100$ au which are strongly affected by various resonances with the giant planets, and orbits with $a>500$ au where we have no observational data (with only a few exceptions). 

The distribution of these orbits is compared with DES detections. \citet{b2022} searched for outer solar-system objects during the six years of DES operations between 2013 and 2019. They found 817 TNOs, including over 200 objects in the scattered disk with $a > 50$~au. There were 33 detections satisfying our selection criteria ($100 < a < 500$~au and $q>35$~au). The relatively small number of distant TNOs observed by DES reflects the difficulty of detecting objects with large perihelion distances.

The DES team developed a survey simulator, which is publicly available on {\tt GitHub}. The DES simulator enables comparisons between test models and the DES data by simulating the discovery probability of each 
member of the test population, that is, the model is biased in the same way as the data.  

The DES simulator needs an absolute magnitude distribution of the distant TNOs as input. Following previous publications on the subject \citep[e.g.,][]{bern2004,ossos29}, we adopted a piece-wise power-law model for the size distribution with two break points at diameters $D_1$ and $D_2$ ($D_2>D_1$). The small ($D<D_1$), intermediate ($D_1<D<D_2$) and large ($D>D_2$) TNOs were assumed to have cumulative size distributions $N(>\!D)$ that satisfied $dN(>\!D)/dD\propto D^{-\alpha-1}$
with different slopes $\alpha$. The free parameters are $D_1$, $D_2$ and the three $\alpha$'s. These were adjusted, after accounting for observational bias, 
to provide an adequate match to the absolute magnitude distribution of distant TNOs detected by DES. As DES reported their detections in the red absolute magnitude $H_r$, the bolometric absolute magnitudes $H$ are mapped to $H_r$ using the relation $H_r=H-0.6$ \citep{b2022}. In order to translate the absolute magnitudes to sizes we assumed a visual albedo $p_{\rm V}=0.05$. We adjusted the parameters of the size distribution such that the observationally biased magnitude distribution of distant TNOs, as it comes from the DES simulator, matches the actual DES detections. We obtained $D_1=100$ km, $D_2=250$ km, $\alpha_{\rm small}=2.1$, $\alpha_{\rm interm}=5.0$ and $\alpha_{\rm large}=2.0$. The shape of this size distribution is 
consistent with constraints from observations of Jupiter Trojans and TNOs with $a<100$ au \citep[e.g.,][]{nes2018}. The results reported in Sec.~\ref{res1} do not depend sensitively on the adopted size distribution.

\smallskip

\noindent{\it Long-period comets.-- } The method of generating the
synthetic population of LPCs from our simulations has been already described in Sec.~\ref{lpc}. For
general information on the orbits of observed LPCs we refer the reader to several review chapters published in the past decade \citep[e.g.,][]{detal2015}, as well as a detailed description in Sec.~2 of \citet{vnd2019}. Here we provide just a brief overview. 

The determination of the orbital distribution of the extreme LPCs with the weakest binding energies to the solar system (also called the nearly parabolic comets), has been a long-term project of the
Warsaw school founded by Grzegorz Sitarski in the 1970s. A recent series of papers from this school
\citep{k2014,ketal2014,kd2017} culminated in \citet{warsaw2020}, where a catalog of 277 nearly parabolic comets was presented. We use the  Class~1 subsample of these orbits, in total 228 comets, for which the astrometric observations provide the most accurate determination of the orbital parameters. In order to minimize the role of observational biases, we further restrict the sample to LPCs with perihelion distance $q\leq 4$~au, representing finally 116 comets.

The determination of the orbital distribution of all LPCs has also a long tradition. It started already in the late 1960s and for several decades was coordinated by Brian Marsden and his collaborators.  Here we use the 17th edition of the catalog of \citet{mw2008}. As in \citet{vnd2019}
we use orbital data from 318 LPCs with the most accurately determined solutions (about $60$\% of the whole catalog). Once again we focus on orbits with perihelia $q\leq 4$~au. This is a compromise between the number of available orbital solutions and their completeness \citep[see the discussion in][]{wt1999}. While the dataset we are using may still suffer some degree of incompleteness, we believe that this does not affect our conclusions. 
 
\section{Results}\label{res}
Table~\ref{sims} provides an overview of our core set of simulations. For brevity we use the identifiers ${\cal S}0$ to ${\cal S}3$. 

We first analyze a class of reference simulations, denoted ${\cal S}0$, based on Newtonian dynamics. We discuss two flavors of these simulations: the simpler ones are based on the present Galactic environment of the solar system, while a second set includes the  effects of its birth cluster \citep{adams2010}.
These provide context for our simulations using MOND.

All other simulations assume the same MOND acceleration scale $a_0=1.2\times 10^{-10}$ m~s$^{-2}$. The simulations sample two different values of the external acceleration $g_{\rm e}$, and two transition functions from the $\mu_n(x)$ family (Eq.~\ref{e2}). 

We use the nominal (present-day) value $g_{\rm e}=2.32\times 10^{-10}$ m~s$^{-2}$ in ${\cal S}1$ and
${\cal S}2$, and a larger value $g_{\rm e}=3.35\times 10^{-10}$ m~s$^{-2}$ in ${\cal S}3$. The larger value corresponds to the centripetal acceleration at a radius that is 70\% of the Sun's current distance from the Galactic center, a plausible estimate for the distance over which it may have migrated since its birth. 

In ${\cal S}1$ we use the gradual function $\mu_2(x)$ popular in many applications of MOND to Galactic and extra-galactic dynamics, and in ${\cal S}2$ and ${\cal S}3$ we use the sharper transition function $\mu_{10}(x)$, which unlike $\mu_2(x)$ is consistent with constraints from planetary ephemerides (see Sec.\ \ref{efe1}). 

In all of these simulations we included the initial migration of Uranus and Neptune, as described in Sec.~\ref{inte}. While many lines of evidence indicate that migration must have taken place, the details remain somewhat uncertain. Therefore we also performed a smaller scale variant of the simulation ${\cal S}2$ without planetary migration; that is, the planetary orbits were static throughout the simulation. This simulation, called ${\cal S}2^\prime$, is discussed in Appendix~\ref{jobs2p}. Its purpose is to show that our conclusions do not depend on the details of planetary migration. 

We start with a discussion of the population of planetesimals detached from the scattered disk and its consistency with DES observations (Sec.~\ref{res1}). Next, we discuss constraints from the binding energy distribution of the observed LPCs (Sec.~\ref{res2}). In Appendix~\ref{aph} we provide
a brief analysis of possible MOND\-ian effects on the distribution of the aphelion directions of nearly parabolic LPCs.
% Tab 2 %%%%%%%%%%%%%%%%%%%%%%%%%%%%%%%%%%%%%%%%%%%%%%%%%%%%%%%%%%%%%%%%%%%%%%%%%%%%%%%%%%%
\begin{deluxetable}{l|cc}[t] 
 \tablecaption{\label{sims}
  List of the principal simulations}
 \tablehead{
  \colhead{} & \colhead{transition function} & \colhead{external acceleration} \\ [-7pt]
  \colhead{} & \colhead{Eq.~(\ref{e2})}      & \colhead{ $g_{\rm e}$ (m~s$^{-2}$)} }
%\decimals
\startdata
 \rule{0pt}{3ex}
 & \multicolumn{2}{c}{{\it -- Reference simulation --}} \\ [1pt]
 ${\cal S}0$\tablenotemark{a}  & --  & -- \\ [3pt]
 & \multicolumn{2}{c}{{\it -- Nominal simulations --}} \\ [1pt]
 ${\cal S}1$  & $\mu_2(x)$      & $2.32\times 10^{-10}$ \\
 ${\cal S}2$  & $\mu_{10}(x)$   & $2.32\times 10^{-10}$ \\ [3pt]
 & \multicolumn{2}{c}{{\it -- Extended simulations --}} \\ [1pt]
 ${\cal S}3$  & $\mu_{10}(x)$   & $3.35\times 10^{-10}$ \\ [2pt]
\enddata
\tablenotetext{a}{A Newtonian simulation using the results in \citet{clu2023} and \citet{vnd2019}.}
\tablecomments{The first column is the simulation label, the second column specifies the transition function, and the third column gives the value of the external acceleration. In all runs $a_0=1.2\times 10^{-10}$ m~s$^{-2}$.}
\end{deluxetable}
%%%%%%%%%%%%%%%%%%%%%%%%%%%%%%%%%%%%%%%%%%%%%%%%%%%%%%%%%%%%%%%%%%%%%%%%%%%%%%%%%%%%%%%%%%% 
% FIG 1 %%%%%%%%%%%%%%%%%%%%%%%%%%%%%%%%%%%%%%%%%%%%%%%%%%%%%%%%%%%%%%%%%%%%%%%%%%%%%%%%%%%%%%%%%%%%%%%
\begin{figure*}[t!]
 \begin{center}
 \begin{tabular}{cc}
  \includegraphics[width=0.45\textwidth]{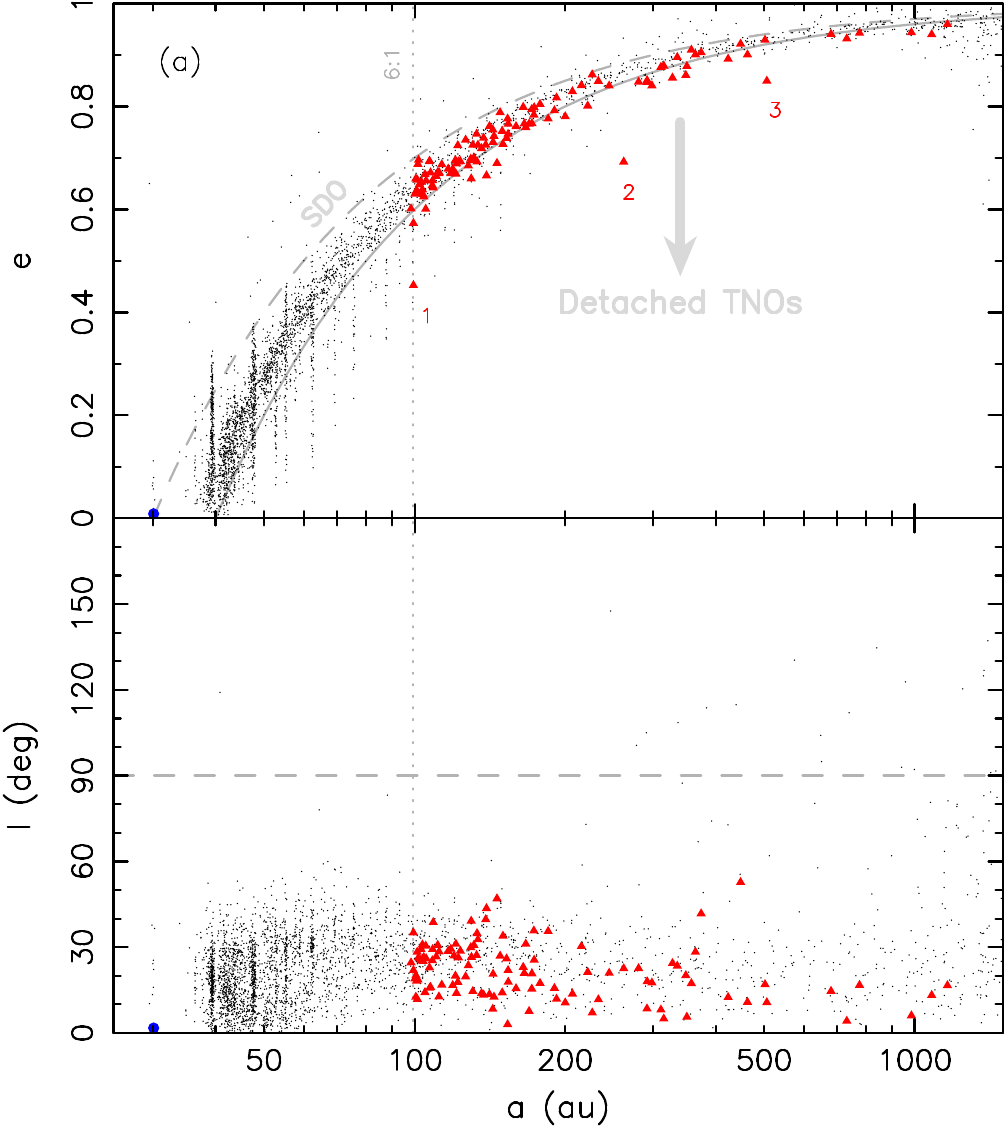} &
  \includegraphics[width=0.45\textwidth]{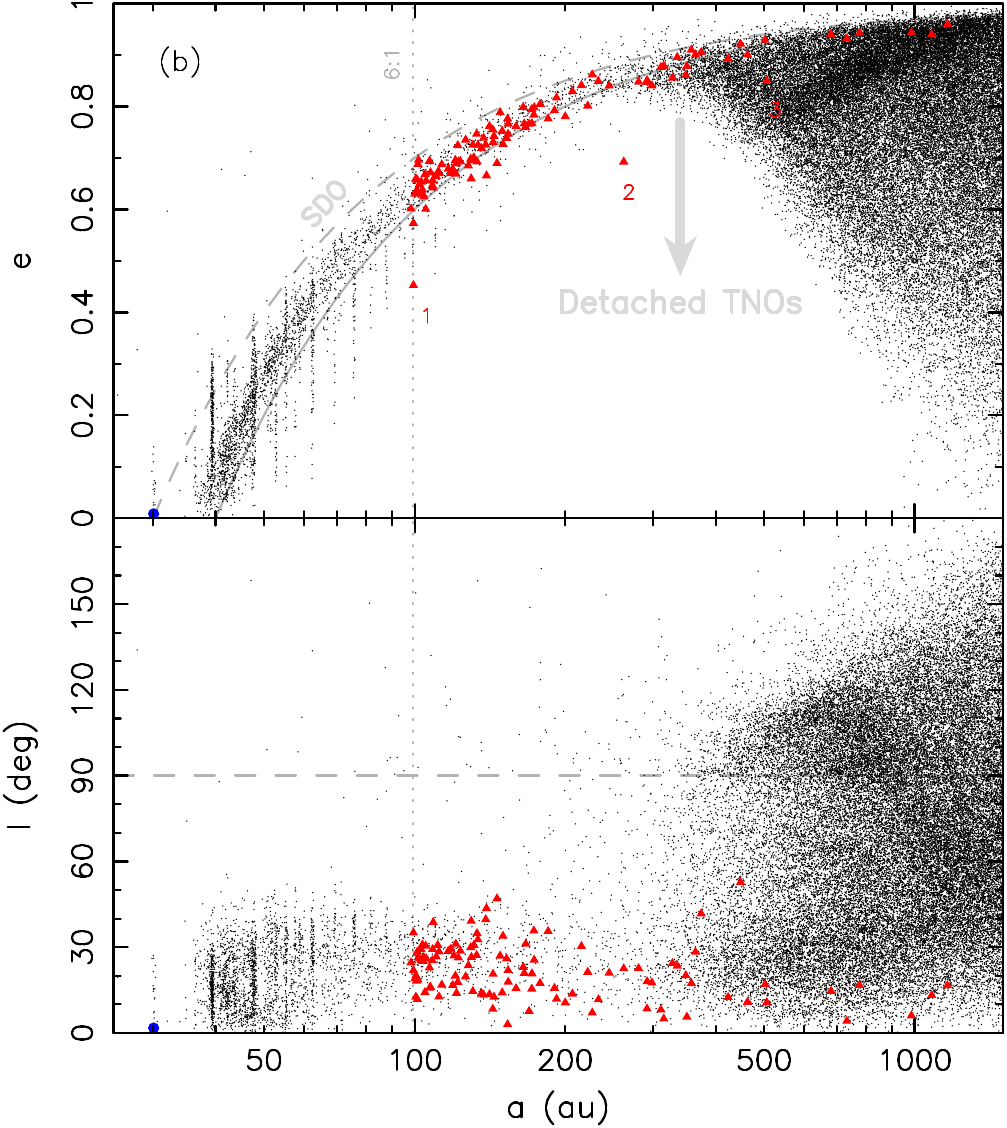} \\ 
  \includegraphics[width=0.45\textwidth]{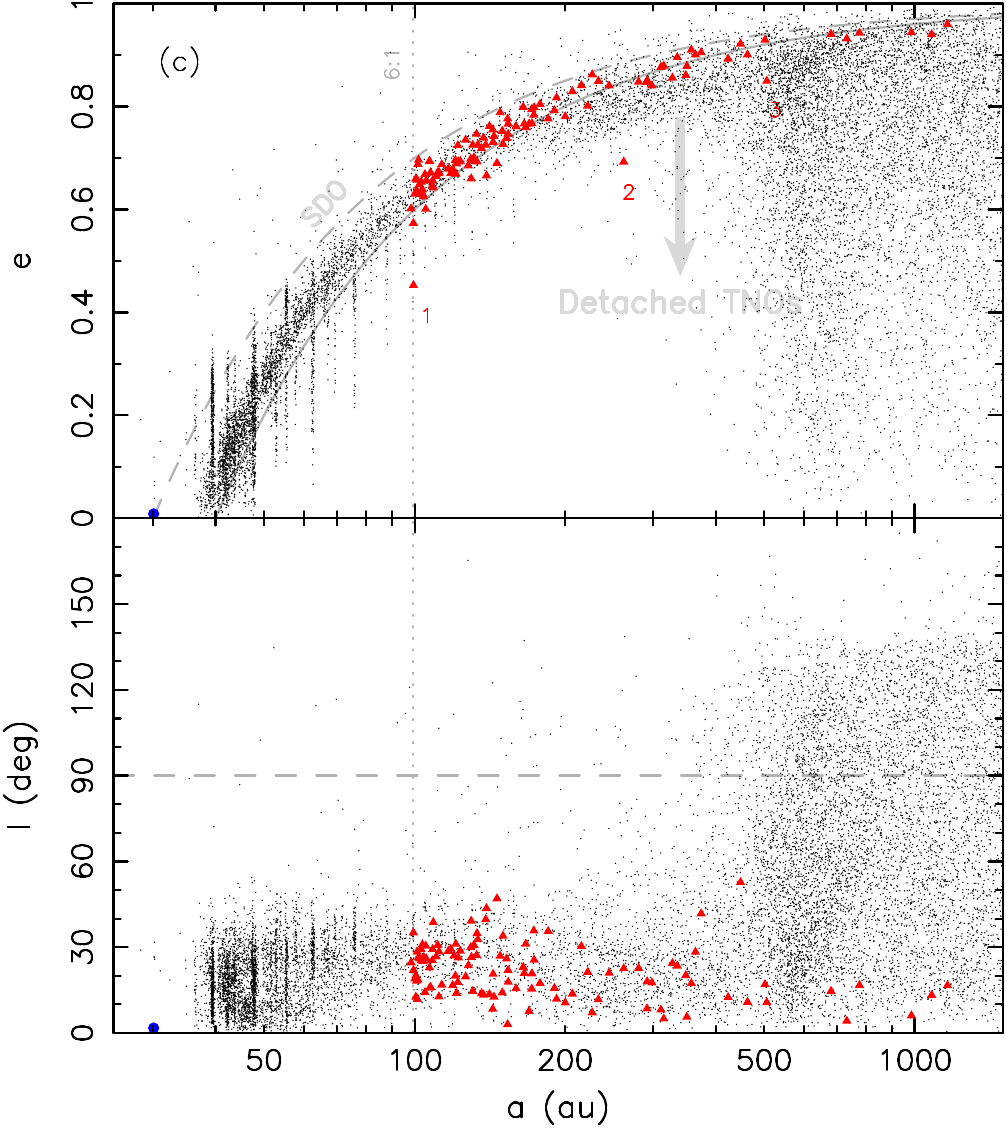} &
  \includegraphics[width=0.45\textwidth]{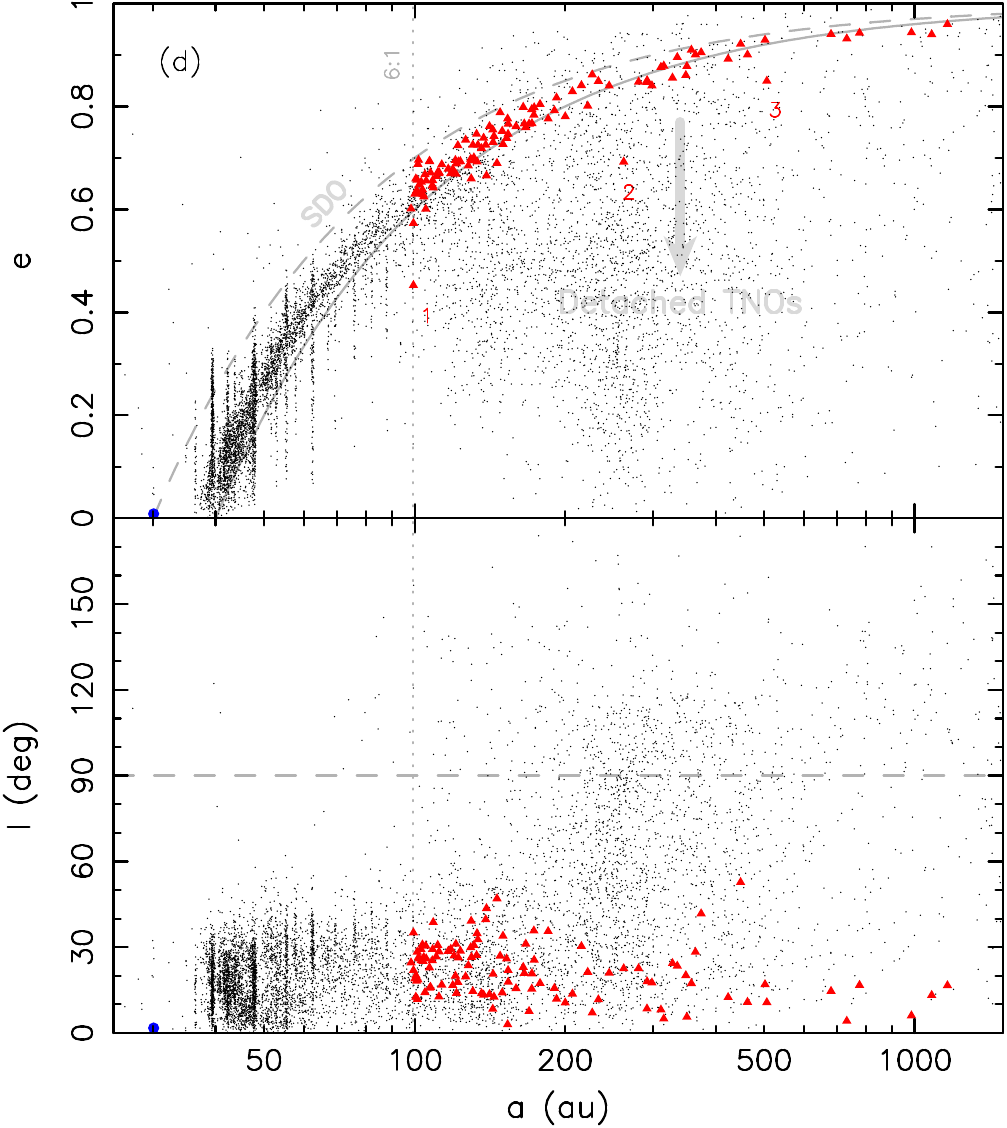} \\   
 \end{tabular}
 \end{center}  
 \caption{Distribution of barycentric orbital elements -- eccentricity $e$ vs.\ semimajor axis $a$ (top parts), and inclination $I$ relative to the ecliptic vs.\ semimajor axis $a$ (bottom parts) -- at the final epoch of four simulations: (a) Newtonian dynamics, similar to the results in \citet{vnd2019} or the ``Galaxy'' simulation in \citet{clu2023}; (b) Newtonian dynamics, with the effects of a solar-system birth cluster included \citep[from the Cluster1 simulation in][]{clu2023}; (c) MOND\-ian dynamics with the sharp transition function $\mu_{10}(x)$ (the ${\cal S}2$ simulation); (d) MOND\-ian dynamics with the gradual transition function $\mu_{2}(x)$ (the ${\cal S}1$ simulation). The plots focus on the region of the distant TNOs ($a\leq 2,000$~au); red triangles are the presently known objects in multi-opposition orbits with $a\geq 100$~au and perihelion
 $q\geq 30$~au. The blue dot is the orbit of Neptune. A few notable objects are labeled in the $a$--$e$ plots:  2014~US277 (label~1), 2012~VP113 (label~2), and (90377) Sedna (label~3). The
 dashed and solid gray lines in the $a$--$e$ plots show perihelion distances $q=30$ and $q=40$~au, roughly delimiting the region occupied by the Neptune scattered disk objects (SDO). The orbits with larger perihelia, such as the labeled cases, belong to the detached disk.  The dotted vertical line indicates the position of the exterior 6:1 mean-motion resonance with Neptune.}
 \label{fig_detach}
\end{figure*}
%%%%%%%%%%%%%%%%%%%%%%%%%%%%%%%%%%%%%%%%%%%%%%%%%%%%%%%%%%%%%%%%%%%%%%%%%%%%%%%%%%%%%%%%%%%%%%%%%%%%%%%
% FIG 1 %%%%%%%%%%%%%%%%%%%%%%%%%%%%%%%%%%%%%%%%%%%%%%%%%%%%%%%%%%%%%%%%%%%%%%%%%%%%%%%%%%%%%%%%%%%%%%%
\begin{figure}[t!]
 \begin{center}
% \begin{tabular}{c}
%  \includegraphics[width=0.8\textwidth]{fAa.eps} \\
%  \includegraphics[width=0.8\textwidth]{fAb.eps} \\ 
% \end{tabular}
  \includegraphics[width=0.47\textwidth]{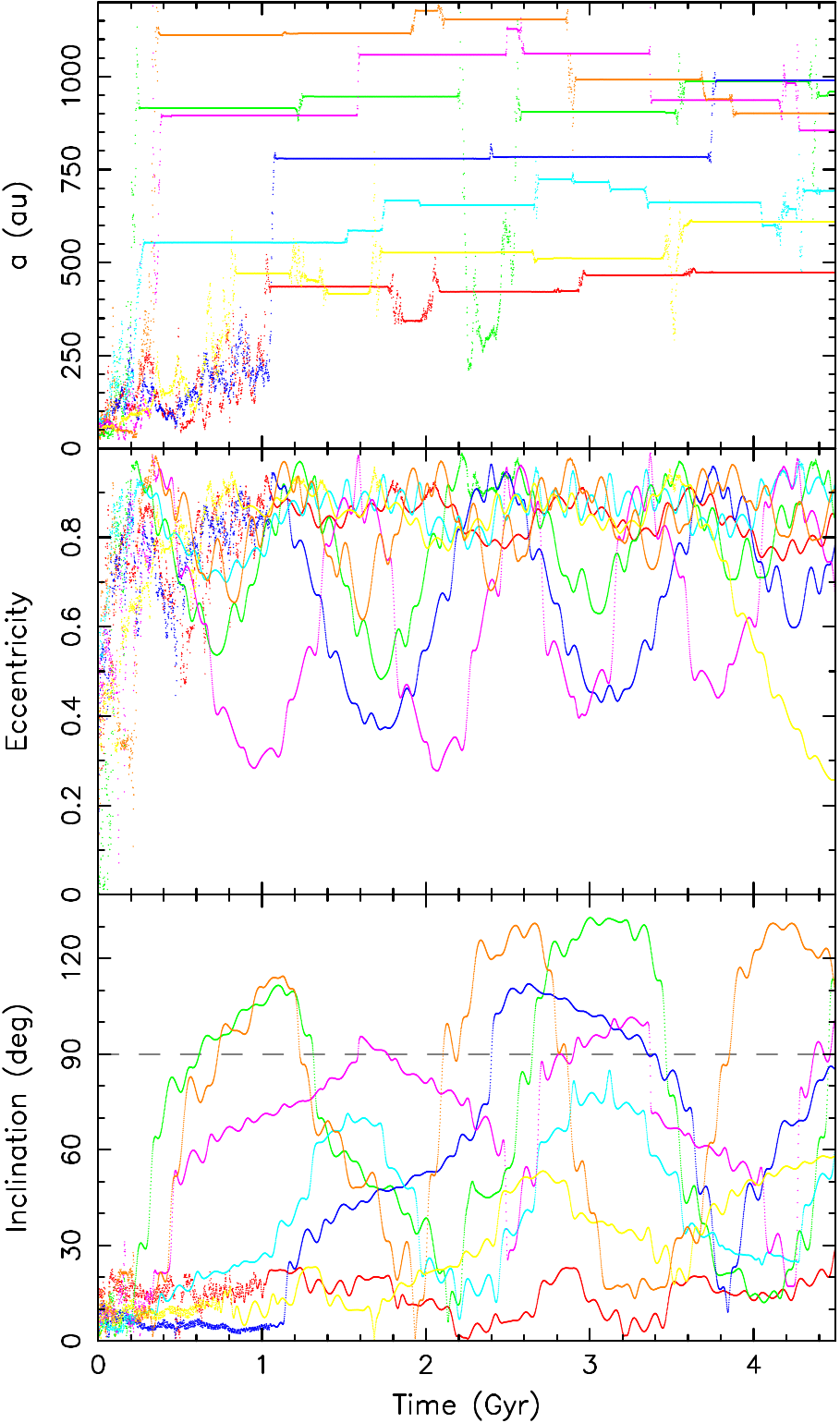} 
 \end{center}  
 \caption{Seven orbits from the simulation ${\cal S}2$ that detached from the SD. Barycentric orbital elements -- semimajor axis, eccentricity and inclination from top to bottom -- are shown as a function of time. While spending most of the time with perihelia $\geq 38$~au, the orbits periodically visit the SD region of phase space, where close encounters with Neptune may excite a random walk in semimajor axis (if soft enough), or extract the 
  particles to the comet population (if stronger).}
 \label{fig_detach1}
\end{figure}
%%%%%%%%%%%%%%%%%%%%%%%%%%%%%%%%%%%%%%%%%%%%%%%%%%%%%%%%%%%%%%%%%%%%%%%%%%%%%%%%%%%%%%%%%%%%%%%%%%%%%%%

\subsection{Implications for distant TNOs}\label{res1}
Figure~\ref{fig_detach} provides a first glimpse at the results, which we use to highlight the main dynamical processes. For that purpose, we show two Newtonian simulations by \citet{clu2023} (top panels), and two of our MOND simulations, 
${\cal S}1$ and ${\cal S}2$ (bottom panels).

Consider first the basic simulation named ``Galaxy" in \citet{clu2023}, equivalent to those in \citet{vnd2019}, shown on the left top panel. This simulation assumes that the Galactic environment at all times is the same as the current environment of the solar system.
The properties of the distant heliocentric populations are basically determined by (i) Neptune's radial migration through the planetesimal disk, and (ii) its terminal semimajor axis of $30$~au. During this evolution, a fraction of the planetesimals were scattered by Neptune into the scattered disk (SD), characterized by large values of semimajor axis but perihelia remaining in the range $30$--$38$~au. Unless the planetesimal orbit becomes unbound to the solar system, which is unlikely in the lifetime of the solar system because of Neptune's low mass, the planetesimals in the SD population remain there for a long time. In this simple model, it is not easy to transfer planetesimals to orbits with larger perihelia than $\simeq 38$~au, namely into the detached disk.%
\footnote{In this work we define the detached disk by simple orbital constraints: $q\geq 38$~au and $a\geq 50$~au, close to but not exactly the same as the proposed terminology in \citet{glad2008}.}
The interaction with Neptune
mean-motion resonances is one possible process that can do so. In Fig.~\ref{fig_detach} this possibility of a transfer to the detached disk is illustrated by 2014~US277 (identified by label~1), an object residing in the $6:1$ mean-motion resonance with Neptune at $a \simeq 100$~au. However, the efficiency of this process decreases with the order of the resonance, such that detached orbits with $a\geq 150$~au have a very small chance to be produced by this resonant mechanism \citep[e.g.,][]{ossos25}. 

% FIG 1 %%%%%%%%%%%%%%%%%%%%%%%%%%%%%%%%%%%%%%%%%%%%%%%%%%%%%%%%%%%%%%%%%%%%%%%%%%%%%%%%%%%%%%%%%%%%%%%
\begin{figure*}[t!]
 \begin{center}
% \begin{tabular}{c}
%  \includegraphics[width=0.8\textwidth]{fAa.eps} \\
%  \includegraphics[width=0.8\textwidth]{fAb.eps} \\ 
% \end{tabular}
  \includegraphics[width=\textwidth]{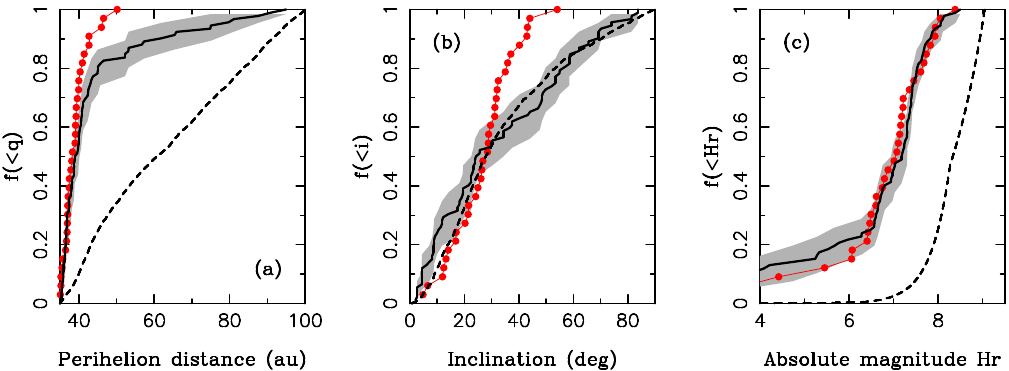} 
 \end{center}  
 \caption{A comparison between the orbital distributions of the observationally biased MOND-based ${\cal S}1$ model (black lines) and DES observations (red dots) of distant
TNOs ($100<a<500$ au and $q>35$ au). The observationally biased model sample was determined using the DES simulator.
The intrinsic model distributions (i.e., without the DES observational bias) are shown as dashed lines.
The shaded areas are 1$\sigma$ envelopes. From
left to right, the panels show the distributions of perihelion distance, orbital 
inclination relative to the ecliptic, and red absolute magnitude.}
 \label{fig_detach_s1}
\end{figure*}
%%%%%%%%%%%%%%%%%%%%%%%%%%%%%%%%%%%%%%%%%%%%%%%%%%%
% FIG 1 %%%%%%%%%%%%%%%%%%%%%%%%%%%%%%%%%%%%%%%%%%%%%%%%%%%

Recent wide-angle surveys have revealed a significant population of non-resonant detached TNOs beyond $a=100$--$150$~au 
\citep[see, e.g., ][]{ossos29}. Something beyond the basic model of this simulation is needed to explain this population. This challenge has resulted in a vast literature, which we do not intend to review here, that mainly discusses two distinct models: (i) if Neptune is not the most distant planet in the solar system, and one or more planetary-mass objects are hidden on distant heliocentric orbits, their perturbations can pull the scattered disk objects onto orbits with larger perihelia \citep{gladman2006,silsbee2018,p92019}, or (ii) close stellar encounters during the period when the young solar system was still in its natal cluster of stars can excite orbits in the scattered disk to larger perihelia \citep[e.g.,][as a representative of early studies of this possibility]{ml2004}. The results shown on the right-hand top panel of Fig.~\ref{fig_detach} illustrate the second class of models. They are based on the ``Cluster1" simulation in \citet{clu2023}, where a simplified model for stellar encounters in the natal cluster was considered. During the brief period before the cluster dissolves, close stellar encounters form a fossil extension of the Oort cloud down to several hundred au heliocentric distance. The orbit of the archetypal detached object (90377) Sedna \cite[identified by label~3; e.g.,][]{bro2004} may be readily explained by this mechanism (if a more violent initial phase is assumed, one could even generate an object with the orbit of 2012~VP113, identified by label~2). For what follows, it is important to stress that this innermost part of the Oort cloud -- to which the Sedna-type orbits belong -- is inactive today in this model. This means that both planetary perturbations and Galactic tides are very weak in this zone; thus, once they are detached from the scattered disk, Sedna-type objects remain in more-or-less fixed orbits for Gyr timescales or more. For that reason they do not contribute to the observed cometary populations today. As confirmation of this argument, both the ``Galaxy" and ``Cluster1" simulations, very different in some aspects as seen on top panels of Fig.~\ref{fig_detach}, provide a population of LPCs with the same orbital characteristics.

As already demonstrated by \citet{pk2016}, \citet{p2017}, and more recently discussed by \citet{bm2023}, MOND offers yet another mechanism for transferring planetesimals from the scattered disk to the detached disk. Our simulations confirm this conclusion. Results from the ${\cal S}1$ and ${\cal S}2$ runs are shown in the bottom panels of Fig.~\ref{fig_detach}. The analysis in Sec.~\ref{efe1} showed that the MOND\-ian perturbation is weaker, and spatially isolated to a narrower range of heliocentric distance, in the case of the sharp transition function $\mu_{10}(x)$. As a result, the torque exerted by the EFE that lifts scattered-disk orbits to larger perihelia in the ${\cal S}2$  simulation is limited to orbits with semimajor axes $\gtrsim 400$~au; in contrast, with the gradual transition function $\mu_2(x)$ used in the ${\cal S}1$ simulation, the detached disk is populated down to semimajor axes $\sim 150$ au. 

The orbit distribution in the detached disk formed in the ${\cal S}2$ simulation is comparable to that of the detached disk formed in the ``Cluster1'' simulation by  stellar encounters in the solar-system birth cluster. The main difference, though, is that the MOND\-ian perturbation is active until the present epoch, providing a continued and vigorous exchange between the detached disk and the scattered disk (see Fig.~\ref{fig_detach1} for a few examples of orbital evolution of the detached particles in the simulation ${\cal S}2$). 
In the next section we show that this process produces a mismatch between the predicted and observed distribution of the LPC binding energies. This problem is only amplified in the ${\cal S}1$ simulation,  which allows a stronger MOND\-ian perturbation that penetrates to smaller heliocentric distances. The bottom right panels in Fig.~\ref{fig_detach} show that in this case the scattered disk extends to orbits with semimajor axes as small as $150$--$200$~au. In fact, in this model the EFE is strong enough beyond $\simeq 1,000$~au that this region -- the inner Oort cloud -- is depleted of particles by the current age of the solar system. 

% FIG 1 %%%%%%%%%%%%%%%%%%%%%%%%%%%%%%%%%%%%%%%%%%%%%%%%%%%%%%%%%%%%%%%%%%%%%%%%%%%%%%%%%%%%%%%%%%%%%%%
\begin{figure}[t!]
% \plottwo{.eps}{.eps}
 \begin{center}
  \includegraphics[width=0.45\textwidth]{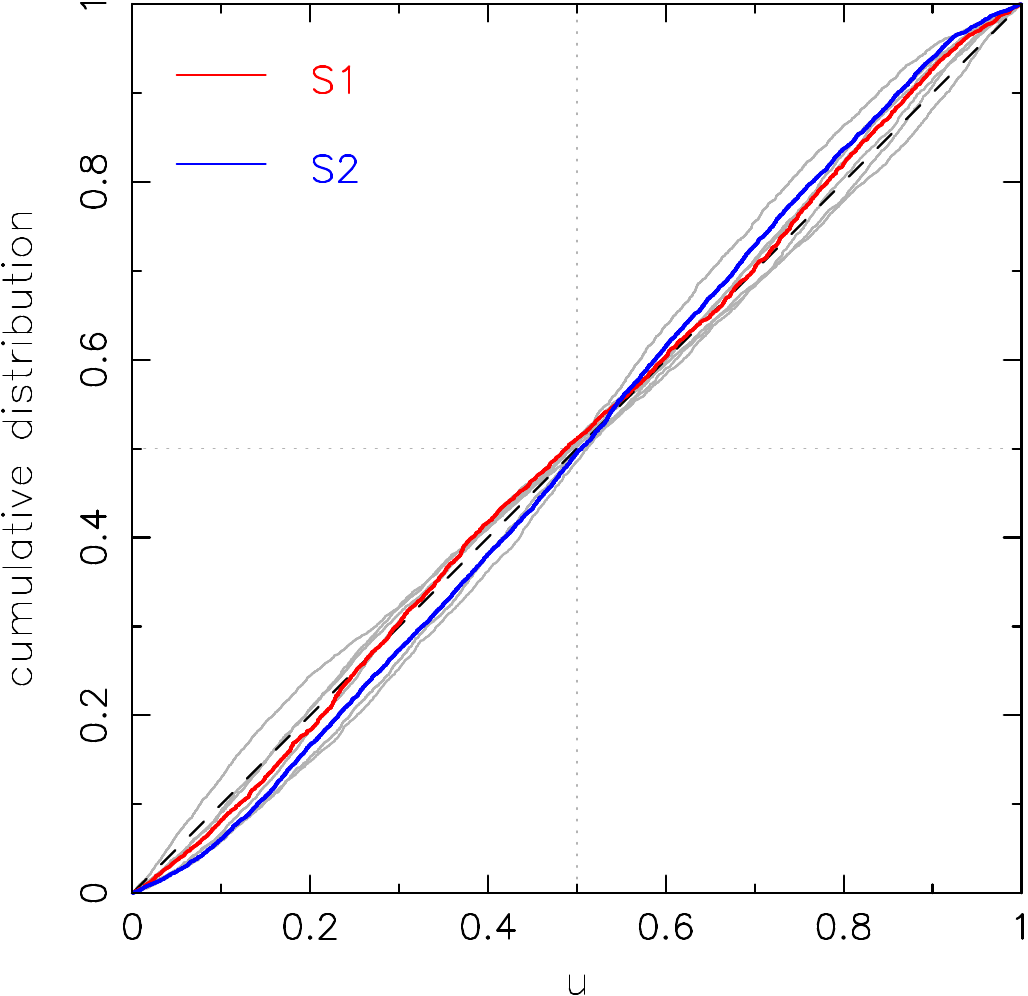}
 \end{center}  
 \caption{Cumulative distribution of the $u$ parameter defined in \citet{bm2023}, normalized to unity, for our core simulations ${\cal S}1$ (red) and ${\cal S}2$ (blue). We used particles at the end of those simulations, epoch $4.5$~Gyr, that were detached from the scattered disk ($q\geq 38$~au) in heliocentric orbits with semimajor axes in the range $100\leq a\leq 1,000$~au. The dashed black diagonal is the uniform distribution in $u$.
 The gray curves are distributions of the $u$ parameter from the two simulations at five equally spaced times in the last $500$~Myr.}
 \label{fig_u}
\end{figure}
%%%%%%%%%%%%%%%%%%%%%%%%%%%%%%%%%%%%%%%%%%%%
\begin{figure*}[t!]
 \begin{center}
% \begin{tabular}{c}
%  \includegraphics[width=0.8\textwidth]{fAa.eps} \\
%  \includegraphics[width=0.8\textwidth]{fAb.eps} \\ 
% \end{tabular}
  \includegraphics[width=0.9\textwidth]{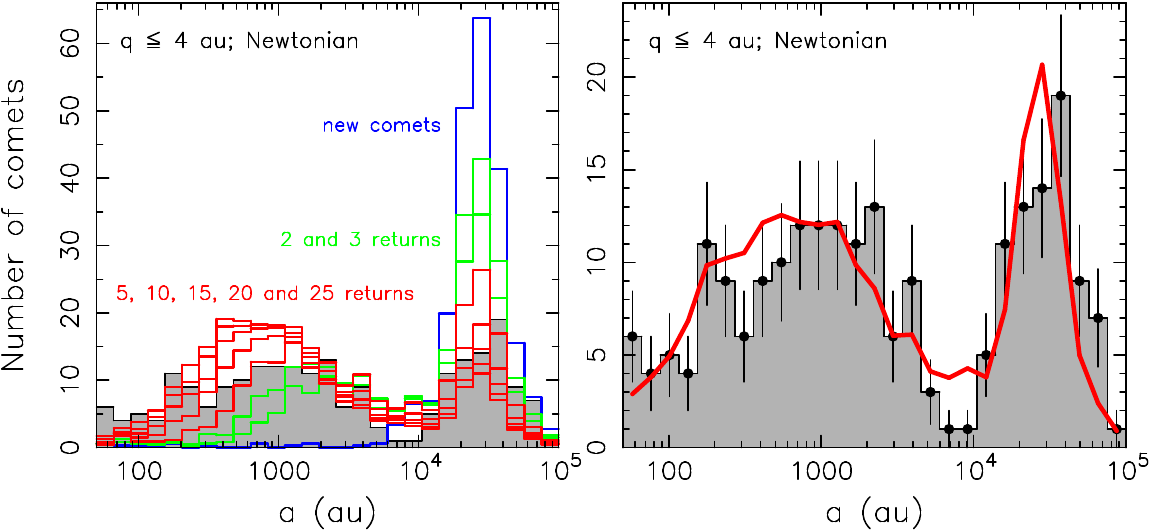} 
 \end{center}  
 \caption{The gray histograms in both panels show the original semimajor axes $a$ (in au) for a sample of $224$ LPCs with perihelion $q\leq 4$~au, from the Marsden--Williams catalog. The error bars on the right panel are simply $\sqrt{N}$. The colored lines show the results from the Newtonian simulation ${\cal S}0$, as described in Sec.\ \ref{res2}. Left panel: The histograms show the distribution of observed LPCs with if we assume that they disappear after $N$ perihelion passages with $q\leq 4$~au: (i) new comets, $N=1$ (${\cal C}_1$, blue curve), (ii) $N=2,3$ (${\cal C}_2$ and ${\cal C}_3$; green curves), and (iii) $N=5,10,15,20,25$ (${\cal C}_5$, etc; red curves). As $N$ increases the fraction of comets in the Oort peak ($a\geq 10^4$~au) declines, and the fraction in the returning hump ($a\leq 10^4$~au) grows. All histograms are normalized to the number of observed comets. Right panel: The results shown in the left panel are combined using the fading law (\ref{fading}) \citep[e.g.,][]{whipple1962,wt1999}, with its parameters $\kappa$ and $c$ adjusted to fit the observations (red curve). The best-fit parameters are given in Table \ref{sims_fit}.}
 \label{fig_lpc_ret_newt}
\end{figure*}
%%%%%%%%%%%%%%%%%%%%%%%%%%%%%%%%%%%%%%%%%%%%%%%%%%%%%%%%%%%%%%%%%%%%%%%%%%%%%%%%%%%%%%%%%%%%%%%%%%%%%%%
% FIG 1 %%%%%%%%%%%%%%%%%%%%%%%%%%%%%%%%%%%%%%%%%%%%%%%%%%%%%%%%%%%%%%%%%%%%%%%%%%%%%%%%%%%%%%%%%%%%%%%
\begin{figure}[t!]
 \begin{center}
  \includegraphics[width=0.47\textwidth]{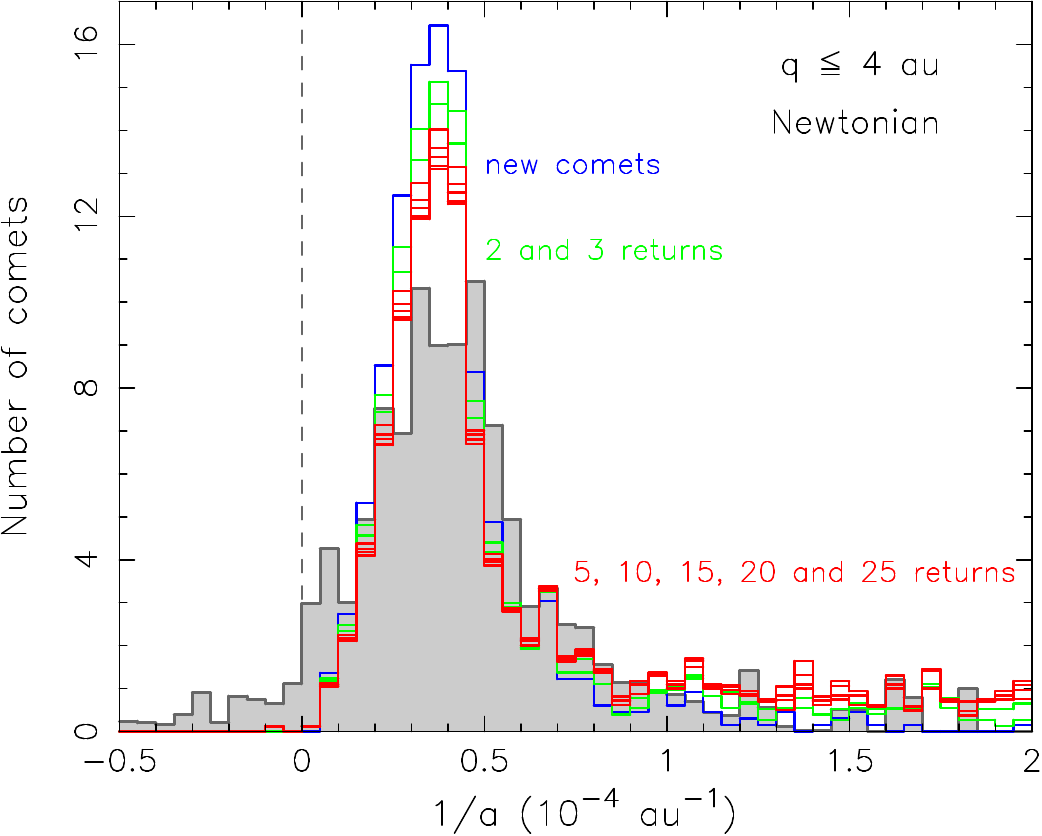} 
 \end{center}  
 \caption{The gray histogram shows the inverse of the original semimajor axes $1/a$ (in units of $10^{-4}$~au$^{-1}$) for a sample of $116$ nearly parabolic LPCs with perihelion $q\leq 4$~au, from the Warsaw catalog. The colored lines show the results from the Newtonian simulation ${\cal S}0$. As in Fig.~\ref{fig_lpc_ret_newt}, the colored histograms show the  simulated distributions of new comets (blue), comets returning up to two and three times (green), and multiply visiting
  comets (red), all normalized to the number of observed comets in the data. The new comets contribute 
  about $80$\% of the total number, and nearly $90$\% in the Oort peak with $1/a\leq 10^{-4}$~au$^{-1}$.}
 \label{fig_lpc_oort_newt}
\end{figure}
%%%%%%%%%%%%%%%%%%%%%%%%%%%%%%%%%%%%%%%%%%%%%%%%%%%%%%%%%%%%%%%%%%%%%%%%%%%%%%%%%%%%%%%%%%%%%%%%%%%%%%%

While this qualitative discussion contributes to our understanding of the main processes, the predictions from a specific simulation can be confirmed or refuted only by a careful quantitative comparison with observations. This is not an easy task, as the observed population of distant TNOs (represented by the red triangles in Fig.~\ref{fig_detach}) is heavily biased by the limitations and selection effects of existing surveys. Thus the synthetic population resulting from our simulation must be culled by an accurately known detection probability before it is compared with the observations. Here we conduct such a comparison between the results of our MOND simulations ${\cal S}1$ to ${\cal S}3$, and TNO discoveries made during the six years of operations of the DES \citep{b2022}, using the approach described in Sec.\ \ref{data}.

Without MOND, the simulations accurately reproduce 
%\st{should we show a figure for this, analogous to Figure \ref{fig_detach_s1} %(I know there's one in Nesvorny et al., but it might be good to have one here %as well)? 
% DN: We can add one, you decide David, or just refer to figures 8 and 9 
%in Nesvorny et al., as now done below 
%Should these figures also show the semimajor axis distribution? 
%DN: we did not want to confuse the argument by discussing 
%the semimajor axis distribution which was the subject of Nesvorny et al. 
%The semimajor axis distributions with and without MOND are the same. 
%Perhaps we can just say it somewhere.  
the orbital distribution of distant TNOs (see figures 8 and 9 in  \citep{clu2023}), including their perihelion distances and orbital inclinations.
With MOND, as implemented in the ${\cal S}$1 model, the observationally biased orbital distributions do not give a good fit to the DES discoveries (Fig.~\ref{fig_detach_s1}). In particular, the simulation contains relatively too many TNOs with large perihelia and large inclinations. This problem arises because torques from the EFE cause distant TNOs to evolve to very large perihelion distances (Fig.~\ref{fig_detach}), so if simulation ${\cal S}_1$ were correct, DES should have detected many more objects with $q>38$ au. Similarly, the inclination distribution of distant TNOs is very extended in the ${\cal S}$1 model. DES observations are not strongly biased in orbital inclination; note that the solid and dashed lines in panel (b) of Fig.~\ref{fig_detach_s1}, representing the observationally biased and intrinsic inclination distributions, are quite similar. Yet, only 10\% of the distant TNOs 
detected by DES have $i>40^\circ$. In contrast, about 40\% of the TNOs in simulation ${\cal S}1$ are detected with $i>40^\circ$. 
A Kolmogorov--Smirnov test applied to the inclination distribution shows only a 0.7\% probability that the 
distributions shown in panel (b) of Fig.~\ref{fig_detach_s1} can be obtained from the same underlying distribution. This argument is already sufficient, on its own, to rule out the ${\cal S}$1 model.

%\st{This analysis only discusses the relative distribution of orbital %elemenst of the extreme TNOs (e.g., there are too many TNOs with $q>50$ au in %the simulation, relative to those with $q$ between 35 and 50 au. But in the %conclusions you say ``Current wide-field optical
%surveys would have detected far more of these objects
%than have actually been observed,'' which is different. So are there too many %detached TNOs in the simulation, or is it just the relative distributions of %perihelion and inclination that are wrong? Can we make a statement like %``Simulation ${\cal S}_1$ predicts that DES should have found $x$ members of %the detached disk, but it only found $y$?}
%DN: Good point. Indeed the argument is based on the relative distribution
%and the statement in the conclusions was misleading. Now corrected.

The ${\cal S}$2 and ${\cal S}$3 simulations, with a sharper MOND transition function, cannot be ruled out from the same argument.
The perihelion distance and orbital inclination distributions obtained in these models are consistent with DES detections of distant TNOs. This is because with the sharp transition function the effects of MOND are negligible at the semimajor axes explored by the detectable TNOs, and we already know that Newtonian models fit the distribution of distant TNOs reasonably well \citep{clu2023}. 
% Tab 3 %%%%%%%%%%%%%%%%%%%%%%%%%%%%%%%%%%%%%%%%%%%%%%%%%%%%%%%%%%%%%%%%%%%%%%%%%%%%%%%%%%%
\begin{deluxetable}{l|ccc|ccc}[t] 
 \tablecaption{\label{sims_fit}
  Best-fit parameters of the fading laws for different simulations}
 \tablehead{
  \colhead{} & \colhead{$\kappa$} & \colhead{$c$} & \colhead{$\chi^2$}
   & \colhead{$\lambda$} & \colhead{$f$} & \colhead{$\chi^2$} \\ [-7pt]
  \colhead{} &  & \colhead{Eq.~(\ref{fading})} & & &
      \colhead{Eq.~(\ref{fading1})} &  }
%\decimals
\startdata
 \rule{0pt}{3ex}
 & \multicolumn{6}{c}{{\it -- Reference simulation --}} \\ [1pt]
 ${\cal S}0$\tablenotemark{a}  & $0.82$ & $1.1$ & $1.5$ 
   & $0.18$ & $0.08$ & $1.5$ \\ [3pt]
 & \multicolumn{6}{c}{{\it -- Nominal simulations --}} \\ [1pt]
 ${\cal S}1$  & $0.67$ & $0.0$ & $3.6$ & $0.44$ & $0.09$ & $3.3$ \\
 ${\cal S}2$  & $0.85$ & $7.8$ & $4.8$ & $0.09$ & $0.19$ & $4.8$ \\ [3pt]
 & \multicolumn{6}{c}{{\it -- Extended simulations --}} \\ [1pt]
 ${\cal S}3$  & $0.60$ & $0.0$ & $3.6$ & $0.35$ & $0.12$ & $3.6$ \\ 
 ${\cal S}2^\prime$\tablenotemark{b}  & $0.47$ & $0.1$ & $4.2$ 
  & $0.17$ & $0.18$ & $4.1$ \\ [2pt]
\enddata
\tablenotetext{a}{Labels of the simulations from Table~\ref{sims}.}
\tablenotetext{b}{See Appendix~\ref{jobs2p}.}
\tablecomments{The parameters $(\kappa,c)$ and $(\lambda,f)$ of the fading models of Eqs.\ (\ref{fading}) and (\ref{fading1}) respectively, corresponding the best fit to the distribution of original semimajor axes for LPCs with $q\leq 4$~au. The last column gives the minimum reduced $\chi^2$ of each fit.}
\end{deluxetable}
%%%%%%%%%%%%%%%%%%%%%%%%%%%%%%%%%%%%%%%%%%%%%%%%%%%%%%%%%%%%%%%%%%%%%%%%%%%%%%%%%%%%%%%%%%% 

Finally, we briefly comment on the results of \citet{bm2023}, who suggest that MOND produces an anti-alignment between the direction ${\bf e}$ towards the Galactic center and the perihelion direction $\hat{\bf p}$ of the detached TNOs. In order to quantify this effect, \citet{bm2023} analyzed the distribution of the parameter $u=(1+ {\bf e}\cdot\hat{\bf p})/2$, determined for a small sample ($N=6$) of objects in the detached disk. They observed a statistical pile-up of $u$ near zero, which they interpreted as capture in a stable Kozai-type island from a secular quadrupole that arose through the EFE. In order to test
this hypothesis, we computed the value of the $u$ parameter for all particles in the detached disk at the end of the simulations ${\cal S}1$ to ${\cal S}3$ (orbits with barycentric semimajor axis
in the range 100~au~$\le a\leq 1,000$~au, with perihelion $q\geq 38$~au). 
In all of these simulations we found that $u$ has a distribution inconsistent with the strong asymmetry found by \citet{bm2023} in the limited observational data ($\langle u\rangle=0.16$ for $N=6$ bodies in the detached disk) and predicted by them if MOND is correct. The red and blue curves in Fig.~\ref{fig_u} show the distribution of $u$ for particles at the end of simulations ${\cal S}1$ and ${\cal S}2$. While not strictly uniform -- we do not expect this, because of the EFE -- the differences from the uniform model are very small. The KS distance of the ${\cal S}1$ and ${\cal S}2$ distributions from the uniform model is $0.04$ at maximum. The gray curves in Fig.~\ref{fig_u} provide the $u$ distributions during the last $500$~Myr in these integrations with a step of $100$~Myr. Overall, there is no systematic and substantial deviation from an uniform distribution. Other concerns with the analysis of \cite{bm2023} are discussed in Sec.\ \ref{intro}.

\subsection{Implications for the Oort cloud and LPCs}\label{res2}

We first briefly review results from Newtonian simulations. We used a reference simulation performed by \citet{clu2023}, which provides results equivalent to those discussed
in \citet{vnd2019}. Both references used a very similar methodology to this paper, the
planetary migration model in particular, but they are not identical. For instance, \citet{vnd2019} used the same initial trans-Neptunian disk model as we use in the MOND\-ian simulations described
below, while \citet{clu2023} had a broader initial disk of planetesimals extending from Jupiter's initial location to $30$~au heliocentric distance. Additionally, several simulations in \citet{clu2023} modeled the dynamical effects from the solar system's birth cluster. The presence or absence of the birth cluster does not affect the distribution
of binding energies of currently observed LPCs, at least in the simulations of \citet{clu2023}.
% FIG 1 %%%%%%%%%%%%%%%%%%%%%%%%%%%%%%%%%%%%%%%%%%%%%%%%%%%%%%%%%%%%%%%%%%%%%%%%%%%%%%%%%%%%%%%%%%%%%%%
\begin{figure*}[t!]
 \begin{center}
% \begin{tabular}{c}
%  \includegraphics[width=0.8\textwidth]{fAa.eps} \\
%  \includegraphics[width=0.8\textwidth]{fAb.eps} \\ 
% \end{tabular}
  \includegraphics[width=0.9\textwidth]{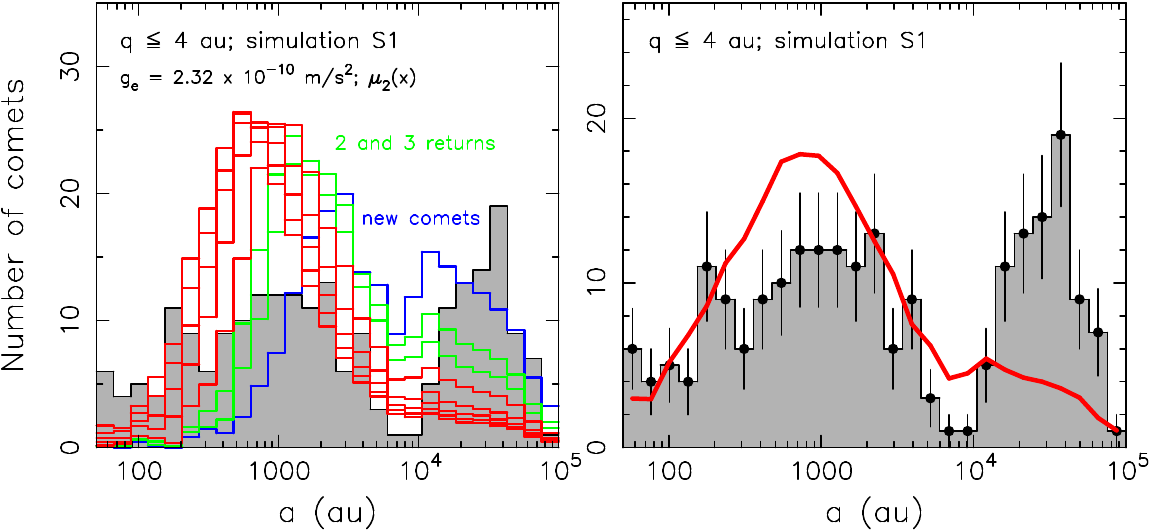} 
 \end{center}  
 \caption{The same as Fig.~\ref{fig_lpc_ret_newt}, but now for the simulation ${\cal S}1$ based on the gradual transition function $\mu_2(x)$ and the nominal external acceleration
  $g_{\rm e}=2.32\times 10^{-10}$ m~s$^{-2}$. Because of the strength of the MOND\-ian torques at heliocentric distances $\sim r_{\rm M}\simeq 7,000$~au (Figs.~\ref{fig_dg_global} and
  \ref{fig_dg_global1}), the distribution of new comets (class ${\cal C}_1$; blue histogram) extends down to semimajor axes as small as $10^3$~au. With more returns (classes ${\cal C}_m, m\geq 2$; green and red histograms) the distribution spreads to even smaller heliocentric distances. The observed Oort peak of
  LPCs is not reproduced. This flaw is highlighted for the best-fit distribution in the right panel.}
 \label{fig_lpc_ret_s1}
\end{figure*}
%%%%%%%%%%%%%%%%%%%%%%%%%%%%%%%%%%%%%%%%%%%%%%%%%%%%%%%%%%%%%%%%%%%%%%%%%%%%%%%%%%%%%%%%%%%%%%%%%%%%%%%
% FIG 1 %%%%%%%%%%%%%%%%%%%%%%%%%%%%%%%%%%%%%%%%%%%%%%%%%%%%%%%%%%%%%%%%%%%%%%%%%%%%%%%%%%%%%%%%%%%%%%%
\begin{figure}[t!]
 \begin{center}
  \includegraphics[width=0.47\textwidth]{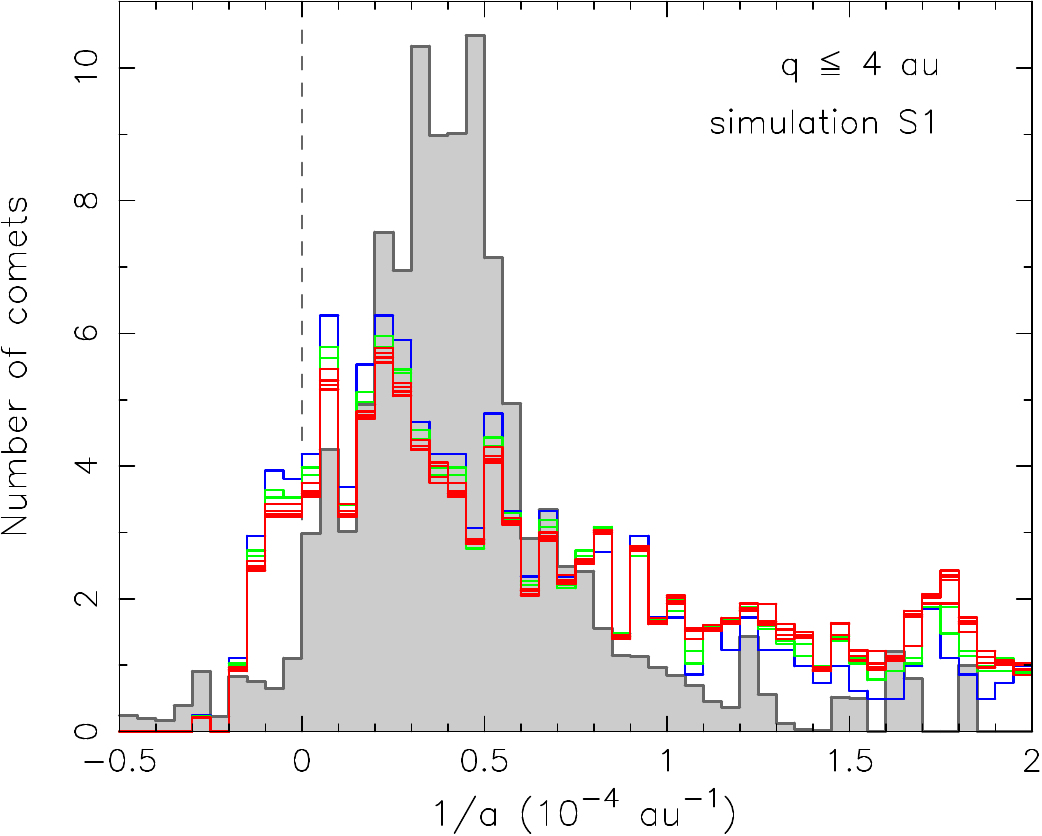} 
 \end{center}  
 \caption{The same as Fig.~\ref{fig_lpc_oort_newt}, but now for the simulation ${\cal S}1$ based on the gradual transition function $\mu_2(x)$ and the nominal external acceleration
  $g_{\rm e}=2.32\times 10^{-10}$ m~s$^{-2}$. The Oort peak is too broad in the model. In addition, a population of LPCs on (formally) hyperbolic orbits emerges in this model.}
 \label{fig_lpc_oort_s1}
\end{figure}
%%%%%%%%%%%%%%%%%%%%%%%%%%%%%%%%%%%%%%%%%%%%%%%%%%%%%%%%%%%%%%%%%%%%%%%%%%%%%%%%%%%%%%%%%%%%%%%%%%%%%%%

Figure~\ref{fig_lpc_ret_newt} shows the simulated distribution of original semimajor axes $a$ for LPCs having perihelion $q\leq 4$~au, compared with the same orbits in the Marsden--Williams catalog. In the left panel, we show the $a$-distribution for a set of models in which the observable comets are allowed to perform a maximum number of perihelion passages with $q\le 4$~au before they disappear from the sample. This family starts with the class ${\cal C}_1$ of comets at their first visit (``new'' comets), continuing with the class ${\cal C}_2$ of comets visiting the target zone at most twice, and so on. In each case we normalized the simulated distribution to contain the same number of LPCs as the data. We see from the figure that the classes with too few visits (${\cal C}_1$, ${\cal C}_2$, or ${\cal C}_3$) over-exaggerate the strength of the Oort peak ($a\geq 10^4$~au), while those for classes with many visible returns (${\cal C}_m$ with $m\geq 15$, say) exaggerate the strength of the hump of returning LPCs ($a\leq 10^4$~au). Clearly a compromise is needed, in which a combination of results from the ${\cal C}_m$ families with a weighting function depending on $m$ would provide a final model that fits the data properly. This was known already to \citet{o1950}, and since
then models with various weighting schemes have been developed. Here we use the simple but successful model of \citet{whipple1962}, in which the probability that a comet survives $m$ returns (i.e., $m$ perihelion passages with $q\le 4$ au) is
the offset power law given by Eq.~(\ref{fading}). The best-fit solution is shown by the red curve on the right panel of Fig.~\ref{fig_lpc_ret_newt} and has $\kappa=0.82$ and $c=1.14$ \citep[this solution is only slightly better than $\kappa\simeq 0.6$--$0.7$ and $c=0$, the power-law model that was used, e.g., by][]{whipple1962,wt1999,vnd2019}. The normalized $\chi^2$ for this model is $1.5$, somewhat larger than would be expected from Poisson statistics (a summary of the best-fit solutions for all models is given in Table~\ref{sims_fit}). 
Taken literally, the likelihood of the best-fitting model is rather small, only $0.02$ \citep[see the $Q$ parameter in Ch.~15.1 of][]{nr2007}.
However, the model is not expected to fit the data perfectly because it does not account for non-gravitational accelerations due to cometary activity (e.g., outgassing; see \citealt{k2020}), and any fading model based solely on the number of perihelion passages is oversimplified. Despite these limitations, the model is able to correctly reproduce the principal characteristics of the distribution of observed LPC orbits, namely (i) the Oort peak composed of very weakly bound cometary orbits ($a=10^4$--$10^5$ au); (ii) the hump of more strongly bound returning comets ($a<10^4$ au), and (iii) the correct relative proportion of these two classes.

Figure~\ref{fig_lpc_oort_newt} provides a zoomed-in view of the LPC population with semimajor axes $a>5,000$ au. Following the tradition since \citet{o1950}, we use the inverse of the original
semimajor axis $1/a$ as the abscissa, instead of $a$ directly. The gray histogram shows $116$ comets from the Warsaw catalog with $q\leq 4$~au, and the color-coded histograms represent the sequence of ${\cal C}_m$ ($m\geq 1$) models from our simulation, each normalized to the total number of comets in the data. Here the new comets (${\cal C}_1$ class; blue histogram) dominate, representing nearly $90$\% of the signal in the Oort peak with $1/a\leq 10^{-4}$~au$^{-1}$. As a result, the properties of the Oort peak are only weakly dependent on the uncertain modeling of the fading process. The match to the data is
satisfactory. The main discrepancy is that the simulated Oort peak is slightly too narrow. Previous studies have shown that non-gravitational accelerations (which we ignore in our model) are the most likely explanation of the differences between the model and the data \citep[e.g.,][]{wt1999,k2020,kd2023}. Non-gravitational forces are likely important for all of the small number of comets on 
apparently hyperbolic orbits ($1/a<0$).
% FIG 1 %%%%%%%%%%%%%%%%%%%%%%%%%%%%%%%%%%%%%%%%%%%%%%%%%%%%%%%%%%%%%%%%%%%%%%%%%%%%%%%%%%%%%%%%%%%%%%%
\begin{figure*}[t!]
 \begin{center}
% \begin{tabular}{c}
%  \includegraphics[width=0.8\textwidth]{fAa.eps} \\
%  \includegraphics[width=0.8\textwidth]{fAb.eps} \\ 
% \end{tabular}
  \includegraphics[width=0.9\textwidth]{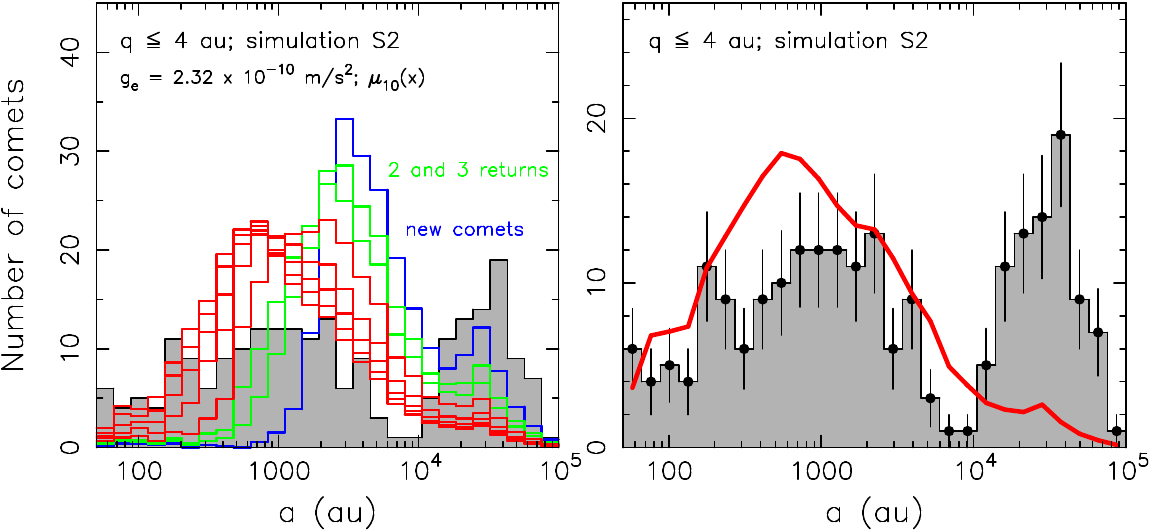} 
 \end{center}  
 \caption{The same as Fig.~\ref{fig_lpc_ret_s1}, but now for the simulation ${\cal S}2$ based on the sharp transition function $\mu_{10}(x)$ and the nominal external acceleration
  $g_{\rm e}=2.32\times 10^{-10}$ m~s$^{-2}$. The problems in matching the distribution
  of the observed LPCs are the same as for simulation ${\cal S}1$.}
 \label{fig_lpc_ret_s2}
\end{figure*}
%%%%%%%%%%%%%%%%%%%%%%%%%%%%%%%%%%%%%%%%%%%%%%%%%%%%%%%%%%%%%%%%%%%%%%%%%%%%%%%%%%%%%%%%%%%%%%%%%%%%%%%
% FIG 1 %%%%%%%%%%%%%%%%%%%%%%%%%%%%%%%%%%%%%%%%%%%%%%%%%%%%%%%%%%%%%%%%%%%%%%%%%%%%%%%%%%%%%%%%%%%%%%%
\begin{figure}[t!]
 \begin{center}
  \includegraphics[width=0.47\textwidth]{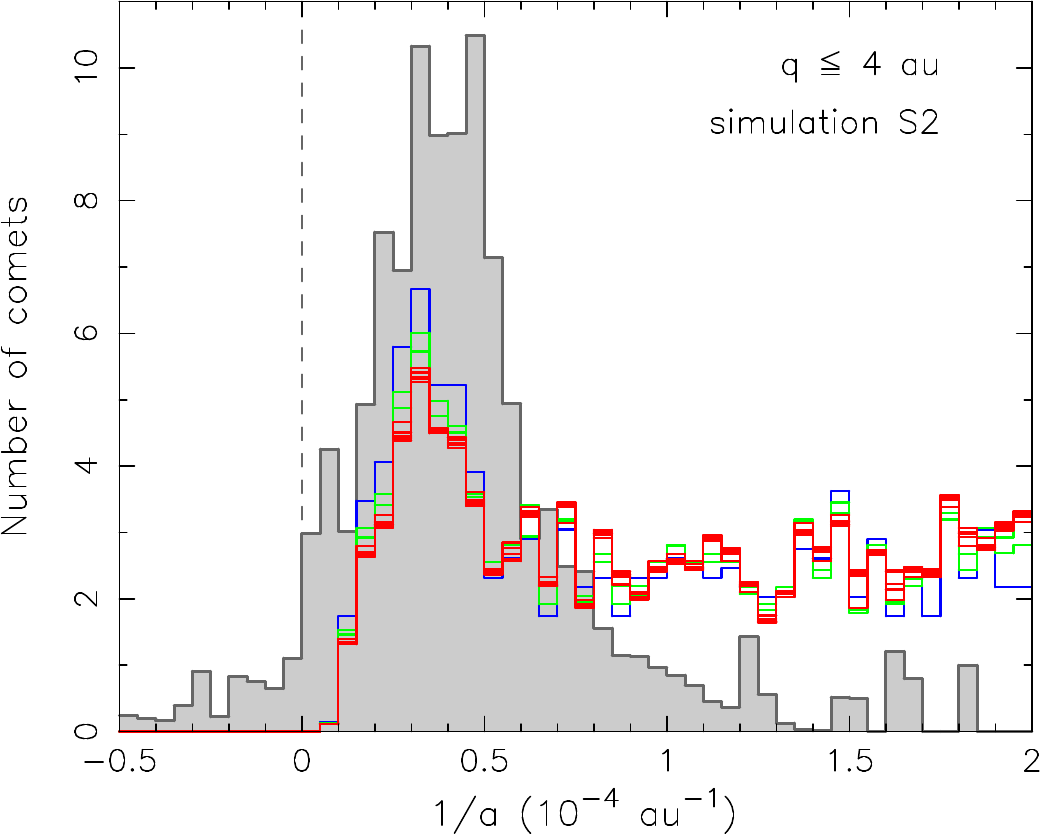} 
 \end{center}  
 \caption{The same as Fig.~\ref{fig_lpc_oort_s1}, but now for the simulation ${\cal S}2$ based on the sharp transition function $\mu_{10}(x)$ and the nominal external acceleration
  $g_{\rm e}=2.32\times 10^{-10}$ m~s$^{-2}$. The signature of the Oort peak is
  absent in the simulated data.}
 \label{fig_lpc_oort_s2}
\end{figure}
%%%%%%%%%%%%%%%%%%%%%%%%%%%%%%%%%%%%%%%%%%%%%%%%%%%%%%%%%%%%%%%%%%%%%%%%%%%%%%%%%%%%%%%%%%%%%%%%%%%%%%%

The Newtonian models we have described provide a benchmark for the success that can be achieved in matching the data. The agreement with the data is rather good although not perfect. If MOND is the correct theory of gravity, then our MOND\-ian simulations of the evolution of LPCs ought to match the data at least as well and probably better. 
\smallskip

\noindent{\it Simulations with MOND and the nominal external acceleration (${\cal S}1$ and
${\cal S}2$).-- }Next, we discuss simulations assuming the nominal value of the external acceleration, 
$g_{\rm e}=2.32\times 10^{-10}$ m~s$^{-2}$.  We use the same
analysis as we did above for the Newtonian simulation.

% FIG 1 %%%%%%%%%%%%%%%%%%%%%%%%%%%%%%%%%%%%%%%%%%%%%%%%%%%%%%%%%%%%%%%%%%%%%%%%%%%%%%%%%%%%%%%%%%%%%%%
\begin{figure*}[t!]
 \begin{center}
% \begin{tabular}{c}
%  \includegraphics[width=0.8\textwidth]{fAa.eps} \\
%  \includegraphics[width=0.8\textwidth]{fAb.eps} \\ 
% \end{tabular}
  \includegraphics[width=0.9\textwidth]{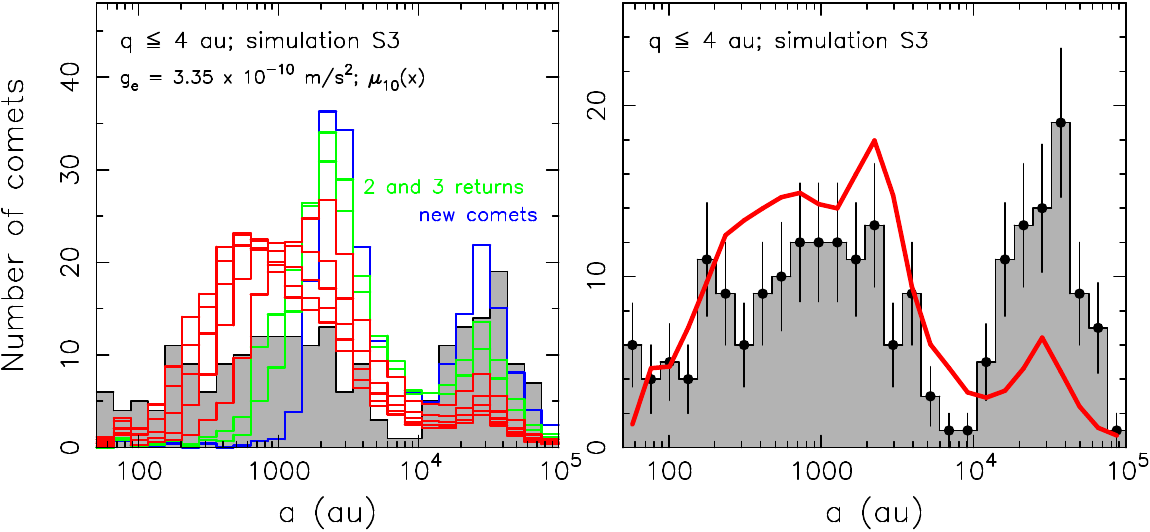} 
 \end{center}  
 \caption{The same as Fig.~\ref{fig_lpc_ret_s2}, but now for the simulation ${\cal S}3$ based on the sharp transition function $\mu_{10}(x)$ and a larger external acceleration, $g_{\rm e}=3.35\times 10^{-10}$ m~s$^{-2}$. Due to the stronger external acceleration the MOND\-ian perturbation is weaker and the signature of the Oort peak emerges more clearly. However, even in the class ${\cal C}_1$ of new comets there is a large  population of LPCs with semimajor axes as small as $10^3$ au. The best-fit model using Whipple's fading law (\ref{fading}), under-represents the Oort peak.}
 \label{fig_lpc_ret_s3}
\end{figure*}
%%%%%%%%%%%%%%%%%%%%%%%%%%%%%%%%%%%%%%%%%%%%%%%%%%%%%%%%%%%%%%%%%%%%%%%%%%%%%%%%%%%%%%%%%%%%%%%%%%%%%%%
% FIG 1 %%%%%%%%%%%%%%%%%%%%%%%%%%%%%%%%%%%%%%%%%%%%%%%%%%%%%%%%%%%%%%%%%%%%%%%%%%%%%%%%%%%%%%%%%%%%%%%
\begin{figure}[t!]
 \begin{center}
  \includegraphics[width=0.47\textwidth]{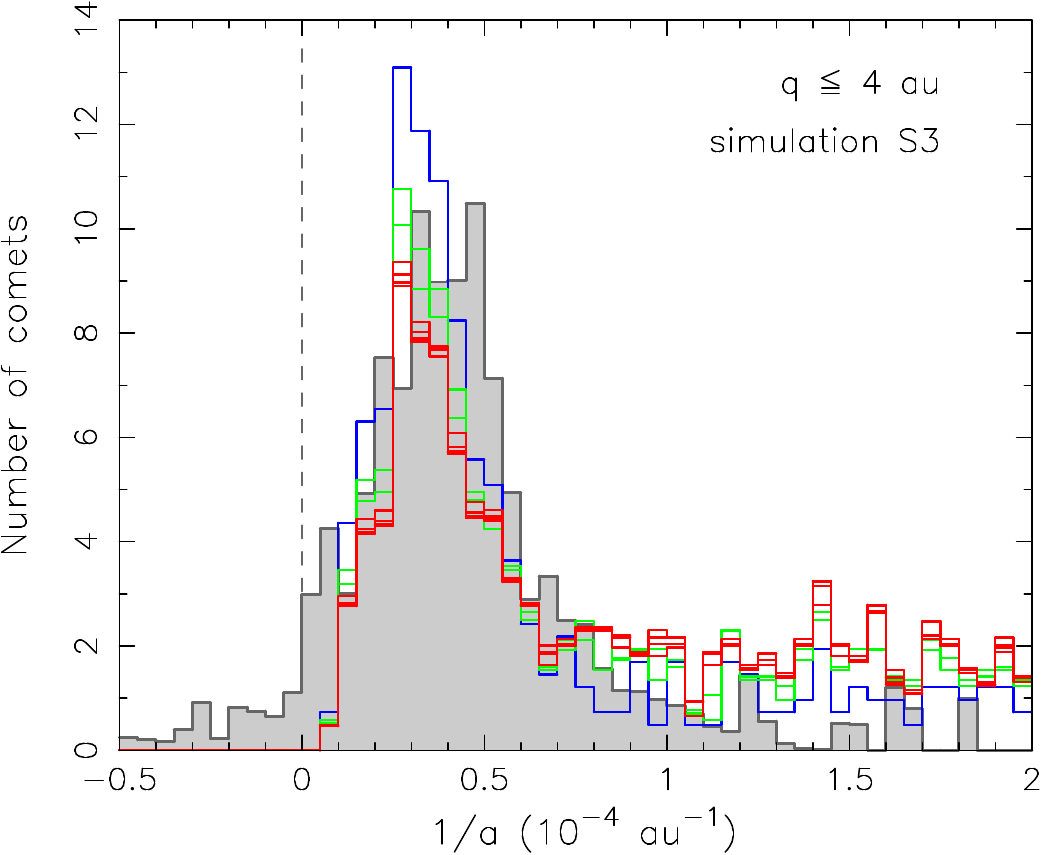} 
 \end{center}  
 \caption{The same as Fig.~\ref{fig_lpc_oort_s2}, but now for the simulation ${\cal S}3$ based on the sharp transition function $\mu_{10}(x)$ and a larger external acceleration, $g_{\rm e}=3.35\times 10^{-10}$ m~s$^{-2}$. Here the correspondence between the model and the observations of the distribution of binding energy $1/a$ for nearly parabolic LPCs appears satisfactory (i.e., comparable to that in the Newtonian simulation, Fig.~\ref{fig_lpc_oort_newt}). The problem with the simulation, namely the large population of new comets with semimajor axes of $10^3$--$10^4$~au (Fig.~\ref{fig_lpc_ret_s3}),
  is beyond the range displayed here.}
 \label{fig_lpc_oort_s3}
\end{figure}
%%%%%%%%%%%%%%%%%%%%%%%%%%%%%%%%%%%%%%%%%%%%%%%%%%%%%%%%%%%%%%%%%%%%%%%%%%%%%%%%%%%%%%%%%%%%%%%%%%%%%%%

Figure~\ref{fig_lpc_ret_s1} compares the distribution of the original semimajor axes $a$ of the observed LPCs with $q\leq 4$~au to the results of our simulation ${\cal S}1$. The changes due to MOND are best illustrated by the distribution of new comets (${\cal C}_1$ class; blue histogram). While previously the outer Oort cloud ($a \ge 15,000$ au) was the only source of new comets, the vigorous MOND\-ian torques at heliocentric distances of a few thousand au (roughly the MOND scale $\sim r_{\rm M}$) provide an additional source of new comets. The ${\cal C}_1$ class thus is dominated by much smaller semimajor axes. At subsequent returns (classes ${\cal C}_m$, $m\geq 2$), the comets tend to bind more strongly to the solar system, corresponding to even smaller semimajor axes. In these classes there is no concentration of comets in the Oort peak between $10^4$ and $10^5$ au.  A combination of the classes ${\cal C}_m$, weighted by Whipple's fading law (\ref{fading}), may again be used to find the formal best fit to the data (right panel on Fig.~\ref{fig_lpc_ret_s1}, and Table~\ref{sims_fit}). The best fit, with parameter values $\kappa=0.67$ and $c=0$, is poor, as the minimum reduced $\chi^2$ is $\simeq 3.6$. The fit is not improved significantly with our alternative fading law (Table \ref{sims_fit}). 

Figure~\ref{fig_lpc_oort_s1} provides a zoomed-in view of the  nearly parabolic LPCs. The simulation shows only modest excursions from a nearly uniform distribution in $1/a$, even for the new comets (class ${\cal C}_1$), and is inconsistent with the data because it does not reproduce the width and height of the Oort peak. Moreover, the simulated distribution now includes far more LPCs on hyperbolic orbits ($1/a<0$) than the data. In the simulation, these comets are bound to the solar system, but their orbits have been significantly affected by MOND\-ian effects. Because the original semimajor axis value is computed by a Newtonian model (the same way as the data are processed), these objects appear be approaching the inner solar system on unbound hyperbolic orbits. We have verified, using smaller scale simulations, that this apparent hyperbolic population of new LPCs becomes increasingly significant for smaller values of the external acceleration value $g_{\rm e}$ and for more gradual MOND\-ian transition functions (see Fig.~\ref{fig_ge_b} for context). 

The simulation ${\cal S}2$, with a sharp transition function, is shown in Figs.~\ref{fig_lpc_ret_s2} and \ref{fig_lpc_oort_s2}. This simulation is also inconsistent with the data. The fit using Whipple's fading model (\ref{fading}) has parameter values $\kappa=0.85$ and $c=7.8$, and the normalized $\chi^2 \simeq 4.8$. The statistical likelihood of a valid model to have this level of $\chi^2$ is now in the land of despair $3\times 10^{-16}$.
The principal mismatch between the model and observations is in the height and width of the Oort peak. No plausible fading model can build the Oort peak out of nothing.

The simulation ${\cal S}2$ used $\mu_{10}(x)$ as an example of a sharp transition function. Using a smaller simulation we verified that the results are similar for an even sharper transition function, $\mu_{20}(x)$. This is not surprising since Fig.~\ref{fig1} shows that the difference between these two transition functions is not very large.
\smallskip

\noindent{\it Simulation with MOND and a larger external acceleration (${\cal S}3$).-- }Finally,
we performed a simulation ${\cal S}3$ using the sharp transition function $\mu_{10}(x)$ and a 45\% larger value of the external acceleration, $g_{\rm e}=3.35\times 10^{-10}$ m~s$^{-2}$. This test was motivated by current ideas about the migration of the solar system in the Galaxy: the solar system probably formed closer to the Galactic center and subsequently evolved outward to its current location. This would imply that the external acceleration $g_{\rm e}$ was larger in the past. Ideally, we should conduct simulations in which $g_{\rm e}$ has a time dependence determined by the Sun's migration history. Unfortunately, (i) the solar migration history is still poorly constrained; (ii) such simulations would require the determination of a continuous set of radial functions $\upsilon_\ell(r,g_{\rm e})$ (Eq.\ \ref{e14}). Given these difficulties, we decided to run a single additional simulation (${\cal S}_3$) in which the external acceleration was constant, and set to the largest value likely to have been experienced by the solar system during its history. Recall that with larger values of $g_{\rm e}$ the MOND\-ian effects are generally reduced, which might help to remove the inconsistencies between the simulated orbital distribution of the LPCs and the observational data. Using the maximum possible value thus provides a strong test of whether MOND is consistent with the orbits of LPCs.

Figures~\ref{fig_lpc_ret_s3} and \ref{fig_lpc_oort_s3} show the simulated distribution of the original semimajor axes of LPCs with $q\le 4$ au, and the $1/a$ distribution for nearly parabolic comets. While the latter distribution is in relatively good agreement with the data, the problem with the distribution of semimajor axes in the whole LPC population persists. Even the new comets (class ${\cal C}_1$; blue histogram) still have  a significant maximum below the $10^4$~au inner boundary of the Oort peak. For that reason, an attempt to fit the observations using Whipple's fading law (Fig.~\ref{fig_lpc_ret_s3}, right panel) still under-represents the Oort peak. The normalized $\chi^2=3.6$, while still much larger than in the Newtonian simulation ($\chi^2=1.5$), is lower than in simulation ${\cal S}_2$, confirming that the predictions of MOND become less inconsistent with the observations when the external field $g_{\rm e}$ is pushed to higher values.

\section{Discussion and conclusions}\label{concl}
% MOND disproved in the solar system

The basic conclusion from this work is that the predictions of the MOND theory of gravity are in conflict with solar-system observations. However,
there are several layers in this statement, which deserve to be discussed in more detail. This is because MOND has more degrees of freedom than conventional theories of gravity (Newtonian gravity and general relativity). 

The most severe conflicts with the observational data occur in the versions of MOND with gradual transition functions. In this paper we have mostly studied a particular example of a gradual transition function, $\mu_2(x)$ (Eq.\ \ref{e2}). MOND theories with this transition function disagree with solar-system data in at least three ways: (i) They predict planetary ephemerides that disagree with observations \citep{bn2011,hetal2014,f2018}. (ii) The torques from the EFE produce relatively broad distributions of perihelion distances and orbital inclinations of distant TNOs, which are inconsistent with DES observations.  
%\st{As I said in an earlier comment, this conclusion is not supported by the %paper, which plots only the relative distributions of perihelion, %inclination, etc.} 
%Current wide-field optical surveys would have detected far more of these objects than have actually been observed. 
(iii) The distribution of binding energies of the long-period comets with perihelia $q\le 4$ au is inconsistent with the data, because orbits with original semimajor axes $>10,000$~au (the Oort peak) are under-represented relative to those with smaller semimajor axes. This statement holds for any fading law that depends only on the number of perihelion passages close to the Sun. 

In spite of its multiple failures, variants of MOND with slow transition functions are still popular in studies of galaxies because they fit galaxy rotation curves more accurately \citep[see, e.g.,][]{fm2012,can2015,lug2015,nag2021,desmond2024}. This result may indicate that the transition function is asymmetric, i.e., $\mu(x)\to 1$ rapidly when $x\gg1$, but $\mu(x)\to x$ gradually when $x\ll1$ (see for example the transition function favored by \citealt{desmond2024}). Such a transition function would not affect our conclusions since we have shown that MOND is incompatible with the LPC data even if we only consider the regime $x\lesssim 1$, i.e., $r\gtrsim r_{\rm M}$. 

The conflicts with the data are less severe for versions of MOND with sharp transition functions. We illustrate this class using the transition function
$\mu_{10}(x)$. Adopting the external acceleration $g_{\rm e}=2.32\times 10^{-10}$ m~s$^{-2}$ from \citet{klio2021} \citep[an independent estimate from][is only 7\% smaller]{mcmil2017}, this model is consistent with planetary ephemerides as well as with the currently available survey constraints on the population of
TNOs detached from the scattered disk. However, the model still predicts a semimajor axis distribution of LPCs that is inconsistent with the data between a few hundred au and $10^5$ au -- the Oort peak at $a>10,000$ au is still too small compared to the population of comets on more tightly bound orbits (see Figs.\ \ref{fig_lpc_ret_s2} and \ref{fig_lpc_oort_s2}). 

We find that larger values of the external acceleration $g_{\rm e}$ reduce the inconsistencies between MOND models with sharp transition functions and the observational data. Our results imply that in order to be consistent with the data, $g_{\rm e}$ must be substantially more than 50\% larger than its current value during much of the history of the solar system. It is unlikely that this constraint can be satisfied in plausible models of the migration history of the solar system; however, (i) the migration history is not well-understood and (ii) simulations that would take into account a fully consistent model of solar-system migration have not yet been done. 

Our simulations have other limitations, although we do not believe that any of these compromise our conclusions:

\begin{itemize}

    \item We have only tested MOND using a specific one-parameter family of models for the transition function, $\mu_n(x)$ (Eq.~\ref{e2}). However, we believe that this family is general enough that there are no plausible transition functions that would be consistent with the observational data on the LPCs. 

    \item We only tested two fading laws (Eqs.\ \ref{fading} and \ref{fading1}), each with two free parameters that we fit to the data. Both assume that the probability of fading depends only on the number of perihelion passages close to the Sun. Other fading laws, which might depend on parameters such as the size of the nucleus rather than the number of perihelion passages \citep[e.g.,][]{jewitt2022}, might offer better fits to the data. However, we note that the MOND\-ian simulations of the distribution of semimajor axes of LPCs already over-predict the number of comets with $a<10^4$~au relative to the number in the Oort peak for comets on their first perihelion passage (``new'' comets) and any fading law that allows comets to remain visible in subsequent passages makes the disagreement worse. 

    \item Perturbations due to non-gravitational forces have been neglected in our analysis. These forces arise from outgassing, which typically begins at a heliocentric distance of a few au, where water ice starts sublimating from the cometary surface. While the inclusion of the effects of outgassing on the orbit are important for modeling the details of the binding energy distribution of LPCs, especially for the Oort-cloud component, their inclusion is unlikely to cure the differences between the MOND\-ian simulations and the observational data. The work of \citet{k2020} suggests the average effect of non-gravitational forces on the orbits of LPCs is a slight shift toward a more bound distribution in $1/a$ ($\Delta(1/a)\lesssim 10^{-5}$~au$^{-1}$), which is much too small and in the wrong direction to make the results in our ${\cal S}1$ to ${\cal S}3$ simulations agree with the observed data.
    
    \item All of our simulations based on MOND\-ian dynamics ignore the possibility that tidal fields and stellar encounters were much stronger during the brief period when the Sun was still part of its birth cluster. However, the effects of the birth cluster were examined in our Newtonian models (Fig.\ \ref{fig_detach}); the presence or absence of the birth cluster did  not affect the distribution of binding energies of the currently observed LPCs in the Newtonian simulations and we believe that it is unlikely to do so in MOND.
    
    \item We have only examined a single value of the critical acceleration, $a_0=1.2\times 10^{-10}$ m~s$^{-2}$. Fits to galactic
    rotation curves have led to $a_0$ values that vary from the one we have used by up to $\pm40\%$ \citep[e.g.,][]{heetal2016}. However, (i) the best-fit value of $a_0$ depends on the transition function, and for the sharp transition functions that we have argued are required by solar-system observations the allowed range of $a_0$ is smaller; (ii) the characteristic MOND radius $r_{\rm M}$ varies only as the inverse square root of $a_0$, so the range of possible values of $r_{\rm M}$ remains much smaller than the range of semimajor axes in the Oort cloud. 

\end{itemize}

% Wide binaries from GAIA
It is not our intention to review the many tests of MOND on the scale of galaxies and clusters of galaxies (see for example \citealt{fm2012}). We note, however, that another potential test of Newtonian dynamics and gravity, on comparable length and mass scales to the LPCs, is provided by wide binary stars \citep{chae2023,chae2024,her2023,heretal2023}.
These papers argue for a breakdown of Newtonian dynamics in binary systems when the semimajor axis exceeds a few thousand astronomical units, which for these systems corresponds to a 
mutual acceleration $\lesssim a_0$. The analysis is challenging, however, because (i) the masses of the stars must be estimated from spectral fitting; (ii) only two of the three components of the spatial separation of the two stars can be measured; (iii) only two of the three relative velocity components can be measured (the referenced papers derive the kinematics from Gaia measurements of the relative proper motion), and the observational errors in the velocity are significant; (iv) the binary systems may contain unseen third components, either bound or unbound  \citep[e.g.,][]{clarke2020}.  Indeed, other authors argue from similar data that the dynamics of this class of binaries \emph{is} well described by Newtonian mechanics \citep[e.g.,][but see \citealt{hc2023}]{pitt2023,banik2024,desmond2024}.  \citet{chae2023} reports a statistically significant acceleration excess $g/g_{\rm N}-1$ of $\simeq (12\pm3)$\% at $g_{\rm N}=10^{-8.91}$ m~s$^{-2}$, or $g_{\rm N}=10.25\,a_0$. If (i) this excess is due to MOND, (ii) the AQUAL formulation is correct, and (iii) the transition function belongs to the family $\{\mu_n(x)\}$ (Eq.\ \ref{e2}), then the transition function must be gradual, with $n\lesssim 1$  (see Fig.~\ref{fig1}). As summarized in Sec.\ \ref{efe1} such gradual transition functions are excluded on
solar-system scales from planetary and comet dynamics, although it is easy to construct an  \emph{ad hoc} transition function, slowly varying for $x\lesssim 10$ and rapidly approaching unity for larger values of the argument, that would be compatible with both sets of observations. 

The heliocentric acceleration of comets in the Oort cloud is significantly less than the critical MOND acceleration $a_0$, and comparable to the acceleration measured in the outer parts of the rotation curves of galaxies. However, the Oort cloud differs from galaxies in at least three relevant aspects: the mass scale is much smaller ($1M_\odot$ vs.\ $10^{11}M_\odot$), the spatial scale ($0.1\,$pc vs.\ $10^4\,$pc), and the strength of the
external field $g_{\rm e}$ ($1.9a_0$ vs.\ a few per cent of $a_0$). If MOND is correct, one or more of these differences may explain why the theory works on galaxies but fails in the Oort cloud. 

\smallskip

% k-mouflage mechanism?
\noindent{\it Other theories of MOND.-- }AQUAL is only one of a number of theories of gravity and dynamics that reproduce the most important features of MOND on galactic scales: the breakdown of Newtonian dynamics and/or gravity at small accelerations, the asymptotic flatness of galaxy rotation curves, etc. The simplest alternative is QUMOND \citep{m2010}, which we discussed briefly around Eq.\ (\ref{eq:qqq}), and we believe that simulations based on QUMOND would give very similar results. More elaborate alternatives include modified-inertia theories \citep{mil2011,mil2023b}, generalized QUMOND \citep{mil2023}, tripotential MOND \citep{mil2023c}, a MOND adaptation of Galileon-k-mouflage theory \citep{bab2011}, etc. 

Ideally, we should repeat our simulations for these other theories. However, there are obstacles to doing so:

\begin{itemize}

    \item In some cases, the theory contains an adjustable screening length $R_{\rm sc}$ that suppresses MOND\-ian effects on small scales. For example, \citet{bab2011} (see also \citealt{mil2023} and similar comments in the concluding section of \citealt{banik2024}) proposed a relativistic theory of gravitation of the tensor-scalar class, for which MOND is the weak-field limit.%
\footnote{In fact, the work of \citet{bab2011} builds on a more general framework, in which theorists aim at constructing gravity theories modified at large, usually cosmological, distances, while preserving the validity of well-tested general relativity on small spatial scales 
 \citep[e.g.,][]{kmouf2009,ishak2019}.} The model has negligible deviations from general relativity on scales below the screening length $R_{\rm sc}$. Although the screening phenomenon
is constructed only for this purpose, the resulting theory is self-consistent and admissible. Generalizations of QUMOND that depend on higher derivatives of the potential may also have screening lengths below which the MOND\-ian effects are strongly suppressed \citep{mil2023}. In such theories our simulations would not test MOND, but would only increase the lower limit to $R_{\rm sc}$ from the size of the planetary system (a few au) to the size of the Oort cloud (5--10 $\times 10^4$ au).

\item In some cases, the theory contains enough degrees of freedom that it is impractical to test with a limited set of numerical simulations. For example, in tripotential MOND \citep{mil2023c}, the transition function depends on three variables, not one. 

\item As discussed already by \citet{m1983a}, some MOND theories preserve Poisson's equation and instead modify the law of inertia. Modern variants of this approach are developed in \citet{mil1994}, \citet{mil2011} and \citet{mil2023b}. The more complicated modified-inertia theories bring a time-nonlocal (memory) effect into the dynamics.  For that reason, a thorough analysis of modified-inertia models in specific astronomical problems has not been carried out so far; in particular, standard $N$-body techniques could not be used for simulations of orbit evolution. 

\end{itemize}

The simulations in this paper appear to rule out AQUAL, one of the simplest and most well-defined MOND theories. All existing MOND theories are heuristic, and thus will eventually be superseded by a ``fundamental'' MOND theory (if one exists). Whatever form this theory takes, it will be strongly constrained by the requirement that it reproduce the orbital distributions of the long-period comets and trans-Neptunian objects. 

\acknowledgments
 We thank Benoit Famaey and Mordehai Milgrom for thoughtful and constructive comments. 
 The simulations were performed on the NASA Pleiades Supercomputer. We thank the NASA NAS
 computing division for continued support. DN was supported by the NASA Emerging 
 Worlds program. 
 ST was supported in part by the Natural Sciences and Engineering Research Council of Canada (NSERC), funding reference number RGPIN-2020-03885.
 
%%%%%%%%%%%%%%%%%%%%%%%% REFERENCES %%%%%%%%%%%%%%%%%%%%%%%%%%
%\newpage
\bibliography{lit}{}
\bibliographystyle{aasjournal}

\appendix

\section{Testing the accuracy of the numerical solution for the MOND potential with
 analytical results}\label{check}
 
The radial functions $\upsilon_\ell(r)$ of the MOND potential $u({\bf r})$ (\ref{e14}) are determined 
numerically in terms of Chebyshev polynomials, segmented in a number of radial zones between the origin and spatial infinity (see Sec.\ \ref{efe} and Appendix \ref{details}). We seek tests of the accuracy of our solutions using comparisons to previously published numerical results 
and analytic asymptotic behaviors at $r\rightarrow \infty$ (tests of the solution
near the origin $r=0$ have already been discussed in Sec.~\ref{efe1}).
\smallskip

\noindent{\it Testing the far zone: $\propto 1/r$ asymptotic behavior.-- }At spatial infinity, \citet{bn2011} re-derive in their Appendix~A an earlier result of \citet{bm1984}, namely an asymptotic behavior of $u$ 
\begin{equation}
 u\left(r,\theta\right) = \frac{GM}{r\mu_e\sqrt{1+\lambda_e\sin^2\theta}}+
  O\left(\frac{1}{r^2}\right) \; , \label{e13}
\end{equation}
where $\mu_e=\mu(x_e)$ and $\lambda_e=[d\log\mu(x)/d\log x]_{x_{\rm e}}$, with $x_e= g_{\rm e}/a_0$. To connect this asymptotic approximation of $u$ with its multipole series representation (\ref{e14}), we rewrite this in terms of Legendre polynomials (recall $\tau=\cos\theta$)
\begin{equation}
 u\left(r,\tau\right) = \frac{GM}{r} \sum_{\ell\geq 0} w_{2\ell}\,P_{2\ell}(\tau)+
  O\left(\frac{1}{r^2}\right)\; . \label{e13bis}
\end{equation}
Straightforward algebra yields
\begin{equation}
  w_{2\ell} = \frac{\kappa_{2\ell}}{\mu_e \alpha_e^{\ell-1} \sqrt{\alpha_e \lambda_e}} \left[  \mathbb{A}_{2\ell}\left(\alpha_e\right)\frac{{\rm arcsin}\sqrt{\alpha_e}}{\sqrt{\alpha_e}}- \mathbb{B}_{2\ell}\left(\alpha_e\right)\sqrt{1-\alpha_e}\right]\; , \label{wel}
\end{equation}
where $\alpha_e=\lambda_e/(1+\lambda_e)$, 
$\kappa_{2\ell}$ are numerical coefficients, and $\mathbb{A}_{2\ell}\left(\alpha_e\right)$ and
$\mathbb{B}_{2\ell}\left(\alpha_e\right)$ are polynomials of order $\ell$ and $\ell-1$. For the first few terms%
\footnote{\citet{m2010}, Eq.~(65), derives within the QUMOND variant of MOND a result equivalent to our monopole term given in (\ref{w0asym}) with (\ref{wel}). In particular, he shows that the effective dynamical mass of the system is equal to $M\,w_0$ (in our notation), whose excess over $M$ arises due to the MOND\-ian phantom mass. However, this interpretation must be taken with some caution, since the asymptotic acceleration is not  spherically symmetric (see Eq.\ \ref{e13}).}
we obtained $\kappa_0=1$, $\kappa_2=15/4$, $\kappa_4=945/64$, $\kappa_6=15015/256$, $\kappa_8=3828825/16384$, $\kappa_{10}=61108047/65536$, and the $\mathbb{A}$ and $\mathbb{B}$ polynomials read
\begin{eqnarray}
 \mathbb{A}_{0}\left(\alpha_e\right)&=& 1\; , \;\;\, \mathbb{B}_{0}\left(\alpha_e\right) = 0\; ,
  \label{w0asym} \\
 \mathbb{A}_{2}\left(\alpha_e\right)&=& 1-\frac{2\alpha_e}{3}\; , \;\;\,
 \mathbb{B}_{2}\left(\alpha_e\right) = 1\; , \label{w2asym} \\
 \mathbb{A}_{4}\left(\alpha_e\right)&=& 1-\frac{8\alpha_e}{7}+ \frac{8\alpha_e^2}{35}\; , \;\;\,
 \mathbb{B}_{4}\left(\alpha_e\right) = 1- \frac{10\alpha_e}{21}\; , \label{w4asym} \\
 \mathbb{A}_{6}\left(\alpha_e\right)&=& 1-\frac{126\alpha_e}{77}+\frac{8\alpha_e^2}{11}-
  \frac{16\alpha_e^3}{231}\; , \;\;\,
 \mathbb{B}_{6}\left(\alpha_e\right) = 1-\frac{32\alpha_e}{33}+\frac{28\alpha_e^2}{165}\; ,
  \label{w6asym} \\
 \mathbb{A}_{8}\left(\alpha_e\right)&=& 1-\frac{32\alpha_e}{15}+\frac{96\alpha_e^2}{65}-
  \frac{256\alpha_e^3}{715} +\frac{128\alpha_e^4}{6435}\; , \;\;\,
 \mathbb{B}_{8}\left(\alpha_e\right) = 1-\frac{22\alpha_e}{15}+\frac{344\alpha_e^2}{585}-
  \frac{12176\alpha_e^3}{225225}\; , \label{w8asym} \\
 \mathbb{A}_{10}\left(\alpha_e\right)&=& 1-\frac{50\alpha_e}{19}+\frac{800\alpha_e^2}{323}-
  \frac{320\alpha_e^3}{323} +\frac{640\alpha_e^4}{4199} - \frac{256\alpha_e^5}{46189}\; , \;\;\,
 \mathbb{B}_{10}\left(\alpha_e\right) = 1-\frac{112\alpha_e}{57}+\frac{2028\alpha_e^2}{1615}-
 \frac{3232\alpha_e^3}{11305}+\frac{21472\alpha_e^4}{1322685}\; . \nonumber \\ [3pt]
  \label{w10asym} 
\end{eqnarray}
In the limit $x_e\rightarrow 0$ the exterior field becomes negligible, and we obtain $\lambda_e\rightarrow 1$, $\alpha_e\rightarrow 1/2$ and $\mu_e\rightarrow 0$. In this case, however, the limit is discontinuous, as the
assumption $u(r)\propto 1/r$ from (\ref{e13}) does not hold -- recall from Sec.~\ref{theory} that the MOND\-ian potential of an isolated spherical system has a logarithmic
divergence at infinity (Eq.~\ref{emondinf}).
In the opposite limit $x_e\rightarrow \infty$, we have $\lambda_e\rightarrow 0$, $\alpha_e\rightarrow 0$ and $\mu_e\rightarrow 1$ -- in this situation the MOND\-ian effects are removed,  $w_0\rightarrow 1$, $w_{2\ell}\rightarrow 0$ for $\ell\geq 1$; and $u(r,\tau)=GM/r$, the Newtonian result.

We used both the complete asymptotic expression (\ref{e13}), and the coefficients of its lowest multipoles in (\ref{e13bis}) computed using (\ref{wel}) to (\ref{w10asym}), to verify the accuracy of our numerical evaluation of $u(r,\theta)$. First, we computed the potential $u$ at the outermost radial domain $r_{\rm max}^I=R'$, thus $u(R',\theta)$, and we compared it with (\ref{e13}). The values matched with fractional accuracy of $\simeq 10^{-4}$--$10^{-5}$ for all our solutions. Second, we used the behavior of the radial functions $\upsilon_\ell(r)$ at the outermost radial zone to determine the coefficients $w_{2\ell}$ numerically. Their correspondence to the analytic form in
(\ref{wel}) was again very accurate (see Fig.~\ref{fig_ge_b}).
\smallskip

\noindent{\it Testing the far zone: $\propto 1/r^2$ asymptotic behavior.-- }As indicated in Eq.\ (\ref{e13bis}), the above analysis permits us to test our numerical evaluation of the even-degree multipoles of the MOND potential $u(r,\theta)$, which asymptotically vary $\propto 1/r$.
The odd-degree multipoles decay faster ($\propto 1/r^2$). We would like to check their asymptotic behavior, in particular to verify the  dipole and octupole terms which are important in the formulation of the MOND\-ian potential. We generalized Eq.\ (\ref{e13}) using the expression
\begin{equation}
 u\left(r,\theta\right) = \frac{GM}{r\mu_e\sqrt{1+\lambda_e\sin^2\theta}}+
  \frac{K_2}{r^2} \frac{g_2\left(\theta\right)}{1+\lambda_e\sin^2\theta}+ 
  O\left(\frac{1}{r^3}\right)\; , \label{asy1}
\end{equation}
where $g_2(\theta)$ is an unknown function at this moment. Following the approach in the Appendix of \citet{bn2011}, we seek $g_2(\theta)$ such that ${\bf g}=\nabla U$ satisfies $\nabla\cdot(\mu{\bf g}) = 0$ far
from mass sources (Eq.~\ref{e5})-. Switching to a new independent variable $\tau=\cos\theta$, introducing $G_2(\tau)=g_2(\theta)$, and collecting terms straightforward algebra results in a linear, second-order differential equation in the Sturm--Liouville form
\begin{equation}
 \frac{d}{d\tau}\left[\left(1-\tau^2\right)\phi\, \frac{dG_2}{d\tau}\right]
  + \frac{2\alpha_e}{\lambda_e} \frac{G_2}{\phi^3} = 0 \; , \label{asy2}
\end{equation}
with $\phi=\phi(\tau)=\sqrt{1-\alpha_e\tau^2}$. Further simplification follows from a transformation $\tau\rightarrow \zeta=\sqrt{1-\alpha_e}\, \tau/\phi$, which brings Eq.~(\ref{asy2}) to a Legendre differential equation for eigenvalue equal to $2$. Choosing a dipole solution which is bound on the interval $[-1,1]$, explicitly $G_2(\zeta)=\zeta$, we thus finally obtain the remarkably simple form
\begin{equation}
 u\left(r,\theta\right) = \frac{GM}{r\mu_e\sqrt{1+\lambda_e\sin^2\theta}}\left(1+
  \frac{K'_2}{r} \frac{\cos\theta}{1+\lambda_e\sin^2\theta}\right) + 
  O\left(\frac{1}{r^3}\right)\; . \label{asy3}
\end{equation}
All terms asymptotically behaving as $\propto 1/r^2$ are antisymmetric under the $\tau\rightarrow -\tau$ transformation. They become the leading-order odd-degree terms in Legendre series representation of $u(r,\theta)$, namely
\begin{equation}
 u\left(r,\tau\right) = \frac{GM}{r} \sum_{\ell\geq 0} w_{2\ell}\,P_{2\ell}(\tau)+ 
  \frac{K_2}{r^2} \sum_{\ell\geq 0} w_{2\ell+1}\,P_{2\ell+1}(\tau)+
  O\left(\frac{1}{r^3}\right)\; , \label{e13bisbis}
\end{equation}
with
\begin{equation}
  w_{2\ell+1} = \frac{\kappa_{2\ell+1}}{\alpha_e^{\ell-2} \left(\alpha_e \lambda_e\right)^{3/2}}
   \left[\mathbb{C}_{2\ell+1}\left(\alpha_e\right)\frac{1}{\sqrt{1-\alpha_e}}-
     \mathbb{D}_{2\ell+1}\left(\alpha_e\right)\frac{{\rm arcsin}\sqrt{\alpha_e}}{\sqrt{\alpha_e}}
   \right]\; . \label{wel1}
\end{equation}
Here again $\kappa_{2\ell+1}$ are numerical coefficients, while$\mathbb{C}_{2\ell+1}\left(\alpha_e\right)$ and
$\mathbb{D}_{2\ell+1}\left(\alpha_e\right)$ are polynomials of the order $\ell$. For the
lowest terms we obtained $\kappa_1=3$, $\kappa_3=105/4$, $\kappa_5=10395/64$,
$\kappa_7=225225/256$, and the $\mathbb{C}$ and $\mathbb{D}$ polynomials read
\begin{eqnarray}
 \mathbb{C}_{1}\left(\alpha_e\right)&=& 1\; , \;\;\, \mathbb{D}_{1}\left(\alpha_e\right) = 1\; ,
  \label{w1asym} \\
 \mathbb{C}_{3}\left(\alpha_e\right)&=& 1-\frac{11\alpha_e}{15}\; , \;\;\,
  \mathbb{D}_{3}\left(\alpha_e\right) = 1-\frac{2\alpha_e}{5}\; , \label{w3asym} \\
 \mathbb{C}_{5}\left(\alpha_e\right)&=& 1-\frac{11\alpha_e}{9}+\frac{274\alpha_e^2}{945}\; , \;\;\,
  \mathbb{D}_{5}\left(\alpha_e\right) = 1-\frac{8\alpha_e}{9}+\frac{8\alpha_e^2}{63}\; ,
  \label{w5asym} \\
 \mathbb{C}_{7}\left(\alpha_e\right)&=& 1-\frac{67\alpha_e}{39}+\frac{1784\alpha_e^2}{2145}-
  \frac{484\alpha_e^3}{5005}\; , \;\;\,
  \mathbb{D}_{7}\left(\alpha_e\right) = 1-\frac{18\alpha_e}{13}+\frac{72\alpha_e^2}{143}-
  \frac{16\alpha_e^3}{429}\; . \label{w7asym}
\end{eqnarray}
We did not find a constraint on the scaling constant $K_2$ in (\ref{e13bisbis}), because
Eq.~(\ref{exx}) is at the $1/r^3$ level in $\mu {\bf g}$ automatically satisfied by virtue of the
antisymmetry of the odd-degree multipoles. As a result, our procedure can only check the relative (rather than
absolute) values of the multipole coefficients $w_{2\ell+1}$. Nevertheless, this still provides a stringent 
test of our numerical solution for the potential $u(r,\theta)$. For instance, we easily find that $w_1:w_3:w_5:w_7\simeq 1:8.73:115.03:1754.99$ for the transition function $\mu_2(x)$ and the nominal external acceleration $g_{\rm e}=2.32\times 10^{-10}$ m~s$^{-2}$, and we verified these ratios hold exactly in the full-scale solution of $u(r,\theta)$ shown in Fig.~\ref{fig_pot}.
\smallskip

\noindent{\it Testing the far zone: $\propto 1/r^3$ asymptotic behavior.-- }The procedure outlined above may be pushed to further orders in the power $1/r$, but it becomes more algebraically complicated. Here we briefly comment on the $\propto 1/r^3$ asymptotics. Introducing
the corresponding term in the non-Newtonian potential
\begin{equation}
 \delta u\left(r,\theta\right) = \frac{K_3}{r^3} \frac{g_3\left(\theta\right)}{(1+\lambda_e\sin^2
  \theta)^{3/2}} \; , \label{asy4}
\end{equation}
we seek again $g_3(\theta)$ to satisfy the source-free field equation $\nabla\cdot(\mu{\bf g}) = 0$. Choosing $\tau=\cos\theta$ as the independent variable, and $G_3(\tau)=g_3(\theta)$, we find
\begin{equation}
 \frac{d}{d\tau}\left[\left(1-\tau^2\right)\phi\, \frac{dG_3}{d\tau}\right]
  + \frac{6\alpha_e}{\lambda_e} \frac{G_3}{\phi^3} = S_3\left(\tau\right) \; , \label{asy5}
\end{equation}
where again $\phi=\sqrt{1-\alpha_e\tau^2}$. The important novelty 
is the source term $S_3(\tau)$ (compare with Eq.~\ref{asy2}), which is a non-linear function of the potential asymptotics behaving as $\propto 1/r$ from Eq.~(\ref{e13}), known at this moment.
We do not develop it in full detail here, just note that one can easily show that $S_3(\tau)$ is an
antisymmetric function and may be developed in odd-degree Legendre polynomials. The inhomogeneity
of (\ref{asy5}) implies that a general solution is composed of a free part, from the
homogeneous reduction, and a forced part, specific to the source on the right-hand side,
therefore $G_3 = G_3^{\rm free} + G_3^{\rm forced}$. Upon a final transformation to
$\zeta(\tau)=\sqrt{1-\alpha_e}\, \tau/\phi$, as above, one easily finds that (\ref{asy5}) transforms
to an inhomogeneous Legendre equation with eigenvalue $6$, unsurprisingly corresponding now to
the quadrupole solution. As a result
\begin{eqnarray}
 G_3^{\rm free}\left(\tau\right)   & = & \frac{1}{2}\left(3\zeta^2 - 1\right) \; ,  \label{asy6a} \\
 G_3^{\rm forced}\left(\tau\right) & = & \sum_{\ell\geq 0} \gamma_{2\ell+1}\,P_{2\ell+1}(\tau)
   \; .  \label{asy6b} 
\end{eqnarray}
The free part can be multiplied by an arbitrary constant, and the amplitudes $\gamma_{2\ell+1}$
of the forced part can be in principle computed from the decomposition of the source term $S_3(\tau)$. We
did not carry out this calculation here, but we have verified that the odd-multipole terms $\upsilon_{2\ell+1}(r)$,
comprising the general MOND\-ian potential $u(r,\theta)$ in (\ref{e14}), have the corresponding $\propto 1/r^3$
asymptotic behavior after the $\propto 1/r$ part has been eliminated. The free solution $G_3^{\rm free}$ projects
onto the even-multipole terms in $u(r,\theta)$. However, it is easy to see that the contribution to the monopole
is zero. Therefore, the monopole radial function $\upsilon_0(r)$ from (\ref{e14}) must not contain
either the $\propto 1/r^2$ or the $\propto 1/r^3$ asymptotics. We have verified that this conclusion holds in in all our numerical solutions.
\smallskip

\noindent{\it More tests.-- }Other tests follow from integrating the field equation (\ref{e5}) over a sphere of arbitrary radius
$r$ \citep[see also Eq.~31 in][]{bn2011}: the right-hand side is equal to the source mass $M$ multiplied by
a constant factor $-4\pi G$, while the left-hand side may be transformed to an integral
over the surface of the sphere using Gauss's theorem. If ${\cal S}_r$ is used to denote the surface of a sphere with radius $r$, we have 
\begin{equation}
 \int_{{\cal S}_r} \mu\left(g/a_0\right) \,{\bf g}\cdot d{\bf S} = -4\pi\, G M\; , \label{exx}
\end{equation}
or
\begin{equation}
 r^2\int_{-1}^1 d\left(\cos\theta\right)\, \mu\left(g/a_0\right) \,
  \left[\sum_{\ell=0}^L \upsilon^\prime_\ell\left(r\right)\,P_\ell\left(\cos\theta\right)+
  g_{\rm e}\cos\theta\right] =  -2\, G M\; . \label{eyy}
\end{equation}
We may compute the integral in the left-hand side of Eq.~(\ref{eyy}) at the boundaries
of all the radial zones used in our solution (typically 17 of them), and each time verify
how accurately Eq.\ (\ref{eyy}) is satisfied; we find a fractional accuracy better than $10^{-5}$. 

\bigskip

\section{Technical details about the numerical solution of the MOND equation}
\label{details}

We first focus on the determination of the source term $\sigma(u,\rho)$ in Eqs.~(\ref{e9}) and (\ref{e10}). Following \citet{bn2011} we numerically solve for the MOND potential $u$ outside the Sun only, therefore in the region where the mass density
is zero ($\rho=0$). With ${\bf g}=\nabla u + {\bf g}_{\rm e}$, orientation of the external field along the $z$-axis, and axisymmetry of $u$, we have ($\tau=\cos\theta$)
\begin{equation}
 g^2 = g_{\rm e}^2 + \left(\partial_r u + 2\tau g_{\rm e}\right) \partial_r u+
  \frac{1-\tau^2}{r^2}\left(\partial_\tau u + 2r g_{\rm e}\right) \partial_\tau u
  \; , \label{a1}
\end{equation}  
and similarly
\begin{equation}
 g^i \partial_i g^2 = \left(\partial_r u + \tau g_{\rm e}\right) \partial_r g^2+
  \frac{1-\tau^2}{r^2}\left(\partial_\tau u + r g_{\rm e}\right) \partial_\tau g^2
 \; . \label{a2}
\end{equation}  
The source term on the right-hand side of (\ref{e10}), given $u=u_n$ at some stage of the iterative process, reads
\begin{equation}
 \sigma\left(u,0\right) = -\frac{\mu^\prime}{a_0 \mu}\frac{g^i \partial_i g^2}{2g}
  \; , \label{a3}
\end{equation}  
with both $\mu$ and $\mu^\prime$ functions of $x=g/a_0$. Choosing $\mu(x)$ from one of the family $\{\mu_n(x)\}$ in Eq.\ (\ref{e2}), we finally obtain
\begin{equation}
 \sigma\left(u,0\right) = -\frac{1}{1+(g/a_0)^n}\frac{g^i \partial_i g^2}{2g^2}
  \; . \label{a4}
\end{equation}  
Thus if we know $u(r,\tau)$ we can use Eqs.~(\ref{a1}), (\ref{a2}) and (\ref{a4}) to compute $\sigma(r,\tau)$ on a large grid of $r$ and $\tau$. All necessary partial derivatives are determined analytically using the Legendre and Chebyshev representations of the latitudinal and radial dependence.

%-----
The iterative solution of the potential function $u$ (Eq.~\ref{e9}, Sec.~\ref{efe}) starts with the Newtonian model
($u=u_{\rm N}=GM/r$) of a mass monopole at the origin immersed in an asymptotically homogeneous field ${\bf g}_{\rm e}$ along the $z$-axis of the coordinate system (contributing to $U_{\rm N}$ by a term ${\bf g}_{\rm e}\cdot{\bf r}$: $U_{\rm N}=u_{\rm N}+{\bf g}_{\rm e}\cdot{\bf r}$). This simple approximation defines a unique point $P$ on the positive $z$-axis at which the acceleration vanishes. Its distance from the origin is $r_{\rm e}=r_{\rm M}/\sqrt{\eta}=\sqrt{GM/g_{\rm e}}$, with $r_{\rm M}$ from Eq.~(\ref{eq:rmdef}) and
$\eta=g_{\rm e}/a_0$ as in Sec.~\ref{efe1}. Define auxiliary functions ${\cal Z}(r,\tau)=(r_{\rm e}/r)^2-\tau$, $\zeta_1(r,\tau)=1-\tau^2+{\cal Z}^2$ and
$\zeta_2(r,\tau)=1-\tau^2-2{\cal Z}^2$. Then the source term $\sigma$ from Eq.~(\ref{a4}) with $u=u_{\rm N}$ reads
\begin{equation}
 \sigma\left(u_{\rm N},0\right) = \frac{g_{\rm e}}{r}\left(\frac{r_{\rm e}}{r}\right)^2\frac{1}{1+\left(\eta \sqrt{\zeta_1}\right)^n}\,\frac{\zeta_2}{\zeta_1}\; . \label{si1}
\end{equation}
The right-hand side of (\ref{si1}) -- the ratio $\zeta_2/\zeta_1$, in particular -- is well-behaved except at the critical point $P$, where it exhibits a singularity. As the iterations proceed, a critical phenomenon around $P$ develops, resulting eventually in an integrable singularity \citep[compare with similar discussion in][]{m1986a,p2020a,bm2023}. This phenomenon causes some loss of accuracy in our numerical solution for the potential, especially for sharp transition functions such as $\mu_{10}(x)$; see Fig.~\ref{fig_dg_global}. Nevertheless, we find this loss of accuracy is limited to a relatively small region near $P$, and we do not believe that its affects our main conclusions. 

%-----
The solution of the Poisson equation (\ref{e10}) is performed using spectral decomposition, which proceeds in two steps. First we represent the angular part of the potential in a series of Legendre polynomials, Eq.~(\ref{e12}). Next, we
develop all radial functions (the potential and source) using series of Chebyshev polynomials, such as
\begin{equation}
 \upsilon^{(n)}_\ell (r) = \sum_0^N \upsilon^{(n)}_{\ell k} T_k(\xi) \label{a5}
\end{equation}  
and
\begin{equation}
 S^{(n)}_\ell (r) = \sum_0^N s^{(n)}_{\ell k} T_k(\xi) \; ,\label{a6}
\end{equation}
(the relation between $\xi$ and $r$ is described in  Sec.~\ref{efe}); $N$ is the adopted maximum degree of the development. For radial zones of stretching factor $3$ (ratio between outer and inner boundary), we found $N\simeq 99$ is sufficient. The coefficients $s^{(n)}_{\ell k}$ were computed using Gaussian quadrature.

Using the spectral decomposition, the differential operator on the left-hand side of Eq.~(\ref{e12}) is represented with an upper triangular matrix ${\bf M}_\ell$, and the whole differential equation becomes a matrix/algebraic relation between the 
coefficients $\upsilon^{(n)}_{\ell k}$ and $s^{(n)}_{\ell k}$ \citep[useful details can be found in][]{gn2009}. The matrices ${\bf M}_\ell$ are singular, due to the additional degrees of freedom  expressed by an arbitrary solution of the homogeneous equation (\ref{e12}). This singularity
is constrained by the boundary conditions at $\xi=\pm 1$ of each domain where we impose continuity of the potential and its radial derivative. In the innermost radial domain, i.e. inside the Sun, we require $\partial u/\partial r=0$ at the origin $r=0$ (Eq.\ \ref{e8a}), and in the outermost radial zone we require $u=0$ at $r\rightarrow \infty$. In order to implement the set of all these boundary conditions, we use the tau method discussed by \citet[][Sec.~2.5.2]{gn2009}. After this operation
is done, the matrices ${\bf M}_\ell$ are no longer singular, and are easily invertible using singular value decomposition \citep[e.g.,][]{nr2007}.

In order to allow for the large dynamic range of the numerical values of the variables manipulated during the solution, and to ensure that rounding errors do not disturb the method, we conservatively use quadruple precision for representing all real quantities.
% FIG 1 %%%%%%%%%%%%%%%%%%%%%%%%%%%%%%%%%%%%%%%%%%%%%%%%%%%%%%%%%%%%%%%%%%%%%%%%%%%%%%%%%%%%%%%%%%%%%%%
\begin{figure*}[t!]
 \begin{center}
% \begin{tabular}{c}
%  \includegraphics[width=0.8\textwidth]{fAa.eps} \\
%  \includegraphics[width=0.8\textwidth]{fAb.eps} \\ 
% \end{tabular}
  \includegraphics[width=0.9\textwidth]{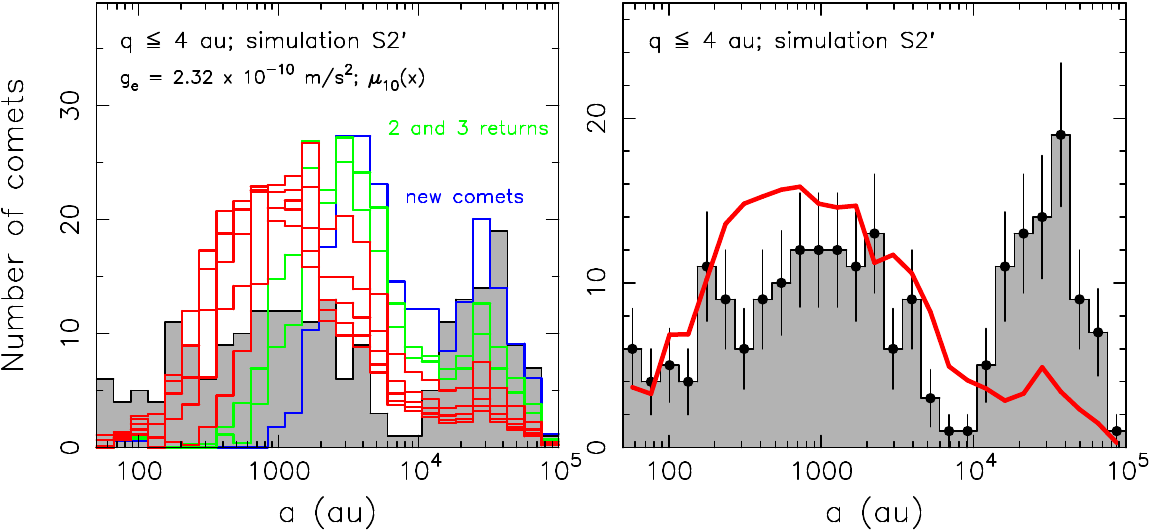} 
 \end{center}  
 \caption{The same as in Fig.~\ref{fig_lpc_ret_s2}, but now for the simulation ${\cal S}2^\prime$, which includes (i) a static configuration of the giant planets, the same as seen at present, and (ii) a limited population of $50,000$ test particles, initially located just exterior to Neptune. Here again the Oort peak, consisting of LPCs with semimajor axes $\geq 10,000$~au, is too small compared to the population of LPCs with semimajor axes $\leq 10,000$~au.}
 \label{fig_lpc_ret_s2p}
\end{figure*}
%%%%%%%%%%%%%%%%%%%%%%%%%%%%%%%%%%%%%%%%%%%%%%%%%%%%%%%%%%%%%%%%%%%%%%%%%%%%%%%%%%%%%%%%%%%%%%%%%%%%%%%
% FIG 1 %%%%%%%%%%%%%%%%%%%%%%%%%%%%%%%%%%%%%%%%%%%%%%%%%%%%%%%%%%%%%%%%%%%%%%%%%%%%%%%%%%%%%%%%%%%%%%%
\begin{figure}[t!]
 \begin{center}
  \includegraphics[width=0.47\textwidth]{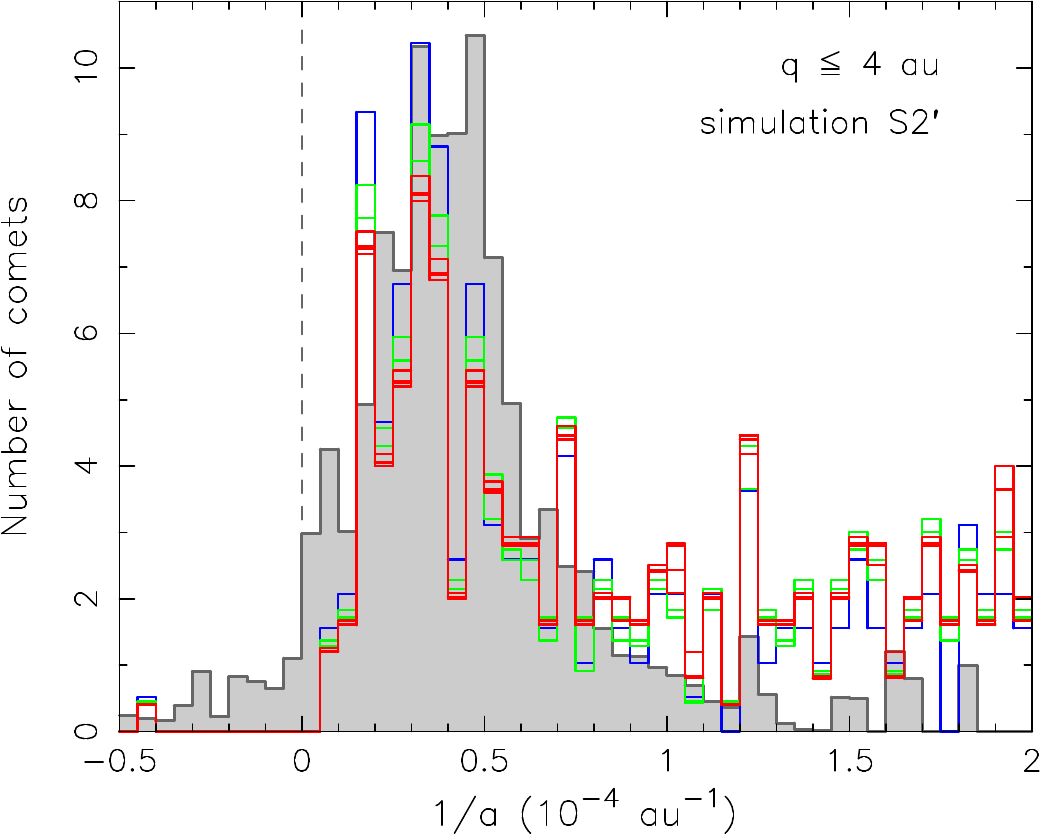} 
 \end{center}  
 \caption{The same as Fig.~\ref{fig_lpc_oort_s2}, but now for the simulation ${\cal S}2^\prime$.}
 \label{fig_lpc_oort_s2p}
\end{figure}
%%%%%%%%%%%%%%%%%%%%%%%%%%%%%%%%%%%%%%%%%%%%%%%%%%%%%%%%%%%%%%%%%%%%%%%%%%%%%%%%%%%%%%%%%%%%%%%%%%%%%%%

\bigskip

\section{Simulation with MOND and no planetary migration}\label{jobs2p}

The core set of simulations discussed in Sec.~\ref{res} was designed to test MOND within a scenario that conforms to our best understanding of the evolution of the solar system. This includes the assumptions that the source of the LPCs was a planetesimal disk initially exterior to the giant planets, which was dispersed when Uranus and Neptune migrated outward. While many lines of evidence indicate that planetary migration must have happened early in the history of the solar system \citep[e.g.][]{nes2018}, the details of the migration history and the architecture of the planetesimal disk are still debated. Therefore, one may ask whether
our conclusions about MOND depend on a particular set of initial conditions for the planetesimal disk and a particular model of planetary migration. Here we test this issue. Instead of considering variants of the planetary migration history we simply assume that all of the giant planets are fixed on their present orbits, and place the comets initially in a narrow range of semimajor axes just outside Neptune. This is our model ${\cal S}2^\prime$.

The giant-planet heliocentric orbits were taken from JPL ephemerides at epoch J2000.0. Since the terrestrial planets were not included, we performed a barycentric correction and added their composite mass to the Sun. We used $50,000$ massless particles (a.k.a. comets or planetesimals) placed initially in a compact and dynamically cold disk just exterior to Neptune -- this is much smaller than the population of $10^6$ particles used in our core simulations, but large enough for our purposes here. Their semimajor axes were uniformly spread between $31$ and $34$~au, the eccentricities and inclinations were drawn from a Rayleigh distribution with scale parameters $0.05$ and $2^\circ$, and all other angular variables were distributed uniformly random between $0^\circ$ and $360^\circ$. We then followed the system for $4.5$~Gyr using the same methods as for our other simulations (Sec.~\ref{met}). The MOND theory was the same as in simulation ${\cal S}2$ , that is, we used the sharp transition function $\mu_{10}(x)$ and a external Galactic field with acceleration $g_{\rm e} = 2.32\times 10^{-10}$
m~s$^{-2}$.

We found that the results are nearly identical to those of simulation ${\cal S}2$ (Sec.~\ref{res}) indicating that the details of the initial distribution of planetesimals and the planetary migration history do not strongly affect the currently observed population of LPCs. We illustrate this conclusion by showing the distribution of LPC binding energies for comets with perihelia $q\leq 4$~au (Figs.~\ref{fig_lpc_ret_s2p} and \ref{fig_lpc_oort_s2p}). These figures are to be compared to  Figs.~\ref{fig_lpc_ret_s2} and \ref{fig_lpc_oort_s2}.
While the details are slightly different, and the noise in simulation ${\cal S}2'$ is larger because of the smaller population of test particles, the principal message is the same. We conclude that the mismatch between the orbital parameters of the observed LPCs and the predictions of MOND cannot be ascribed to particular assumptions about the initial spatial distribution of the planetesimal disk or the planetary migration history.
% FIG 1 %%%%%%%%%%%%%%%%%%%%%%%%%%%%%%%%%%%%%%%%%%%%%%%%%%%%%%%%%%%%%%%%%%%%%%%%%%%%%%%%%%%%%%%%%%%%%%%
\begin{figure*}[t!]
 \begin{center}
 \begin{tabular}{c}
  \includegraphics[width=0.45\textwidth]{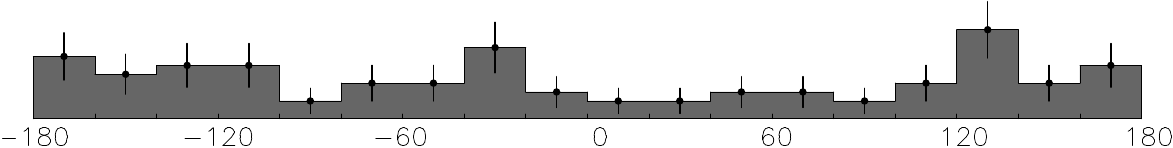} \\
  \includegraphics[width=0.7\textwidth]{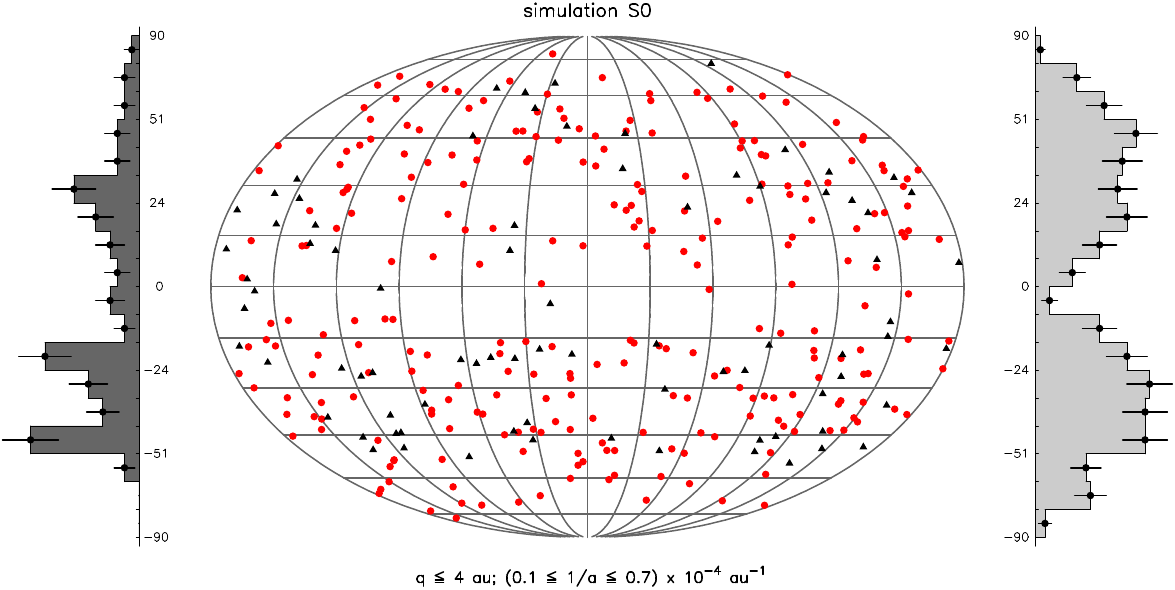} \\
  \includegraphics[width=0.7\textwidth]{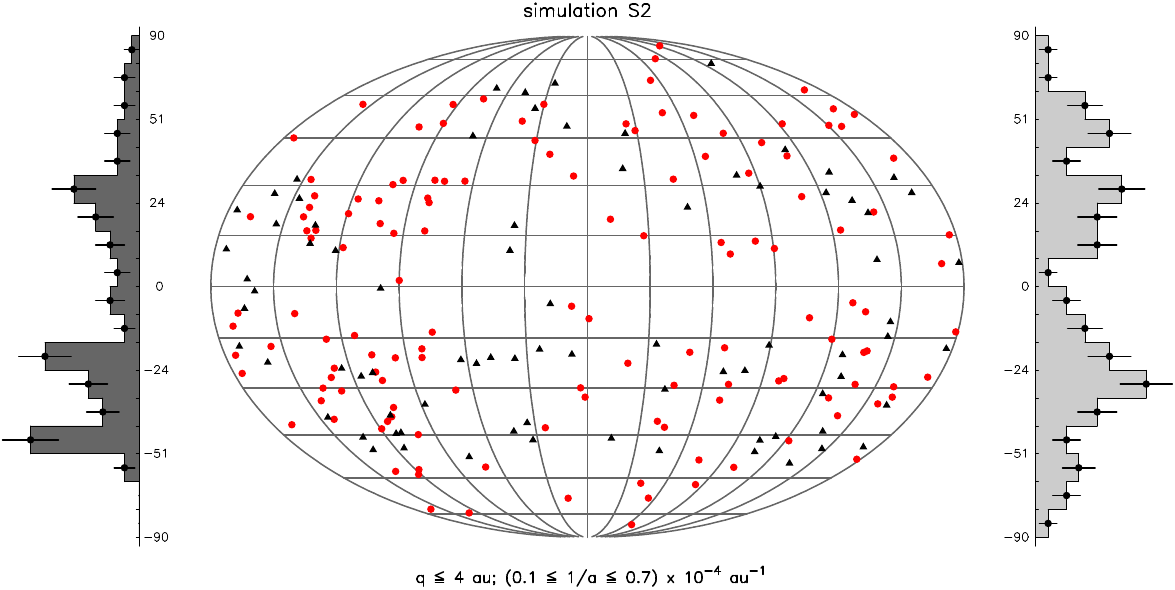} \\ 
  \includegraphics[width=0.7\textwidth]{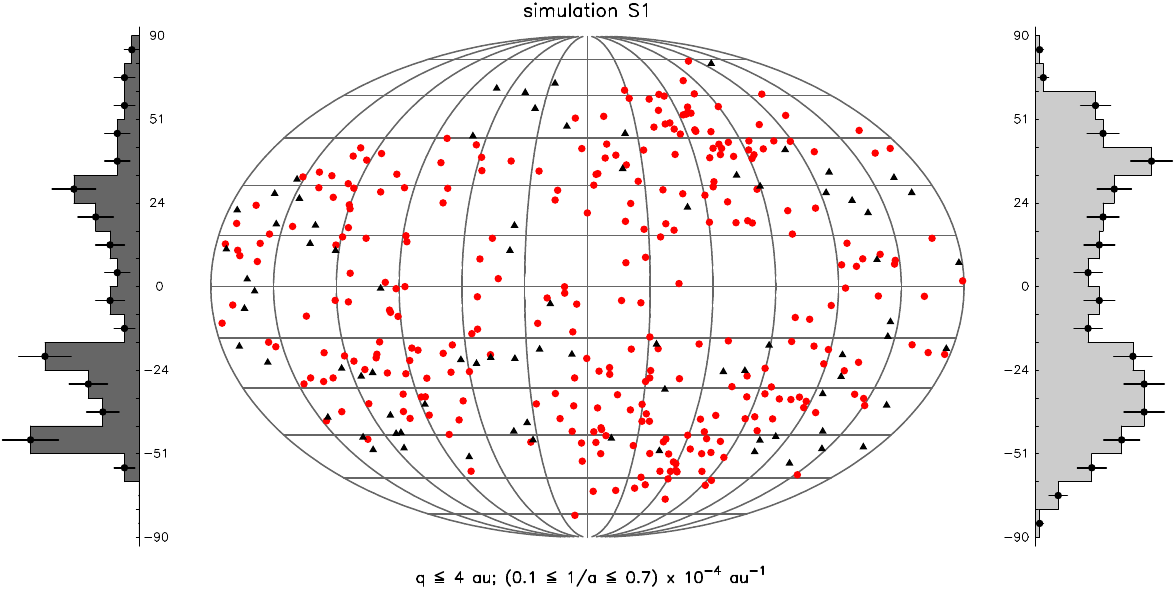} \\ 
 \end{tabular}
 \end{center}
 \caption{The aphelion directions of nearly parabolic LPCs in our simulations ${\cal S}0$ (top), ${\cal S}2$ (middle), and ${\cal S}1$ (bottom) are shown by red symbols in the Galactic coordinate system. This is a Mollweide projection, in which the direction to the Galactic
 center is at the origin and the north Galactic pole is at the top. The plot shows LPCs from our simulations with: (i) perihelion distance $q\leq 4$~au, (ii) inverse value of the original semimajor axis $a$ in the range $0.1\leq 10^4/a\leq 0.7$~au$^{-1}$, (iii) first visit to the solar system (``new comet''). The data are collected from the last $500$~Myr of our simulations. The black triangles show the aphelion direction for 81 class-1 LPCs from the Warsaw catalog using similar selection criteria (see text). The histograms on
 both sides of the figures show the latitude distribution of the aphelia, for observed LPCs on the left, in black, and for the simulated data on the right, in gray. The histogram at the top shows the longitude distribution of the observed LPCs. The bars at the bin center indicate a formal Poissonian uncertainty of the value.}
 \label{fig_lpc_aph}
\end{figure*}
%%%%%%%%%%%%%%%%%%%%%%%%%%%%%%%%%%%%%%%%%%%%%%%%%%%%%%%%%%%%%%%%%%%%%%%%%%%%%%%%%%%%%%%%%%%%%%%%%%%%%%%

\bigskip

\section{Distribution of aphelion directions of nearly parabolic LPCs}\label{aph}

Finally, we briefly analyze the distribution of aphelion directions for the original\footnote{As usual, the ``original'' orbital elements are the barycentric orbital elements that the comets had at a heliocentric distance of $250$~au, before the orbits are influenced by planetary perturbations. Note that the aphelion direction is much less sensitive to planetary perturbations than the semimajor axis.} orbits of
nearly parabolic LPCs. We use this subclass of LPCs as their extremely large aphelion distances make them vulnerable to exterior perturbations. While encounters with individual stars, if strong enough (i.e., high-mass star and/or small impact parameter and/or slow relative velocity), naturally leave their imprint in the LPC population over a few Myr interval \citep[e.g.,][]{fb2014,fb2015}, some authors argued that there might also be persistent anisotropies. For instance, \citet{mw1999} and \citet{mw2011} claimed to identify an anomalous concentration of LPC aphelia along a great circle located at Galactic longitudes $-45^\circ$ and $135^\circ$, and that this concentration was evidence for a distant planetary companion of a few Jupiter masses, residing in the Oort cloud. While it is not our intention to discuss in detail such scenarios, we bring up his topic here because \citet{pk2017} suggested that a similar effect may be produced by MOND itself. 

Here we use the simulations analyzed in Sec.~\ref{res} and describe the aphelion distributions they predict. Similar to the study of the semimajor axis distribution, we compare our results with data in the Warsaw catalog of nearly parabolic LPCs \citep{warsaw2020}. In order to minimize contamination by uncertain orbits and to focus on the largest orbits, we select: (i) class~1 solutions, (ii) with perihelia $q\leq 4$~au, and (iii) with binding energies $1/a$ in the range $0.1\leq 10^4/a \leq 0.7$~au$^{-1}$ (these represent the bulk of the Oort peak). We find 81 comets that satisfy these criteria in the Warsaw catalog. The distribution of aphelion directions in Galactic coordinates is shown by black triangles in Fig.~\ref{fig_lpc_aph}. Without conducting a statistical analysis, we believe that the evidence for a concentration about the great circle discussed in \citet{mw1999} (a polar circle along longitudes $135^\circ$ and $315^\circ$) is rather weak: the maxima in the corresponding longitude bins, shown in the histogram at the top of the figure, are not more than 1$\sigma$ above the level set by a uniform distribution. However, here we focus more on comparison of our simulations.

The upper panel on Fig.~\ref{fig_lpc_aph} shows results from the reference model ${\cal S}0$ which contains only Newtonian perturbations (red circles). The longitude distribution is fairly uniform as expected. There is, however, a clear latitudinal variation with maxima at $\pm 45^\circ$ galactic latitude (see the gray histogram on the right). A similar variation in seen in the observational data (black histogram on the left), although the signal is biased by the uneven geographical distribution of comet discoveries and the difficulty of finding comets near the Galactic plane. This is evidence for the substantial role of the smooth Galactic tide in injecting new comets from the Oort cloud to the inner solar system \citep[e.g.,][]{del1987,rickman2010}. The remaining panels of Fig.~\ref{fig_lpc_aph} show the results from simulations  ${\cal S}2$ (middle) and ${\cal S}1$ (bottom). These are based on MOND\-ian dynamics, with two forms of the transition function $\mu(x)$. While there are some longitudinal variations in the distribution, especially in the case of the ${\cal S}1$ simulation that uses the gradual $\mu_2(x)$ transition function, neither simulation shows the level of aphelion anisotropy that was described in \citet{pk2017}. We believe the methodology in that paper was too simplified to realistically predict the angular distribution of the observed comets.

%\listofchanges
\end{document}